\documentclass[useAMS, usenatbib,letterpaper,10pt]{mn2e}
\title[eROSITA: forecasts]{The X-ray cluster survey with $\eRO$: \\forecasts for cosmology, cluster physics and primordial non-Gaussianity}

\author[Pillepich et al.] {Annalisa Pillepich\thanks{annalisa.pillepich@ucolick.org}$^1$$^,$$^2$,
Cristiano Porciani,$^3$ and Thomas H. Reiprich,$^3$\\
$^1${Institute for Astronomy, ETH Zurich, 8093 Zurich, Switzerland}\\
$^2${UCO/Lick Observatory, University of California, 95064 Santa Cruz, USA}\\
$^3${Argelander-Institut f\"ur Astronomie, Auf dem H\"ugel 71, D-53121 Bonn,Germany}
}

\usepackage[bookmarks,bookmarksnumbered,colorlinks=true, citecolor=blue, linkcolor=black]{hyperref}
\usepackage{amsfonts}
\usepackage{amsmath}
\usepackage{amssymb}
\usepackage{color}
\usepackage[dvips]{graphicx}
\usepackage{verbatim}
\usepackage[table]{xcolor}
\usepackage{times}

\def \BE{\begin{equation}}
\def \EE{\end{equation}}	
\def \BC{\begin{center}}
\def \EC{\end{center}}
\def \BEA{\begin{eqnarray}}
\def \EEA{\end{eqnarray}}
\def \LN{\mathrm{ln}}
\def \LOG{\mathrm{log}_{10}}

\def \eRO{\it eROSITA}

\def \HI{ h^{-1}}
\def \MSUN{M_{\odot}}

\def \LUMUN{\rm erg ~ s^{-1}}
\def \FLUXUN{\rm erg ~ s^{-1} ~ cm^{-2}}

\def \FNL{f_{\rm NL}}
\def \FNLL{f_{\rm NL}^{\rm local}}
\def \SIGMA8{\sigma_{8}}
\def \OM{\Omega_{\rm m}}
\def \OB{\Omega_{\rm b}}
\def \OL{\Omega_{\rm \Lambda}}

\def \SG{\sigma}
\def \SGI{\sigma^{-1}}
\def \SKEW{S_3}

\def \DELTAC{\delta_{\rm c}}

\def \BGAUSS{b_{\rm G}}

\def \MCINQ{M_{500}}

\def \MDELTA{M_{\Delta}}
\def \DELTAT{{\Delta}_{b}}

\def \ERGS{\mathrm{erg ~ s^{-1}}}
\def \ALPHALM{\alpha_{\rm LM}}
\def \BETALM{\beta_{\rm LM}}
\def \GAMMALM{\gamma_{\rm LM}}
\def \SIGMALM{\sigma_{\rm{LM}}}
\def \ALPHATM{\alpha_{\rm TM}}
\def \BETATM{\beta_{\rm TM}}
\def \SIGMATM{\sigma_{\rm{TM}}}
\def \RHOLT{\rho_{\rm LT}}

\def \COV{\boldsymbol{\Sigma}}
\def \ICOV{\boldsymbol{\Sigma}^{-1}}
\def \MUDATA{\boldsymbol{\mu}}
\def \ERR{\sigma_{\alpha}}
\def \FCOUNTS{F^{\rm counts}}
\def \FCLUST{F^{\rm clust}}
\def \COVCOUNTS{\Sigma^{\rm counts}}

\begin{document}

\maketitle

\begin{abstract}
Starting in late 2013, the $\eRO$ telescope will survey the X-ray sky 
with unprecedented sensitivity. 
Assuming a detection limit of 50 photons  in the (0.5-2.0) keV energy band with a typical exposure time of 
1.6 ks, we predict that
$\eRO$ will detect $\sim 9.3 \times 10^4$ clusters of galaxies
more massive than $5 \times 10^{13} \HI \MSUN $, with the currently planned all-sky survey.
Their median redshift will be $z\simeq 0.35$. 
%and the cluster sample will extend up to $z\sim 1.5$.
%
We perform a Fisher-matrix analysis to forecast the constraining power
of $\eRO$ on the $\Lambda$CDM cosmology and, simultaneously, on
the X-ray scaling relations for galaxy clusters.
Special attention is devoted to the possibility of detecting
primordial non-Gaussianity.
% of the local,
%equilateral and orthogonal types.
%
We consider two experimental probes: the number counts 
and the angular clustering of a photon-count limited sample of clusters.
We discuss how the cluster sample should be split to optimize the 
analysis and we show that redshift information of the individual clusters is vital
to break the strong degeneracies among the model parameters.
For example, performing 
a ``tomographic'' analysis based on photometric-redshift estimates
and combining 1- and 2-point statistics
will give marginal 1-$\sigma$ errors of
$\Delta \SIGMA8 \simeq 0.036$ and $\Delta \Omega_{\rm m} \simeq
0.012$ without priors%(where $\SIGMA8$ and $\Omega_{\rm m}$ parameterize the amplitude of the
%linear power spectrum of density perturbations 
%and the mean matter density, respectively)
, and improve the current estimates on the slope of the luminosity-mass
relation by a factor of 3.
Regarding primordial non-Gaussianity, $\eRO$ clusters alone will give 
$\Delta f_{\rm NL} \simeq 9, 36, 144$ for the local, orthogonal and equilateral 
model, respectively.
Measuring redshifts with spectroscopic accuracy would further tighten the constraints
by nearly 40 per cent (barring $f_{\rm NL}$ which displays smaller improvements).
Finally, combining $\eRO$ data with the analysis of temperature anisotropies in the
cosmic microwave background by the Planck satellite
should give sensational constraints on both
the cosmology and the properties of the intracluster medium.
\end{abstract}


\section{Introduction}
$\eRO$\footnote{Extended ROentgen Survey with an Imaging Telescope Array, http://www.mpe.mpg.de/erosita/} \citep{Predehl:2006,Predehl:2010} is the primary
science instrument onboard the Spectrum Roentgen-Gamma ({\it SRG})
satellite\footnote{http://hea.iki.rssi.ru/SRG/en/index.php},
a fully funded mission with a currently planned launch in late 2013.
$\eRO$ will perform an X-ray all-sky survey with a sensitivity
$\sim$30 times better than {\it ROSAT}. 
Assuming a detection limit of 50 photons in the (0.5-2.0) keV energy band and with an average exposure time of $\sim 1.6$ ks, we expect that $\eRO$ will be able to detect  {\it all} clusters of galaxies in the observable Universe with masses higher than  $\sim 3 \times 10^{14} \HI \MSUN $.
This makes it an ideal probe of cosmology.

The evolution in the number density of massive galaxy clusters 
as well as their clustering properties strongly depend on the cosmological
parameters. Cluster number counts trace the normalization and the growth of 
linear density perturbations \citep[e.g][]{Allen:2011}.  
The mass function of clusters detected in X-rays  by {\it ROSAT}
and re-observed with {\it Chandra} has been recently used 
to delve into the mystery of cosmic acceleration and thus constrain the 
equation-of-state parameter of dark energy \citep{Mantz:2010a, Vikhlinin:2009c}.
These studies are based on samples containing a few tens of galaxy 
clusters. 

The advent of $\eRO$ will produce cluster catalogs with $\sim 10^5$ objects.
Photometric redshifts of the X-ray-detected clusters will become available 
thanks to a series of complementary multi-band optical surveys (e.g. PanSTARRS\footnote{Panoramic Survey Telescope \& Rapid Response Systems}, DES\footnote{Dark Energy Survey}, and LSST\footnote{Large Synoptic Survey Telescope}), while planned massive spectroscopic surveys are designed with $\eRO$ 
follow-up as science driver, e.g. 4MOST\footnote{4m Multi-Object Spectroscopic Telescope for ESO}.
All this will vastly improve upon current cosmological constraints. 
$\eRO$ will
likely be the first Stage IV dark-energy experiment \citep[as described in
the US Dark Energy Task Force report,][]{Albrecht:2006} to be realized.

At the same time, better data
will allow us to relax some model assumptions. An interesting example
is given by the statistical properties of the primordial density perturbations
that seeded structure formation in the Universe. They are often modelled with a
Gaussian random field. 
In fact,
the simplest inflationary models predict negligible - i.e. well below the current detection limit - levels of primordial non-Gaussianity. Nevertheless other theories for the generation of the primordial density fluctuations (both inflationary and not) can produce measurable levels of primordial non-Gaussianity: e.g. multiple-scalar-field scenarios, curvaton models, ghost inflation, topological defects \cite[see][and references therein for a complete review]{Bartolo:2004, Chen:2010}. Quantifying the primordial non-Gaussian signal has thus emerged as one of the most powerful tools to test and discriminate among competing scenarios for the generation of primordial density fluctuations. 

Primordial non-Gaussianity is expected to leave imprints on the large-scale structure of the evolved universe, ranging from the abundance of collapsed massive objects to their spatial clustering properties \cite[see][and references therein for an updated review]{Desjacques:2010a}. Clusters of galaxies as tracers of the underlying dark-matter (DM) haloes are thus an optimal target for the search of primordial non-Gaussianity. 

Regrettably, theoretical models provide robust predictions for the halo
abundance and clustering only in terms of DM masses. Therefore some knowledge
of the relation between cluster observables and the corresponding halo masses 
is required to constrain cosmological parameters.
In X-ray astronomy, cluster masses can be determined exploiting a series of scaling relations (luminosity--mass, temperature--mass, gas mass--total mass), whose shape and magnitude can be inferred by dedicated observations or by
invoking assumptions about the properties of the intracluster medium (ICM). 
Once the functional form of these relations is known, large cluster samples can
be used to derive self-consistent constraints on cosmological and scaling-relation parameters from the same data set \citep[the so-called self-calibration approach,][]{Majumdar:2003,Hu:2003}.
This should also provide some insight into the gas physics of the intracluster
medium. Note, however, that dedicated observations will be anyway necessary
to test the range of validity of the assumed scaling relations and provide
an independent test of the retrieved parameters.

The goal of this paper is to assess the power of $\eRO$ in the simultaneous
determination of cosmological and scaling-relation parameters, also accounting
for primordial non-Gaussianity. Forecasts are made using the Fisher 
information matrix for the measurements of the abundance and two-point spatial clustering of X-ray selected clusters.
A series of papers have recently addressed similar topics.
\cite{Oguri:2009} and \cite{Cunha:2010} discussed optical
surveys as the HSC\footnote{Hyper Suprime-Cam on Subaru telescope}, DES, and LSST. Similarly, \cite{Sartoris:2010} focussed on the envisaged X-ray satellite
WFXT\footnote{Wide Field X-ray Telescope}.
%different Fisher matrix approaches are there utilized, whether accounting for 2D or 3D clustering information or performing a  counts-in-cells experiment including the full covariance of cluster counts due to the large-scale structure clustering. 
More qualitative analyses have been carried out by \cite{Fedeli:2009}, \cite{Roncarelli:2010}, and \cite{Fedeli:2011} for the $\eRO$, the South-Pole-Telescope, and the Euclid surveys, yet not providing detailed forecasts, or with no assessment of the effect of the uncertainties in the scaling relations, or only considering a small subset of cosmological parameters. 
Our work differentiates from the previous ones in the sense that it is meticulously tailored around $\eRO$ and the actual way observations will be taken. 
Galaxy-cluster number counts and power spectra are calculated in terms of the 
raw photon counts that will be detected at the telescope.
This is done by adopting observationally motivated X-ray scaling relations, 
taking into account the spectral energy distribution of the ICM emission, 
the photoelectric absorption suffered by the photons along the line-of-sight, 
and the expected telescope response.

The plan of the paper is as follows. In Section \ref{SEC:DEFINITIONS}, we explain how we compute the expected abundances and projected angular clustering of DM haloes in the
presence of primordial non-Gaussianity and in Section \ref{SEC:FIDUCIAL} we define our fiducial model. 
In Section \ref{SEC:CONVERSIONS}, we focus on the transformation between actual halo masses and
observed photon counts.
The $\eRO$ cluster survey is described in detail in Section \ref{SEC:EROSITA}, while the experiments
we consider are discussed in Section \ref{SEC:OBSERVABLES}.
In Section 7, we describe how we compute the Fisher information matrix while, in Section 8, we derive the constraints on the cosmological and scaling-relation parameters for an all-sky survey with $\eRO$, giving special attention to the primordial non-Gaussianity parameter of the local type, $\FNLL$. 
We consider two experiments: the measurement of cluster abundances and of the
corresponding angular power spectrum. For both probes,
we distinguish the cases in which 
redshift information on the individual clusters is available or not.
Eventually we combine the different $\eRO$ measurements and 
also discuss how the model constraints are consolidated by using the (currently forecasted) results
of the Planck satellite as a prior probability distribution.
Table \ref{TAB:ERRORS} summarizes our main findings.
We investigate the effects of modifying the survey strategies and changing the observational cuts in Section 9, while in Section 10 we extend our analysis to other interesting models of primordial non-Gaussianity, complementary to the local type: the so-called orthogonal and equilateral shapes.
A thorough discussion of the assumptions  in our analysis is presented in Section 11 where we quantify the effects of changing the fiducial model and adding extra parameters.
Our conclusions are summarized in Section 12.

\section{Mass function and bias of dark-matter haloes}
\label{SEC:DEFINITIONS}
\subsection{Halo definition}
The densest regions of the cosmic matter distribution are commonly dubbed
``dark-matter haloes''. Their outer boundaries (and thus their masses) are 
somewhat arbitrary depending on the adopted definition. In numerical 
simulations haloes are most commonly identified using either the Friends-Of-Friends (FOF) or the Spherical-Overdensity (SO) algorithms \citep{Einasto:1984,  Davis:1985}. 
While the FOF method approximately tracks isodensity contours \citep{Efstathiou:1985, More:2011}, the SO one fixes the mean density of the structure within a sphere grown around the halo centre. We define $M_{\Delta}$ to be the mass enclosed in a sphere whose mean density is $\Delta$-times the critical density of the universe at the time the halo is considered: 
\BE
\MDELTA = \frac{4}{3} \pi (\Delta \rho_{\rm crit}) R^3_{\Delta}.
\label{EQ:MDELTA}
\EE
This definition is widespread among observers and
slightly differs from what is commonly adopted by the numerical community
where the background density $\bar{\rho}_{\rm {m}} $ is used instead of the critical one.
Thus, for a given population of SO haloes, the values that the parameter
$\Delta$ assumes in the two approaches differ by a factor $\OM(z) = \bar{\rho}_{\rm m}(z)/\rho_{\rm crit}(z)$.

From an observational point of view, it is standard to define cluster masses adopting the SO algorithm. In fact, for X-ray observations of galaxy clusters, better measurements of the gas mass and temperature can be achieved in regions with
high-density contrast where structures are much brighter and 
relatively relaxed with respect to the outer regions (and also the assumption
of hydrostatic equilibrium is more accurate). 
On the other hand, observational results should be compared with numerical
simulations where the finite resolution and force softening might create problems for 
very high values of $\Delta$.
For these reasons, the best compromise is to define the DM haloes in terms of $\MCINQ$.

\subsection{Halo mass function}
The halo mass function, $dn/dM(M,z)$, 
gives the halo abundance per unit volume and
per unit mass as a function of redshift. It is conveniently described in terms
of the function $f$ defined as
\BE
\frac{dn}{dM}(M,z) = f(\SG,z)~ \frac{\bar{\rho}_{\rm {m,0}}}{M}\frac{{d}
~{\rm ln}
[\SG^{-1}(M,z)]}{{d}M}\;.
\label{EQ:MF}
\EE
where $\bar{\rho}_{\rm{
m,0}}$ denotes the mean background matter density today, 
and $\SG^2(M,z)$ is the variance of the smoothed linear density field, $\langle \delta^2_M(z) \rangle$:
\BE
\SG^2(M,z) =  \frac{1}{2 \pi^2} \int^{\infty}_{0} k^2 ~P_{\rm lin}(k,z)~
W^2(k,M)\, {d}k,
\label{EQ:SIGMAOFM}
\EE
with $P_{\rm lin}(k,z)$ the corresponding power spectrum and
$W(k,M)$ a window function with mass resolution $M$; 
%or $R$, $M = \frac{4}{3} \pi \bar{\rho}_{\rm {m}} R^3 $; 
here we use a top-hat filter in real space, which in Fourier space reads
\BE
W(k,M) = 	3 ~\frac{\sin(x)-x \cos(x)}{x^3},
\EE
where $x \equiv kR$ and $M = \frac{4}{3} \pi \bar{\rho}_{\rm {m}} R^3 $.
The function $f$ is calibrated against high-resolution numerical simulations
and different parameterization are available in the literature.
Following \cite{Tinker:2008}, we assume that the function $f$ weakly depends on $z$ and $\DELTAT$, where $\DELTAT$ defines the mean overdensity of SO haloes as in Eq.~(\ref{EQ:MDELTA}) but wrt the evolving mean background density of the universe.
For Gaussian initial conditions, we thus write
\BE
f_{\rm T}(\SG,z,\DELTAT) = A\left[ \left(\frac{\SG}{b}
\right)^{-{a}}+1\right]\exp{\left(-\frac{c}{\SG^2} \right)},
\label{EQ:TINKERMF}
\EE
with
\BEA
A(\DELTAT,z) &=& A^{\rm TMF}_0(\DELTAT)(1 + z)^{-A^{\rm TMF}_z} \nonumber \\
a(\DELTAT,z) &=& a^{\rm TMF}_0(\DELTAT)(1 + z)^{-a^{\rm TMF}_z} \nonumber\\
b(\DELTAT,z) &=& b^{\rm TMF}_0(\DELTAT)(1 + z)^{-\alpha} \nonumber\\
c(\DELTAT,z) &=& c^{\rm TMF}_0(\DELTAT) \nonumber\\
\LOG \alpha(\DELTAT) &=& -\left[ \frac{0.75}{\LOG(\DELTAT/75)}\right]^{1.2}
\label{EQ:TINKERMFPARAM}
\EEA
where $A^{\rm TMF}_z=0.14$ and  $a^{\rm TMF}_z = 0.06$. The zero subscripts indicate the values of the parameters obtained at $z=0$ and listed in Table 2 of \cite{Tinker:2008}. Note that because of the different definition of SO-halo masses, an interpolation of the best-fit parameters of Table 2 in \cite{Tinker:2008} is required to compute these coefficients. For Gaussian initial conditions, 
this formula for the DM halo mass function reproduces to 5 per cent level of accuracy the abundance of SO haloes in N-body simulations for different choices of the standard cosmological parameters and for a 
wide range of masses and redshifts which encompasses what is probed by $\eRO$.
\subsection{Non-Gaussian corrections}
We consider here extensions to the standard cosmological model 
where the primordial curvature perturbations (more precisely the Bardeen's
potential) that generate the large-scale structure are not Gaussian.
In the synchronous gauge and for sub-horizon perturbations
in the mass density (or in the gravitational potential), 
these models are characterized by a non-vanishing bispectrum,
$B(k_1,k_2,k_3) \neq 0$.
The functional form of the bispectrum with respect to the triplet
of wavenumbers ${\bf K}=(k_1,k_2,k_3)$ describes the ``shape'' of primordial 
non-Gaussianity while its overall amplitude (at fixed ${\bf K}$) defines
the ``strength'' of the non-Gaussian signal. The latter is generally quantified
in terms of the non-linearity parameter $\FNL^{\rm i}$ where the index i refers
to a particular non-Gaussian shape.
As a reference case, we first focus on the ``local'' model for primordial
non-Gaussianity, where the linear gravitational potential can be written
as 
\BE
\phi({\bf x})=
g({\bf x})+\FNLL [g^2({\bf x})-\sigma^2_g ]+ \dots
\label{EQ:LOCALNG}
\EE
where $g$ is a Gaussian random field with zero mean and variance $\sigma^2_g$.
In this case, the bispectrum of linear density perturbations reads:
$B(k_1,k_2,k_3)\simeq 2\, \FNLL\, [P_{\rm lin}(k_1)~P_{\rm lin}(k_2)+
\ {\rm 2\ perms.}]$.
Note that we normalize $\FNLL$ by imposing the relation (\ref{EQ:LOCALNG}) 
at very high redshift which sometimes is referred to as the CMB normalization.
We postpone to Section \ref{SEC:NONLOCALFNL} the extension of our analysis to other non-Gaussian models, such as the equilateral and orthogonal shapes.

In general, for non-Gaussian initial conditions,
we compute the halo mass function through a multiplicative correction to the Gaussian one:
\BE
\left( \frac{dn}{dM}\right)_{\rm NG} =  \left(\frac{dn}{dM}\right)_{\rm{T}, M_{500}} R_{\rm NG}(M,z,\FNL),
\label{EQ:NGMF}
\EE
(here the subscript $T, M_{500}$ refers to the fact 
we use Eq. (\ref{EQ:TINKERMF}) and $\Delta=500$).
Among the possible choices for the non-Gaussian correction $R_{\rm NG}(M,z,\FNLL)$, we adopt the prescription by \cite{LoVerde:2008}, which is in rather
good agreement with numerical simulations 
\citep{Pillepich:2010, Giannantonio:2010a}:
\BEA
\nonumber
R_{\rm NG}(M,z,\FNLL) &=&  1 + \frac{1}{6}\frac{\SG^2}{\DELTAC} \left[ \SKEW \left(  \frac{\DELTAC^4}{\SG^4} -2  \frac{\DELTAC^2}{\SG^2} - 1 \right)  \right.\\
&+& \left. \frac{d\SKEW}{d \LN \SG} \left( \frac{\DELTAC^2}{\SG^2} - 1\right) \right]\;.
\label{EQ:NGMFLOVERDE}
\EEA
Here $\delta_c$ is the critical density contrast for halo collapse fixed to 1.686 for SO haloes \citep{Desjacques:2010a}, while $\SKEW$ denotes the standardized third central moment (skewness) 
of the smoothed density field, namely $\SKEW =  \langle \delta_M^3 \rangle/ \sigma^4$:
\BEA
\SKEW \sigma^4 = \frac{\FNLL}{(2 \pi^2)^2} & & \int dk_1 k_1^2 ~W(k_1,M) ~P_{\rm lin}(k_1) \nonumber \\
							           & & \int dk_2 k_2^2 ~W(k_2,M) ~P_{\rm lin}(k_2) \nonumber \\
							           & & \int^{1}_{-1} d\mu ~ W(k_{12},M) \left[ 1 + 2\frac{P_{\rm lin}(k_{12})}{P_{\rm lin}(k_2)}\right ]
\label{EQ:SKEW}
\EEA
where $k_{12}^2 = k_1^2 + k_2^2 + 2 \mu k_1 k_2$.

Note that we do not use the direct fitting formulae for the $\FNLL$-dependent 
mass function presented in \cite{Pillepich:2010} 
as they have been obtained for FOF haloes and there is no trivial mapping
between FOF and SO haloes.

\subsection{Halo bias}
We are not only interested in extracting cosmological information from measurements of the abundance of the X-ray clusters, but also from the angular clustering of the observed objects. From numerical and theoretical work, we know that the clustering of DM
haloes in which the clusters reside is biased relative to that of the underlying mass distribution
by an amount which depends on halo mass, redshift, and the
scale at which the clustering is considered \citep[see e.g.][]{Mo:1996,Catelan:1998,Smith:2007}. 
In general, the bias of DM haloes can be expressed as 
\BE
b(k,M,z) = \sqrt{\frac{P_{\rm hh}(k,M,z)}{P(k,z)}}\;,
\label{BHHDEF}
\EE
where $P_{\rm hh}$ is the power spectrum of the halo density field and $P(k,z)$ is the matter power spectrum. 
Within Gaussian scenarios, the halo bias on large scales ($k \lesssim 0.05~h\, {\rm Mpc}^{-1}$) depends on the halo mass and redshift but is independent of $k$ \citep[e.g.][]{Tinker:2010}. We compute this linear bias using the peak-background-split model \citep{Cole:1989, Mo:1996}:
\BE
\BGAUSS = 1-\frac{1}{\DELTAC} \frac{d \LN f}{d \LN \SGI}\;.
\label{EQ:PBSBIAS}
\EE
On the other hand, a strong $k$-dependence of the bias appears at smaller scales ($k \gtrsim 0.1 ~h\,{\rm Mpc}^{-1}$), e.g. \cite{Smith:2007}, \cite{Manera:2010}. 

\cite{Dalal:2008} have shown that 
primordial non-Gaussianity of the local type introduces a scale-dependent 
correction to the large-scale bias of massive DM haloes.
This is because large and small-scale density
fluctuations are not independent when $\FNLL\neq 0$.
We thus extend the expression for the halo bias to non-Gaussian scenarios following \cite{Giannantonio:2010a}:
\BE
b(k,M,z) = \BGAUSS + \FNLL ~(\BGAUSS - 1)~ \frac{\Gamma}{\alpha(k,z)}.
\label{EQ:NGBIAS}
\EE
which is in good agreement with N-body simulations as shown by
\cite{Giannantonio:2010a} and \cite{Desjacques:2009}. 
In Eq.~(\ref{EQ:NGBIAS}), the different factors read $\Gamma=3\,\delta_{\rm c}\,\Omega_{\rm m}\,H_0^2/c^2$ and $\alpha(k,z)=k^2\,T(k)\,D(z)\,g(0)/g(\infty)$, where $T(k)$ is the linear matter transfer function, $D(z)$ is the linear growth factor of the density perturbations (normalized to unity today), and $g(z)$ is the growth factor of the potential perturbations. Throughout this paper, we use the linear matter transfer function $T(k)$ computed using the {\sc LINGER} code by \cite{Bertschinger:2001}.\\
%We assume that the bias of DM haloes of a given mass $\MCINQ$ is a precise estimate of the bias of the X-ray clusters of mass $\MCINQ$.\\

\section{Fiducial Model}
\label{SEC:FIDUCIAL}
We consider a flat $\Lambda$CDM universe characterized by 5 standard
parameters (3 for the background -- $\Omega_{\rm m}$, $\Omega_{\rm b}$, 
and $h$ --  and 2 for the scalar perturbations -- $n_{\rm s}$ and $\sigma_8$).
On top of this we introduce the extra parameter $\FNLL$.
We list all the model parameters and the choice of their fiducial values in Table~\ref{TAB:PARAM}, where we distinguish among cosmological, cluster-physics (see Section \ref{SEC:CONVERSIONS}), and survey (see Section \ref{SEC:EROSITA}) parameters.
We assume Gaussian perturbations as our fiducial case, i.e. $\FNLL=0$. We choose the other fiducial values by adopting the best-fitting parameters
from the combination of 5-yr data from
the Wilkinson Microwave Anisotropy Probe (WMAP), 
baryonic acoustic oscillations (BAO), 
and supernovae Ia (SN), see \cite{Komatsu:2009}.

\begin{table*}
\footnotesize
\begin{center}
\begin{minipage}{180mm}
\caption{\label{TAB:PARAM} Model and survey parameters plus their fiducial values.
The parameters which are allowed to vary in the Fisher-matrix analysis are typed in boldface. Note that we assume a flat cosmology where $\Omega_\Lambda=
1-\Omega_{\rm m}$ throughout the statistical analysis.}
\begin{center}
\begin{tabular}{| c l c c l|}
\hline
Cosmological Parameter & Description & Fiducial Value & Current error \footnote{WMAP5+BAO+SN, for the Cosmology sector (68.3 per cent credibility interval (CI),
with the exception of $\FNLL$ for which the 95.4 CI per cent is indicated)} 
& Reference \\
\hline
$\FNLL$ & \textbf{Non-linearity Parameter (Local)}& 0 & $-9 \le \FNLL \le +111$ & \cite{Komatsu:2009}\\
$\SIGMA8$ & \textbf{Normalization of $P(k)$}	& 0.817 & $\pm$0.026 & \cite{Komatsu:2009} \\
$\OM$ & \textbf{Dark Matter Fraction}	& 0.279 & $\pm$0.0058 & \cite{Komatsu:2009}\\
$n_s$ & \textbf{Spectral index} & 0.96 	& $\pm$0.013 & \cite{Komatsu:2009} \\
$h$ 	& \textbf{Hubble Constant} & 0.701 & $\pm$0.013 & \cite{Komatsu:2009} \\
$ \OB$ & \textbf{Baryon Fraction}	& 0.0462	& $\pm$0.0015 & \cite{Komatsu:2009} \\
$\OL$ & Dark Energy Fraction	& 0.721 & $\pm$0.015 & \cite{Komatsu:2009} \\
$ w$ & Equation-of-State Parameter (constant)	& -1	& $-0.14 < 1+w < 0.12$	& \cite{Komatsu:2009} \\
\hline
\hline
X-ray Cluster Parameter  & Description & Fiducial Value & Current Error & Reference\\
\hline
$\ALPHALM$ & \textbf{LM relation: Slope} & 1.61 & $\pm$0.14 &  \cite{Vikhlinin:2009b} \\
$\GAMMALM$ & \textbf{LM relation: $z$-dependent Factor} & 1.85 & $\pm$0.42 &  \cite{Vikhlinin:2009b} \\ 
$\BETALM $& \textbf{LM relation: Normalization} & 101.483 & $\pm$0.085 &  \cite{Vikhlinin:2009b} \\
$\SIGMALM$ & \textbf{LM relation: Logarithmic Scatter} & 0.396 & $\pm$0.039 & \cite{Vikhlinin:2009b} \\ 
$\ALPHATM$ & TM relation: Slope & 0.65 & $\pm$0.03 &  \cite{Vikhlinin:2009b} \\
$\BETATM $& TM relation: Normalization & $3.02 \times10^{14} M_{\odot}h^{-1}$ & $\pm 0.11 \times10^{14} $ &  \cite{Vikhlinin:2009b} \\
$\SIGMATM$ & TM relation: Logarithmic Scatter & 0.119 & 0.03\footnote{from hydrodynamical simulations, systematic error encompassing variations for different sub-populations of clusters: relaxed or unrelaxed, at low or high redshifts} 
& \cite{Kravtsov:2006} \\ 
$\RHOLT$ & LT correlation coefficient & 0 & - & -\\ \
$Z_{\rm ICM}$ & Intracluster metallicity &$ 0.3 Z_{\odot}$& -& \cite{Anders:1989}\\
$N_{\rm H} $ & Hydrogen column density along los & $3 \times 10^{20}$ atom/cm$^2$& - & \cite{Kalberla:2005}\\
\hline
\hline
Survey Parameter  & Description & Fiducial Value & - & Reference\\
\hline
& X-ray Energy Band & 0.5-2.0 keV & - &-\\
%$ \rm {z_{min}-z_{max}} $ & Redshift Range & $0.01 \le z \le 1.5$ & - &-\\
$\eta_{\rm min}$&Minimum raw photon count & 50 & - &-\\
$M_{\rm min}$ & Minimum considered mass ($\MCINQ$) & $5\times10^{13} M_{\odot}h^{-1}$& -&-\\
$f_{\rm sky}$ & Sky coverage & 0.658 $\Rightarrow 27,145~ {\rm deg}^2$ \footnote{{\it all-sky} survey excising $\pm 20~ {\rm deg}$ around the Galactic plane}  & - &\cite{Predehl:2010}  \\
$T_{\rm exp}$ &Exposure Time & $1.6 \times 10^3$s (all-sky survey) & - & -\\
\hline
\end{tabular}
\end{center}
\end{minipage}
\end{center}
\end{table*}
\section{From halo mass to X-ray photon Counts}
\label{SEC:CONVERSIONS}
$\eRO$ will perform an X-ray all-sky survey in about four years starting in late 2013, from a L2 orbit.
The instrument will consist of seven identical Wolter-1 mirror modules of 358 mm of diameter, each containing 54 nested mirror shells, and of a fast frame-store pn-CCD detector. This will result in a 1 deg field of view, an effective area of 1500 cm$^2$ at 1.5 keV, and an averaged point spread function (PSF) of 25$\arcsec$-30$\arcsec$ half energy width (HEW; on-axis: 15$\arcsec$ HEW). It will operate in the broad energy range (0.5--10) keV.

In order to convert non-observable quantities such as $\MCINQ$ into the raw photons counts effectively collected in the detector  ($\eta$), a series of transformations have to be considered\footnote{Since the conversion between the energy flux and
the photon counts depends on the energy spectrum, to mimic more closely the
experimental reality we prefer not to adopt a flux limit 
in our analysis.}:
\[
\MCINQ \rightarrow [L_X ,T_X ] \rightarrow  {\rm photon\ counts} = \eta
\]
where $L_X$ denotes the X-ray luminosity and $T_X$ the average temperature of the intracluster gas.
Consequently, the mass function in Eq.~(\ref{EQ:MF}) has to be converted in a raw-count function:
\[
\frac{dn}{d\MCINQ} \rightarrow  \frac{dn}{d \eta}.
\]
In full generality, given two variables $X$ and $Y$, the conversion 
in the differential number counts can be performed as follows:
\BE
\frac{dn}{dY}(Y^*) = \int \frac{dn}{dX}(X)~ P(Y^*|X)~ dX,
\label{EQ:CONVSCATTER}
\EE
where $P(Y|X)$ denotes the conditional probability of getting $Y$ for given 
$X$. When the scatter in the $Y-X$ relation is negligible (i.e. the probability
$P$ is a Dirac-delta distribution), this reduces to
\BE
\frac{dn}{dY} = \frac{dn}{dX} \frac{dX}{dY}. 
\label{EQ:CONVNOSCATTER}
\EE
%
%Notice that taking into account the scatter in the mean relations reveals fundamental at the high-mass tails, where effects of primordial non-Gaussianity are important.\\
%
On the other hand, any function of the variable $X$ can be written
in terms of $Y$ as follows:
\BE
b_Y(Y^*) = \frac{ \int b(X)~ \frac{dn}{dX}(X)~ P(Y^*|X)~ dX}{\int  \frac{dn}{dX}(X)~ P(Y^*|X)~ dX},
\label{EQ:CONVSCATTERF}
\EE
which, for negligible scatter and when $Y = f(X)$, gives 
\BE
b_Y(Y^*) = b(\mathit{f}^{-1}(Y^*)).
\label{EQ:CONVNOSCATTERF}
\EE
In the following, we summarize the specific choices we adopted to express
the cluster number counts and bias in terms of the photon counts.

\subsection{Luminosity--mass relation}
To connect cluster masses to X-ray luminosities, we refer to the ${L_X}-M_{500}$ relation obtained by \cite{Vikhlinin:2009b} through the observation of two sets of Chandra galaxy clusters with median
 redshifts of about 0.05 and 0.5, in the (0.5--2.0) keV rest-frame band and with $L_X$ integrated within 2 Mpc. %, i.e. the total $L_X$. 
The luminosity--mass ($LM$) relation is conveniently
written such that the normalization is taken at the effective mean mass, 
to minimize correlations among parameters:
\BEA
\label{EQ:VIKHLININLM}
\mu_L \equiv \langle \LN L_X \rangle & = & \left[\BETALM + 1.5 (\SIGMALM^2-0.396^2)\right]\\
\nonumber
& + & \ALPHALM~ \LN (M_{500}/3.9 \times 10^{14} \MSUN) \\
\nonumber
& + & \GAMMALM~ \LN E(z) - 0.39 ~ \LN(h/0.72) \pm \SIGMALM,
\EEA
where $L_X$ is measured in erg s$^{-1}$ and $E(z)=H(z)/H_0$.
The slope $\ALPHALM$ reads $1.61 \pm 0.14$, the normalization  $\left[\BETALM + 1.5 (\SIGMALM^2-0.396^2)\right] = 101.483 \pm 0.085$, and the redshift-dependence factor $\GAMMALM = 1.85 \pm 0.42$ (see Table~\ref{TAB:PARAM}). The symbol $\SIGMALM$ on the right hand side indicates the observed scatter in $\LN L_X$ at fixed $M$. The nature of such a scatter is consistent with a log-normal distribution:
\BE
P(\LN L_X|M) = \frac{1}{\sqrt{2 \pi \SIGMALM^2}}~ \exp \left [-\frac{(\LN L_X - \mu_L)^2}{2 \SIGMALM^2} \right]
\label{EQ:LOGNORMAL}
\EE
where $\SIGMALM = 0.396 \pm 0.039$. This corresponds to a (symmetrized) relative error of $\sim $ 40 per cent.

Note that since \cite{Vikhlinin:2009b} determined $\MCINQ$ from the relation between $Y_X$ (the product of $T_X$ and gas mass) and mass, whose evolution was assumed to be self-similar based on results from numerical simulations, the redshift evolution of relation (\ref{EQ:VIKHLININLM}) has not been measured directly.
\subsection{From X-ray luminosity  to the number of detected photons}
\label{xtocounts}
Given $L_{\rm X}$ in the (0.5--2.0) keV rest-frame 
energy band and the redshift $z$
of the clusters, we can predict the number of photons detected by {\it
eROSITA} in the (0.5--2.0) keV observer-frame energy band. To do this,
we require further information: (i) the emitted spectrum, which
depends on the temperature and metal abundance of the intracluster
gas; (ii) the photoelectric absorption suffered by the X-ray photons
along their way; (iii) the $\eRO$ instrumental response and 
effective area; and (iv) the exposure time, $T_{\rm exp}$.

(i) We derive the cluster average X-ray temperature using the results by \cite{Vikhlinin:2009b}, where the relation $\MCINQ$--$T_X$ is parameterized with a power law (as expected if the temperature of the intracluster medium scales 
with the depth of the gravitational potential). 
To measure this relation, \cite{Vikhlinin:2009b} define the cluster temperature as the average spectral temperature integrated within a given radial range. Inverting the scaling-relation of their Table 3 with no pre-fixed slope, we write the temperature-mass relation ($TM$) as
\BEA
\mu_T \equiv \langle \LN T_X \rangle &=& \ALPHATM~ \LN (\MCINQ/ \BETATM) + \ALPHATM~ \LN E(z) \nonumber \\
&+& \LN(5 \rm{keV}) \pm \SIGMATM
\label{EQ:TMRELATION}
\EEA
where $T_X$ is measured in keV,  the slope $\ALPHATM = 0.65 \pm 0.03$ and $\BETATM = (3.02 \pm 0.11) \times 10^{14} \HI M_{\odot}$. The symbol $\SIGMATM$ is the scatter in $\LN T_X$ at fixed mass, and reads $\SIGMALM \simeq 0.119$ (see Table~\ref{TAB:PARAM}). This value is derived from numerical simulations \citep{Kravtsov:2006} and we assume a lognormal distribution also for the conditional probability of getting $T_X$ given $\MCINQ$, in analogy with Eq.~(\ref{EQ:LOGNORMAL}). The fit in 
Eq.~(\ref{EQ:TMRELATION}) holds for relaxed clusters only: to take into account unrelaxed morphologies, the estimated mass from Eq.~(\ref{EQ:TMRELATION}) should be multiplied by a factor of 1.17. We do not implement this correction.
For the metal abundance, we choose a value of 0.3 relative to solar
abundance \citep{Anders:1989}, which is typical
for nearby clusters, with little or no evolution to higher redshifts \citep{Tozzi:2003}.
We use the {\it apec} model \citep{Smith:2001} within {\sc xspec}
\citep[][version 12.5.1n]{Arnaud:1996} to model the emitted spectrum (thermal bremsstrahlung, recombination emission,
 as well as bound-bound transitions from collisionally ionized,
 optically thin plasma), for given
temperature, metal abundance, luminosity, and redshift.

(ii) The photoelectric absorption can be modeled using the hydrogen
column density along the line-of-sight, $N_{\rm H}$, and assuming the
associated abundance of helium and heavy elements. For this, we use the {\it phabs}
model within {\sc xspec} (with the default cross sections), in order to
self-consistently use the same metal abundances as above. The Galactic
$N_{\rm H}$ varies across the sky. We use recent data from 21 cm radio
measurements provided by Peter Kalberla (the Leiden-Argentine-Bonn, LAB,
survey; \citealt{Kalberla:2005}) and determine the $N_{\rm H}$
distribution, excising $\pm 20$ deg around the Galactic plane. We
find that $3\times 10^{20}$ atoms cm$^{-2}$ is a typical value, which we
adopt throughout.

\begin{figure}
\begin{center}
\includegraphics[width=8cm]{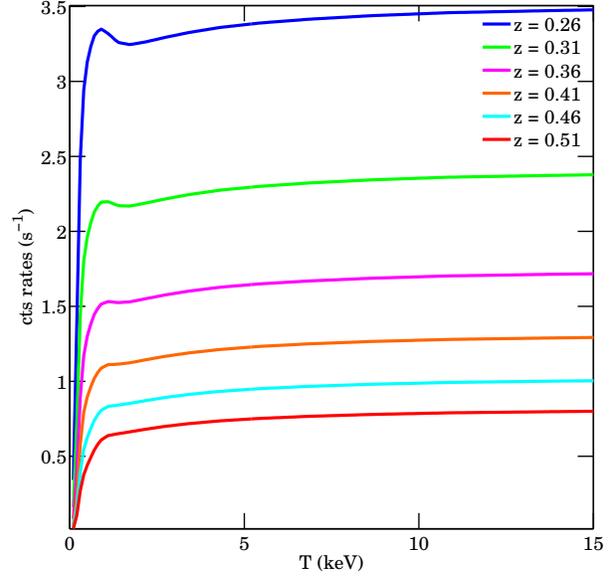}
\caption{\label{FIG:EROSITA} Characteristics of the $\eRO$ telescope: $\eRO$ response (rate of X-ray photons counted in the detector) 
in the (0.5--2.0) keV energy band (observer frame), for a source of luminosity $L_X = 10^{45} \ERGS$ in the (0.5--2.0) keV rest-frame band, at different redshifts, $z = 0.26,~ 0.31, ...$, from top to bottom.}
\end{center}
\end{figure}

(iii) We model the $\eRO$ response using the latest preliminary
matrices provided by Frank
Haberl\footnote{http://www.mpe.mpg.de/erosita/response/}. In
particular, we use the file erosita\_iv\_7telfov\_ff.rsp, which takes
into account the CCD quantum efficiency and filter transmission, as
well as the mirror area of all seven telescopes, averaged over the
field-of-view. The response of $\eRO$ in terms of count rates is shown in Fig.~\ref{FIG:EROSITA} for some characteristic redshifts 
and for a cluster luminosity of $L_X = 10^{45} \ERGS$.

(iv) The exact orbit and scanning strategy of $\eRO$ are currently under investigation. We consider two different versions of the expected four-year all-sky exposure map kindly provided by Jan Robrade. We exclude $\pm20$ deg around the Galactic
 plane.
 %, as well as the area of the Magellanic Clouds and the Virgo
 %cluster since these areas most likely will not be used for
 %extragalactic surveys (note that one of the deep exposure poles will
 %likely fall on the Large Magellanic Cloud). 
 We then calculate the
 average exposure in the remaining area, which results in 2.4 ks, for
 both exposure map versions. This exposure assumes 100 per cent efficiency.
 For a more realistic estimate, we need to multiply 2.4 ks by a factor smaller than 1. This
 factor is rather uncertain at this time. An optimistic value is 0.8,
 a pessimistic one $0.8\times0.67=0.54$ (the factor 0.67 is supposed to
 account for soft proton flares, which are expected at $\eRO$'s L2
 orbit). Therefore, we expect 1.3 ks $< T_{\rm exp} <$ 1.9 ks. We assume
 1.6 ks as a realistic estimate of the average exposure time, unless specifically mentioned
 otherwise.

Once $L_X$, $T_X$ and $z$ are known, we are thus able to compute the
expected photon counts $\bar{\eta}_{\rm fid}(L_X, T_X, z)$ within Xspec for
our fiducial reference cosmology. 
Since $\eta$ depends on the assumed cosmology through
 the luminosity distance $D_L$, we scale the photon counts according to
 $\bar{\eta} = \bar{\eta}_{\rm fid} D_{L, {\rm fid}}^2/D_L^2$, every time the assumed cosmology is varied.
We also account for the statistical uncertainty in the photon counts by
assuming that they follow Poisson statistics. In brief, the
probability distribution of the counts is
\BE
P(\eta|M,z) = \int dL_X ~dT_X ~ P[\eta|\bar{\eta}(L_X,T_X,z)] 
~ P(L_X,T_X | M,z)\;,
\label{EQ:PROB}
\EE
where $\eta$ is a Poisson variate of mean $\bar{\eta}$ and $P$ generically
denotes any probability density function. 
It is interesting to consider the possibility that there is a correlation between 
the X-ray temperature and luminosity for clusters of a given mass.
In fact this would reduce the scatter in the distribution of $\eta$ at fixed 
mass.
We write the joint conditional probability
$P(L_X,T_X | M, z)$ as a bivariate lognormal distribution
\BE
P(\mathbf{X}|M) = \frac{1}{2 \pi  |\mathbf{\Sigma}|^{1/2}} \exp \left[ -\frac{1}{2} (\mathbf{X}-\boldsymbol{\mu})^T \mathbf{\Sigma}^{-1} (\mathbf{X}-\boldsymbol{\mu})\right]
\EE
where the vector $\mathbf{X}$ and its covariance matrix read
\BE
\mathbf{X} = \left ( \begin{array}{c} 
\LN L_X \\
\LN T_X  \end{array}\right) \rm{ and  } ~\mathbf{\Sigma} = \left(   \begin{array}{cc} \SIGMALM^2 & \RHOLT \SIGMALM \SIGMATM \\  \RHOLT \SIGMALM \SIGMATM&\SIGMATM^2 \end{array}  \right)\;.
\EE
Here
$\boldsymbol{\mu}$ refers to the mean values of the scaling relations and 
$\RHOLT$ is the linear correlation coefficient between the residuals in
luminosity and temperature.
Theoretically it is expected that $L_X$ and $T_X$ may be highly
 correlated \citep{Stanek:2010} or anticorrelated \citep{Kravtsov:2006}; on the other hand, data for a set of clusters drawn from the {\it ROSAT} all-sky survey with {\it Chandra} follow-up show no evidence of a
correlation \citep{Mantz:2010b}.
\begin{figure*}
\begin{center}
\includegraphics[width=7.77cm]{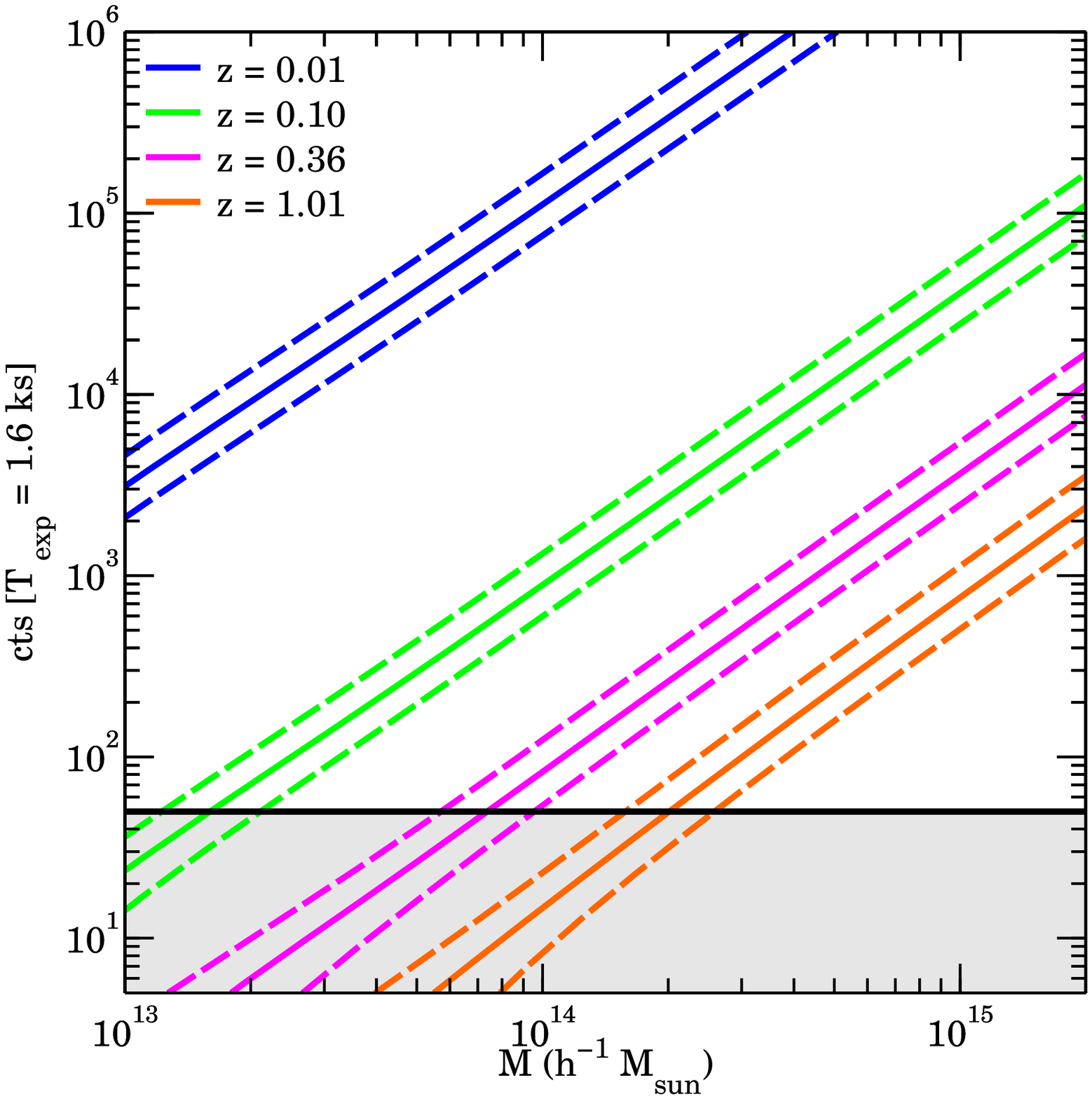}
\includegraphics[width=8cm]{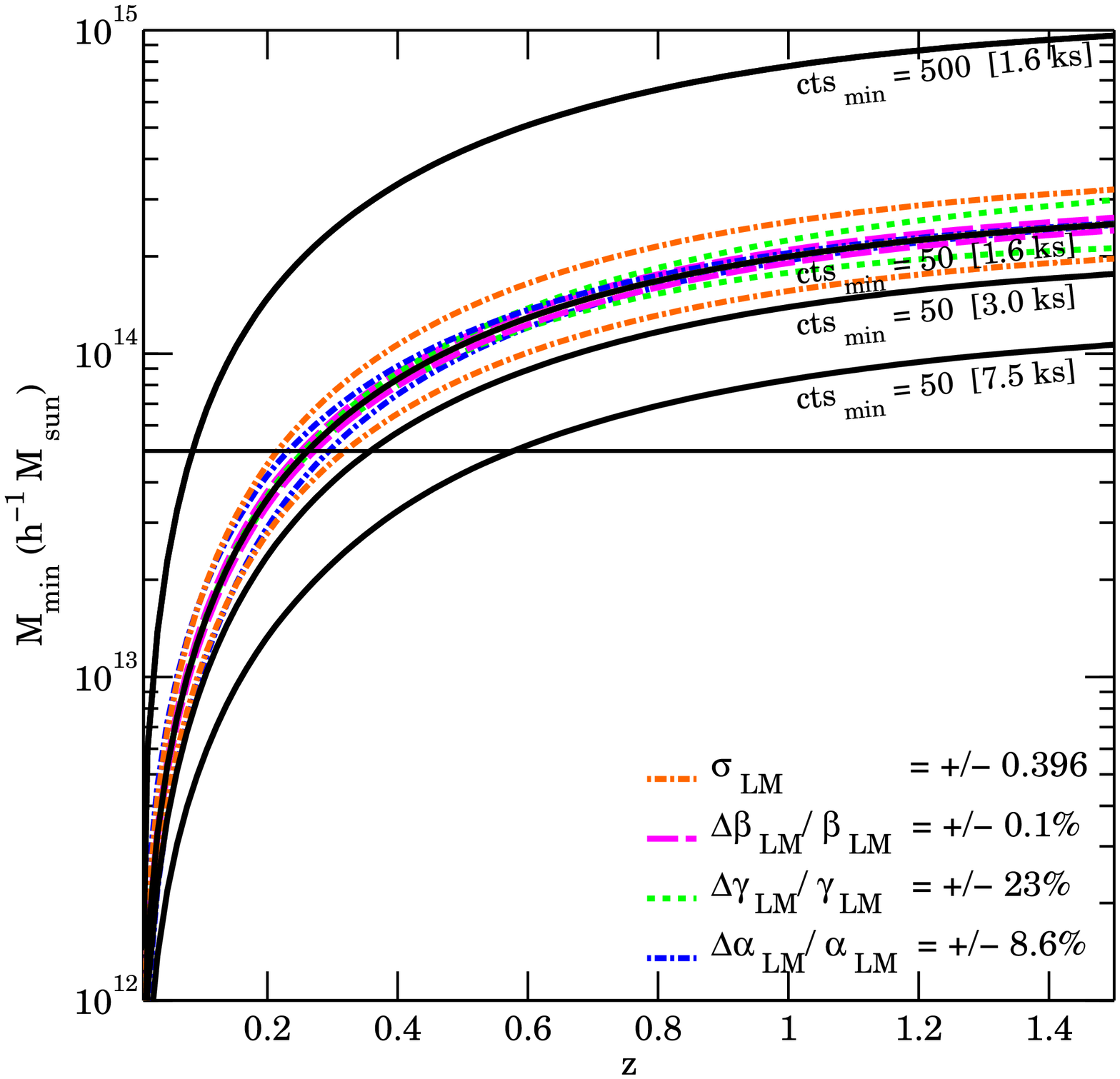}
\includegraphics[width=8.15cm]{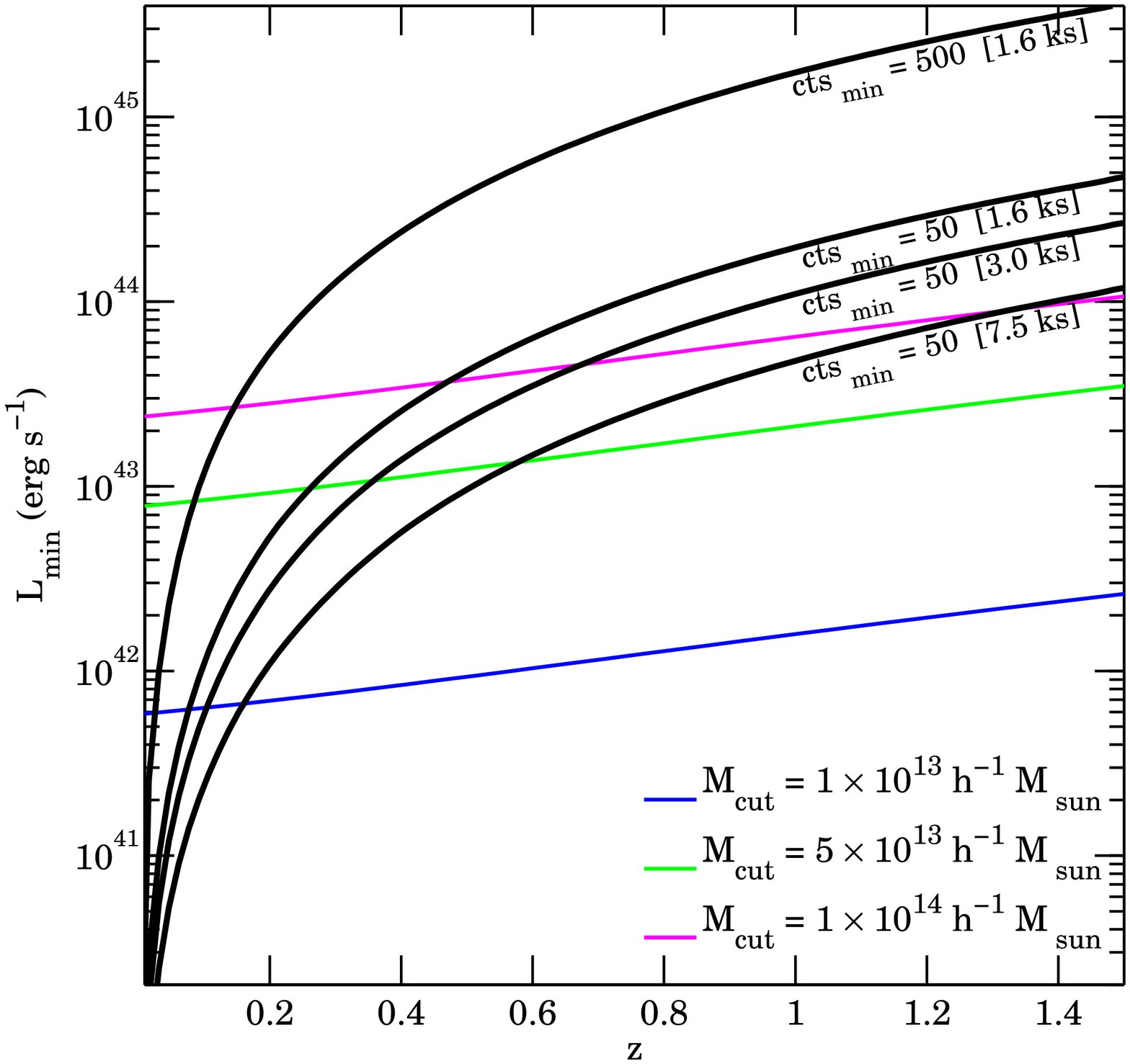}
\includegraphics[width=8cm]{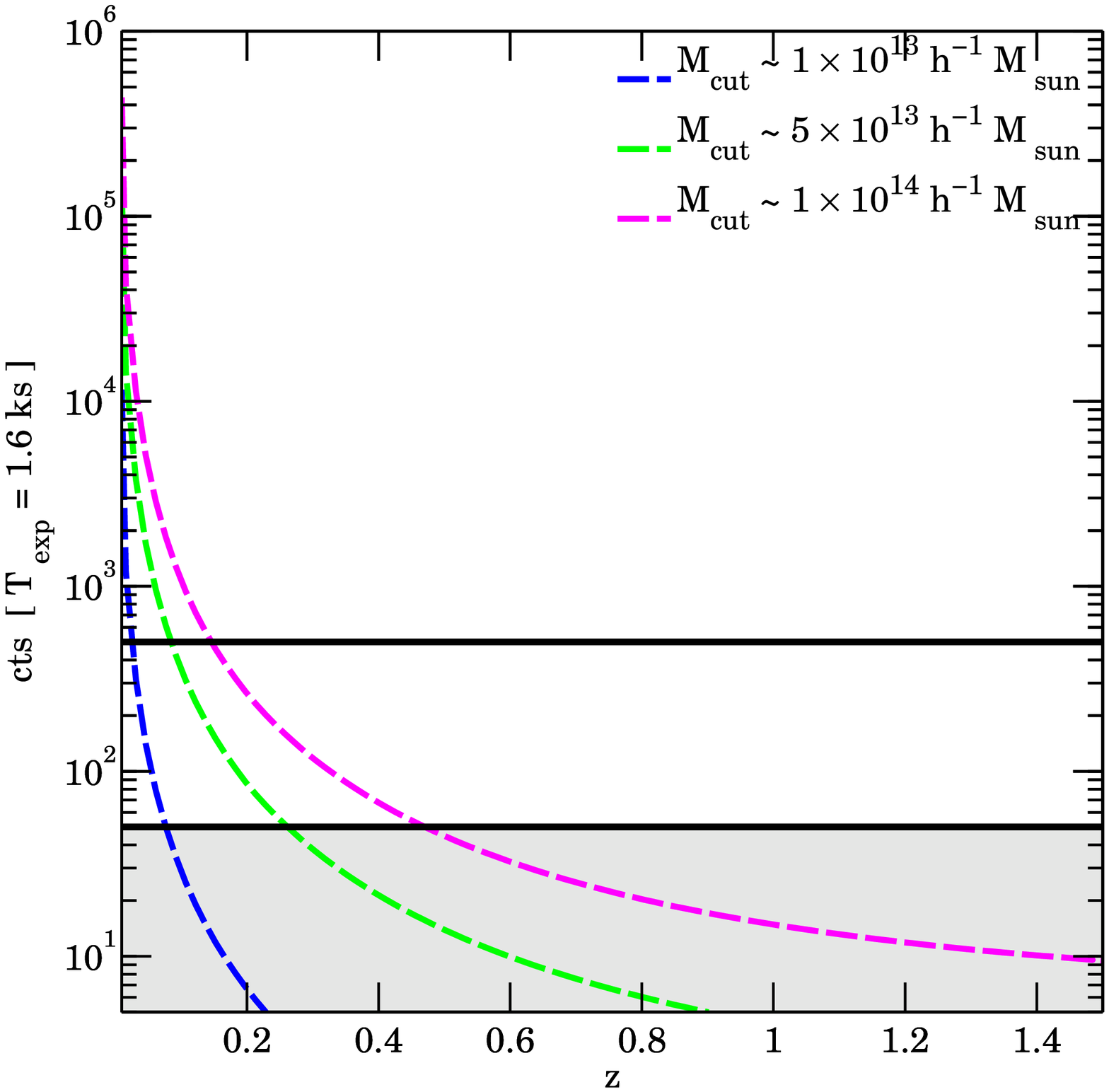}
\caption{\label{FIG:SELECTIONS} Observable--mass relation and sample selections for the $\eRO$ galaxy clusters. Top left: effective photon count vs mass relation ($\eta-\MCINQ$) derived via Eq.~(\ref{EQ:PROB}). The solid curves indicate the mean relation at $z \simeq 0.01,~0.10,~0.36, {\rm and } ~1.51$, from top to bottom, along with a 68 per cent level of scatter encompassed by the dashed lines (corresponding to about $\pm$ 41 per cent symmetrized relative error in the range of interest). The black solid line marks the detection threshold of 50 photons.
Top right: mass threshold ($\MCINQ$) corresponding to different photon-count detection limits as a function of redshift. 
We choose 50 photons with an integration time of $1.6$ ks as our reference choice. 
The thick horizontal line indicates the additional cut at $M = 5 \times 10^{13} \HI \MSUN$ which we impose to avoid considering clusters that possibly do not
obey power-law scaling relations or exhibit a mass-dependent scatter.
The colored curves show the effect that the current uncertainties on the parameters of the luminosity--mass relation (see legend) would have on the determination of the detection limit. Bottom left: threshold luminosity as a function of redshift, for different photon-count detection limits. The colored lines indicates the position of the additional possible mass-cuts in the luminosity-redshift plane (see legend). Bottom right:  selection criteria in the $\eta-z$ plane. The black solid lines correspond to the detection thresholds of 50 and 500 photons; 
the colored curves connect points corresponding to a given mass limit (colors as in the left panel).}
\end{center}
\end{figure*}
If we assume that $L_X$ and $T_X$ are independent variables 
and naively combine the 
mass-luminosity and mass-temperature relations from \cite{Vikhlinin:2009b}
we find that the (intrinsic) scatter in luminosity at fixed $T_X$ should be
$\sim 0.49$ (in the ln-ln plane). 
In case of a positive correlation among the variables, this scatter should
be smaller.
Direct measurements of the scatter from the REXCESS \footnote{REpresentative XMM-Newton ClustEr Structure Survey} survey
\citep{Pratt:2009} give 
0.276 for core-excised clusters and 0.666 when the cluster cores are included. On the other hand, other studies of the 64 brightest clusters in the sky \citep[HIFLUGCS \footnote{HIghest X-ray FLUx Galaxy Cluster Sample}, see ][]{Mittal:2011} suggest a reduced importance of cool cores as scatter contributors and an intrinsic scatter in the $L_X$-$T_X$ relation of 0.455 for the whole sample.
Estimating the degree of correlation by directly
comparing these figures is thus inconclusive and better data are needed. 
We have initially attempted to include $\RHOLT$ in the set of parameters
that we would like to constrain with $\eRO$.
However, we found out
that the weak dependence of the photon counts, $\eta$, on $T_X$ (see Fig.~\ref{FIG:EROSITA}) makes 
all the observables rather insensitive to $\RHOLT$ which remains 
unconstrained by the data.
To simplify the analysis, we thus assume  $\RHOLT \equiv 0$. 

The resulting photon counts as a function of cluster mass and redshift
are shown in the upper-left panel of Fig.~\ref{FIG:SELECTIONS}.
This gives an effective $\eta - \MCINQ$ relation which scales as ~ $M^\alpha$ with
$\alpha \sim 1.60-1.65$ (depending on redshift).
The scatter in our observable-mass relation at fixed mass is dominated
by $\SIGMALM$ but also depends on $\SIGMATM$ and the Poisson noise
in the photon counts. This corresponds to a (symmetrized) relative uncertainty at fixed mass of about 41 per cent at $z\sim0.1$; at $z\sim1$ this ranges between 66 and 41 per cent for objects with $ M \gtrsim 5 \times 10^{13} \HI \MSUN$.

\section{The $\eRO$ experiment}
\label{SEC:EROSITA}
For our purposes, we characterize the $\eRO$ survey using three numbers:
the fraction of the sky covered by the experiment $f_{\rm sky}$,
% redshift range, 
the exposure time $T_{\rm exp}$, and the detection threshold in 
raw photon counts $\eta_{\rm min}$ (see Table~\ref{TAB:PARAM}). 
We take as a reference an all-sky survey covering more than 27,000 deg$^2$ and 
with an average exposure time of $\sim 1.6$ ks; moreover, we limit our analysis to the (0.5--2.0) keV X-ray energy band, such that all the figures in this work have to be considered in that band. This all-sky survey should exquisitely constrain model parameters, since it will simultaneously provide robust statistics ($\sim10^5$ galaxy clusters detected with more than 50 photons) and sample very large spatial scales where primordial non-Gaussianity is expected to leave its strongest imprints ($k\lesssim 0.01 h~ {\rm Mpc}^{-1} $).
\begin{figure*}
\begin{center}
\includegraphics[width=7.6cm]{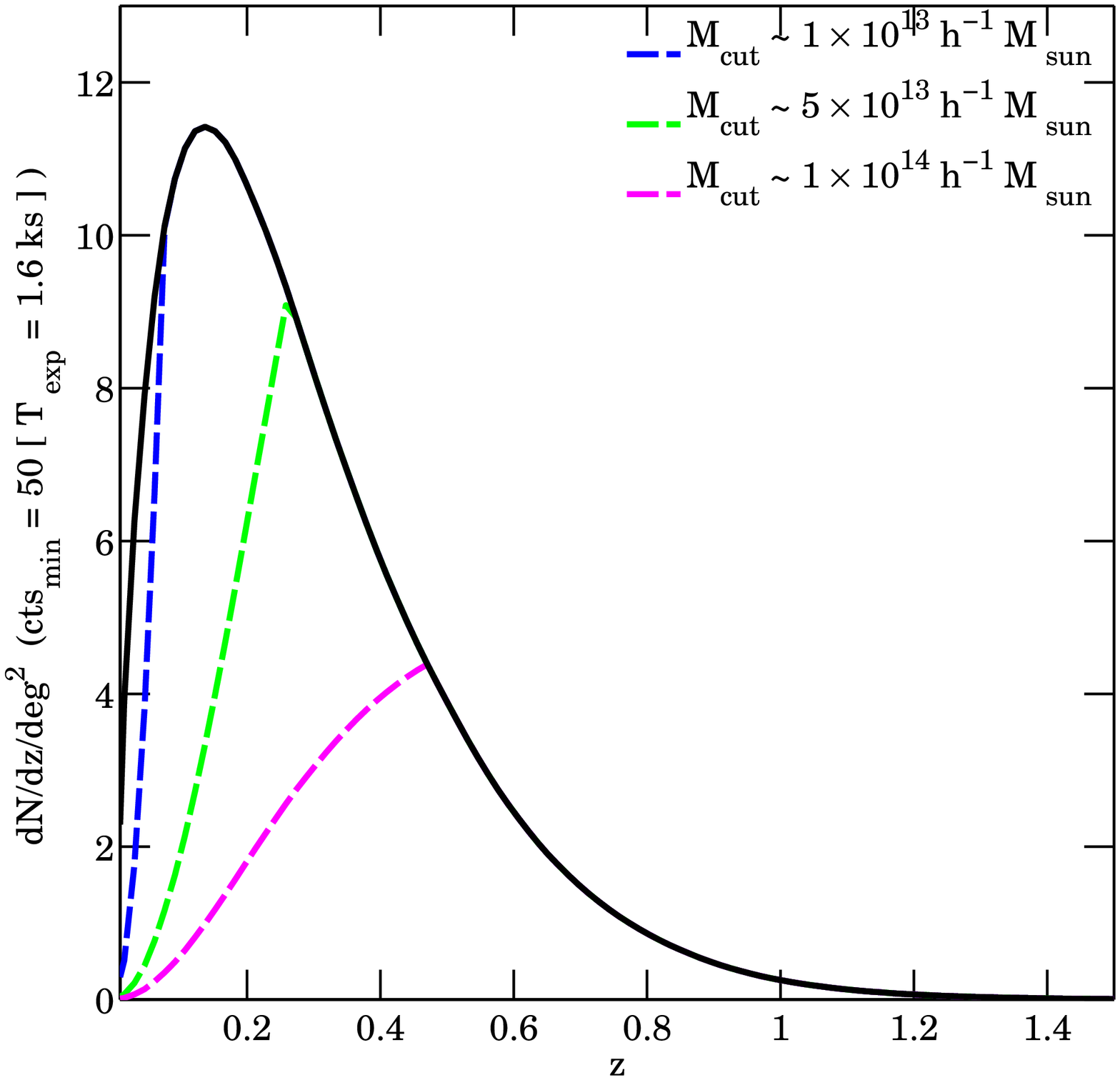}
\includegraphics[width=9.1cm]{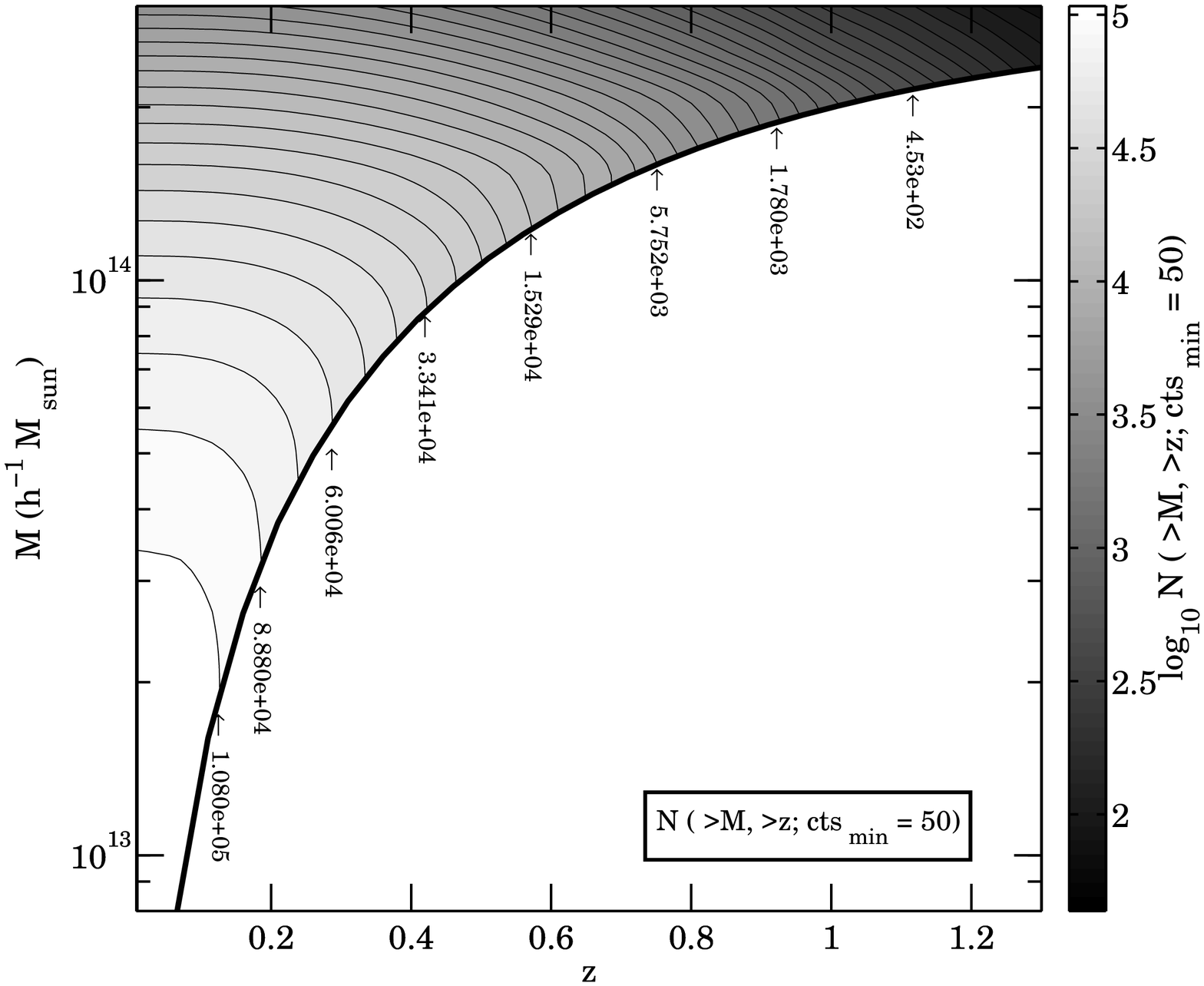}
\caption{\label{FIG:NUMBERS} Left: Redshift distribution of $\eRO$ clusters for different minimum masses (color coded as in the bottom panels
of Fig.~\ref{FIG:SELECTIONS}). 
Right: Number of clusters passing the detection threshold of 50 counts, as a function of redshift and $\MCINQ$, with sky coverage fraction $f_{\rm sky} = 0.658$.}
\end{center}
\end{figure*}
As a reasonable source-detection threshold
we choose $\eta=50$ counts as the default. This number is motivated by the experience
from the {\it ROSAT} All-Sky Survey (RASS), where the count limit has
typically been set at (10--30) photons in scientific analyses of
cluster catalogs, e.g., \cite{Schuecker:2001}. In the
future, this limit will be tested against detailed simulations of the
$\eRO$ survey.
While 50 source counts are sufficient for a luminosity
determination, one could measure the intracluster gas mass already with
about 500 source counts. The gas mass is a better proxy of the total
gravitational mass than the luminosity, so when applying the 500 count
limit one could in practice assume a smaller scatter in the observable--mass relation. While in this Section we discuss both cluster samples determined using these different thresholds, the whole statistical analysis of Sections 7 and 8 will be performed with the cut of 50 photons.

Low-redshift objects which are fainter than $10^{42} ~\LUMUN$ could in 
principle be detected with more than 50 photon counts with an average exposure time of 1.6 ks (Fig.~\ref{FIG:SELECTIONS}, bottom-left panel). This corresponds to masses smaller than $\sim 10^{13} \HI \MSUN$ at $z\lesssim 0.1$ (same Figure, upper-left and upper-right panels). 
Although scaling-relation studies which extend towards masses of $\sim 10^{13} \HI \MSUN$ are available \citep{Sun:2009, Eckmiller:2011OK}, the adopted prescriptions from \cite{Vikhlinin:2009b} are obtained from a cluster sample limited at about $7 \times 10^{13} \HI \MSUN$. Moreover, the scatter in the observed relations increases for
 low mass clusters and groups \citep{Eckmiller:2011OK}: this would
 necessitate to introduce a mass dependent scatter in our analysis. For these reasons, in addition to the photon-cut of 50 photons assumed for cluster detection, we add a redshift-dependent luminosity cut, which is effective in the low redshift part ($z \lesssim 0.25$) to remove low-mass systems ($M < 5 \times 10^{13} \HI \MSUN $): see Fig.~\ref{FIG:SELECTIONS}, upper-right and bottom-left panels. In practice, we pick our fiducial cosmology and our fiducial luminosity--mass relation as a function of $z$ and we look for a redshift-dependent luminosity cut that corresponds to the desired minimum mass: this, in turn, results
into a redshift-dependent $\eta$ cut (Fig.~\ref{FIG:SELECTIONS}, bottom-right panel) which is model independent and which can be easily applied to real data (while a true-mass cut cannot), if redshift information is available \footnote{In order to apply this additional cut redshift measurements are only required for $z \lesssim 0.25$ where, most likely, no new clusters will be discovered by $\eRO$. Therefore this sample selection should be easily feasible in practice.}. Note that neither a luminosity cut nor a mass cut are in practice applied here: low-mass objects are removed solely referring to the unique curve in the $\eta-z$ plane which corresponds to the chosen mass threshold in the fiducial model (Fig.~\ref{FIG:SELECTIONS}, bottom-right panel): such cut is kept fixed throughout the statistical analysis independently of the assumed values for the cosmology and scaling-relation parameters (see Section \ref{SEC:DISC_MCUT}).

In Fig.~\ref{FIG:SELECTIONS}, we compare a series of photon-count thresholds ($\eta=$50 and 500 photons with $T_{\rm exp}=1.6$ ks, and $\eta=$50 photons with $T_{\rm exp}=1.6, 3.0, 7.5$ ks) and additional luminosity cuts (corresponding to masses of $\sim 1, 5, 10 \times 10^{13} \HI \MSUN$), in the $\MCINQ-z$ plane (upper-right panel), in the $L_X-z$ plane (bottom-left panel), and in the $\eta-z$ plane (bottom-right panel). Fig.~\ref{FIG:SELECTIONS} shows that, away from the Galactic plane, $\eRO$ will detect \emph{all} clusters more luminous than $ \sim 5 \times 10^{44} ~\LUMUN $ up to $z \sim 1.5$, namely all massive ($M \gtrsim 3 \times 10^{14} \HI \MSUN$) clusters in the observable Universe.

In the left panel of Fig.~\ref{FIG:NUMBERS}, we show the redshift distribution of the population of clusters detected by $\eRO$ (see Section 5 and Eq.~(\ref{EQ:DNOVERDZ}) for details): the solid black curve is the redshift distribution of all the galaxy clusters above the cut of 50 photons for $T_{\rm exp}=1.6$ ks; the corresponding total number of objects in the whole sky ($f_{\rm sky} \equiv 0.658$) is $\sim 1.37 \times 10^5$ objects. In the same panel, the dashed curves indicate the redshift distributions of the same population of clusters once the additional cuts are applied. The integrated number of objects reduces to $\sim 9.32 \times 10^4$ and the median redshift shifts to $\sim 0.35$ when objects below $5 \times 10^{13} \HI \MSUN$ are removed. We summarize our findings in Table \ref{TAB:NTOT}. To facilitate comparisons with other works using different observables than raw photon counts, we show in the right panel of Fig.~\ref{FIG:NUMBERS} the cumulative mass distribution of clusters above our detection threshold of 50 counts, as function of $z$ and $\MCINQ$.

\begin{figure}
\begin{center}
\includegraphics[width=8cm]{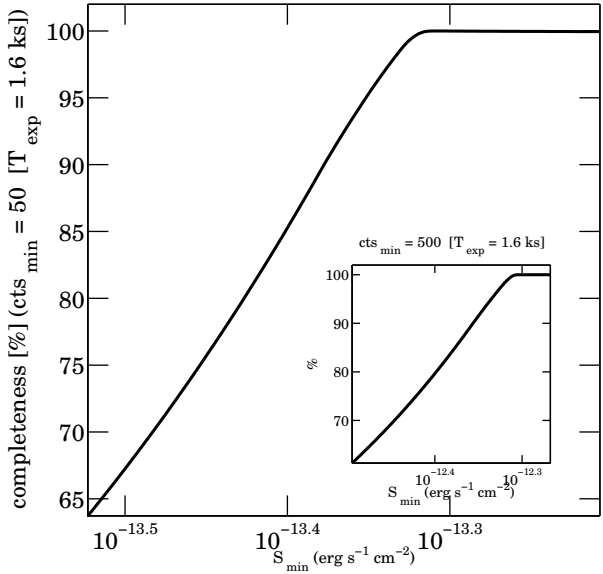}
\caption{\label{FIG:COMPLET} Completeness of a survey limited in flux and photon counts wrt a purely flux-limited selection with $S>S_{\rm min}$ for
a 50-photon limit (main panel) and 500-photon limit (inset).}
\end{center}
\end{figure}

Finally, in Fig.~\ref{FIG:COMPLET}, we compare the selections discussed so far with an alternative flux-based selection, often employed in X-ray cluster surveys. Two cases are shown: a comparison with a 50-photon limit (main box) and with a 500-photon limit (smaller inbox). The solid curves show the fraction of objects which pass a threshold both in flux and in photon counts with respect to the objects which pass a threshold set only in terms of flux. Minimum fluxes of $3.2, 4.2, 4.4 \times 10^{-14} ~\FLUXUN$ in the (0.5--2.0) keV energy band would result, respectively, in 68, 90, 95 per cent completeness levels for $\eta_{\min} = 50$. For $\eta_{\min} = 500$, 68, 90, 95 per cent completeness is reached with flux limits of $3.5, 4.4, 4.6 \times 10^{-13} ~\FLUXUN$, respectively.
\begin{table}
\begin{center}
\begin{tabular}{lcc}
\hline
 					& $N_{\rm clusters}$ & $z_{\rm median}$	\\
					& {\scriptsize $ ~(\eta_{\rm min} = 50, T_{\rm exp}=1.6$ ks)} &\\
\hline
all objects 							& $1.37 \times 10^5$	& 0.25\\
$M \gtrsim 1 \times 10^{13} ~ (\HI \MSUN )$	& $1.31 \times 10^5$ 	& 0.27 \\
$M \gtrsim 5 \times 10^{13} ~ (\HI \MSUN)$	& $9.32  \times 10^4$ 	& 0.35 \\
$M \gtrsim 1 \times 10^{14} ~ (\HI \MSUN)$  	& $5.57 \times 10^4$ 	& 0.46\\
\hline
\end{tabular}
\caption{Number of galaxy clusters detected by $\eRO$ and their median redshift for the all-sky survey, with sky coverage fraction $f_{\rm sky} = 0.658$.} 
\label{TAB:NTOT}
\EC
\end{table}
%Note that these completeness levels in
%terms of a flux limit are optimistic because variations in hydrogen
%column density and exposure time across the survey area are
%neglected. Forecasts employing flux limits should take realistic
%completeness levels into account.

\section{Observables}
\label{SEC:OBSERVABLES}
The measurements that we consider are galaxy-cluster abundances and angular clustering. In this section we study their sensitivity to the 
actual values assumed by the model parameters.

\begin{figure*}
\begin{center}
\includegraphics[width=8cm]{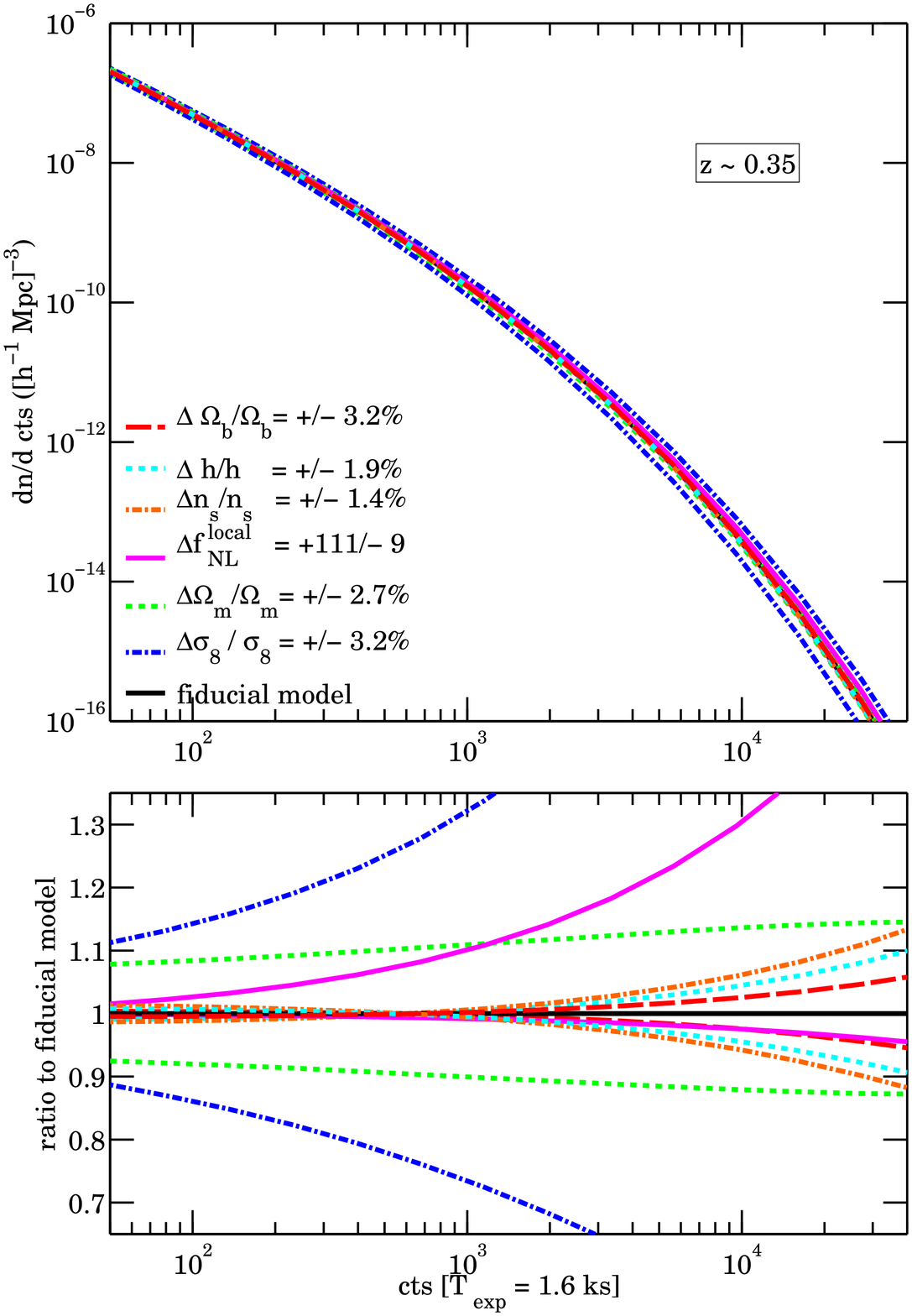}
\includegraphics[width=8cm]{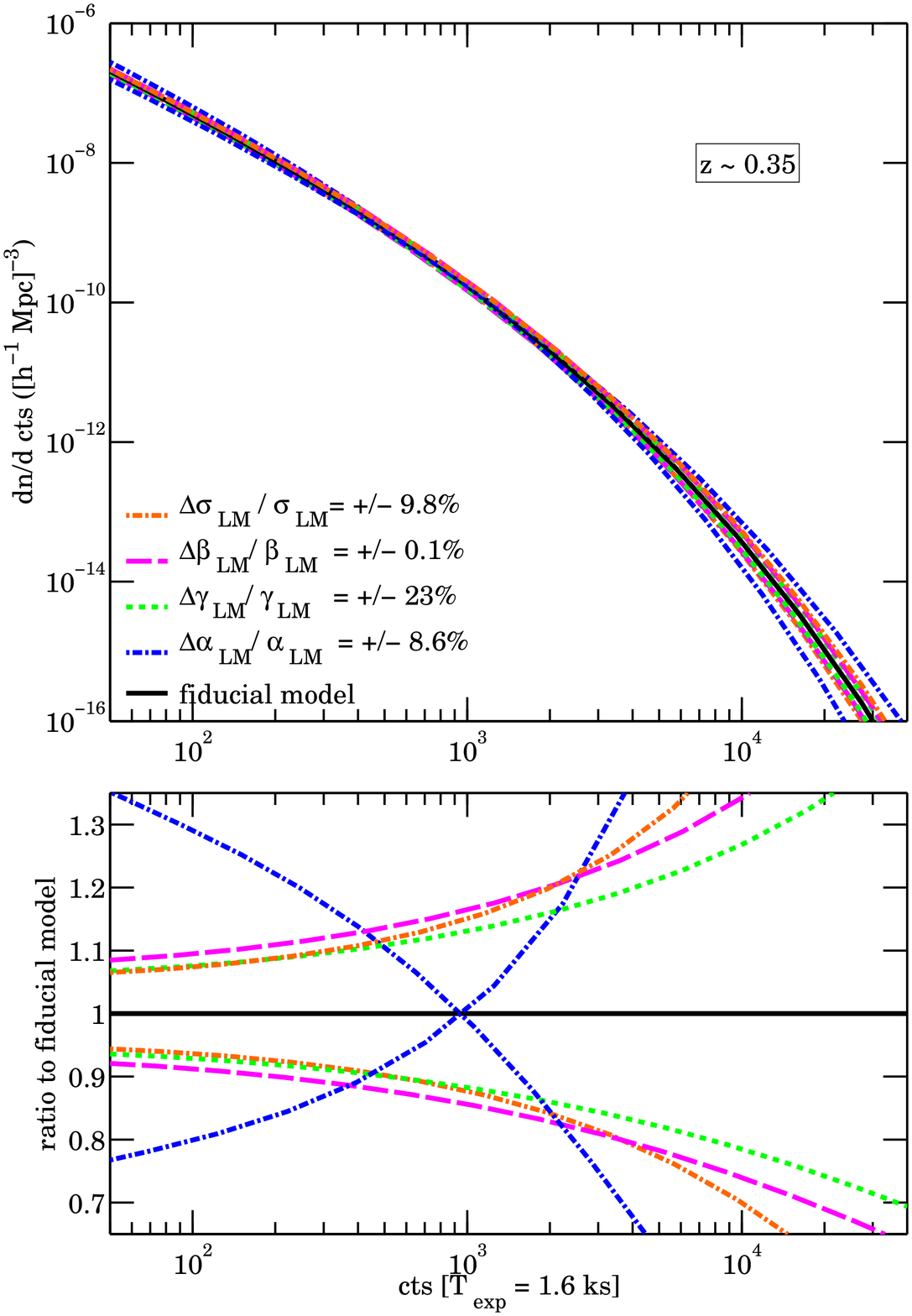}
\caption{\label{FIG:COUNTS_ALL} 
Number counts of $\eRO$ clusters.The black solid line refers to our fiducial model while the other lines show
the dependence on the cosmological parameters (left) and on the parameters of the LM scaling relation (right), varied one at the time of the relative amounts indicated in the legend (variations correspond to the current uncertainties on the parameters as indicated in Table \ref{TAB:PARAM}). }
\end{center}
\end{figure*}

Fig.~\ref{FIG:COUNTS_ALL} shows how the $\eRO$ number counts 
depend on the cosmological parameters and on the parameters characterizing the physics of the ICM. 
In the upper panels of Fig.~\ref{FIG:COUNTS_ALL}, the ``count function'' $\frac{dn}{d \eta}$ at the median
redshift of the sample is shown for our fiducial model (black solid line) 
and for alternative models where the cosmological (left) and ICM (right) parameters are
varied (one at the time) by the current uncertainty (listed in Table \ref{TAB:PARAM}).

The redshift distribution of the clusters above the detection threshold of $\eta_{\rm min}=50$ counts with $T_{\rm exp} = 1.6$ ks relates to the count function as follows
\BEA
\frac{dN}{dz~{\rm deg}^2} (> \eta_{\rm min},z) &=& ~\frac{4 \pi}{\it{A}} f_{\rm sky} \left[\frac{c}{H(z)} D_A^2(z)\right] \nonumber\\
& &\int^{\infty}_{\eta_{\rm min}} \frac{dn}{d\eta}(\eta, z) ~ d\eta,
\label{EQ:DNOVERDZ}
\EEA
where $A$ is the survey area in deg$^2$, $c$ is the speed of light,
and $D_A$ is the comoving angular diameter distance (which coincides with $r$
for flat universes).
The total number of clusters detected above a certain detection threshold 
can be calculated by integrating the equation above over redshift and multiplying the result
by $A$.
%\BEA
%N(> \eta) =A\,\int^{z_{\rm max}}_{z_{\rm min}} 
%\frac{dN}{dz~{\rm deg}^2} (> \eta,z) ~dz\;.
%\EEA
%

The bottom panels of Fig.~\ref{FIG:COUNTS_ALL} give the relative change of the observables wrt the fiducial model. 
The largest deviations are due to changes in
$\SIGMA8$, $\OM$, and $\FNLL$, and all the four parameters of the LM relation. All parameters but $n_{\rm s}$ and $h$ generate larger uncertainties at higher redshifts than at lower redshifts (not shown).
Note that changing $\SIGMA8$ within the current uncertainty modifies
the cluster counts by 20 per cent for objects detected with more than 
$\sim 400$ photons at the median redshift, or for all the objects above $z\gtrsim 0.6$. 
Also primordial non-Gaussianity (with positive $\FNLL$) 
has a stronger impact at higher redshifts and for higher photon counts. 
The effects of the Hubble constant and of the scalar spectral index 
exceed the per cent level only for the brightest clusters, 
e.g. above a few $10^3$ photon counts around the median redshift, 
which already suggests that binning the $\eRO$ data in $\eta$ should better 
constrain these parameters.
\begin{figure*}
\begin{center}
\includegraphics[width=8cm]{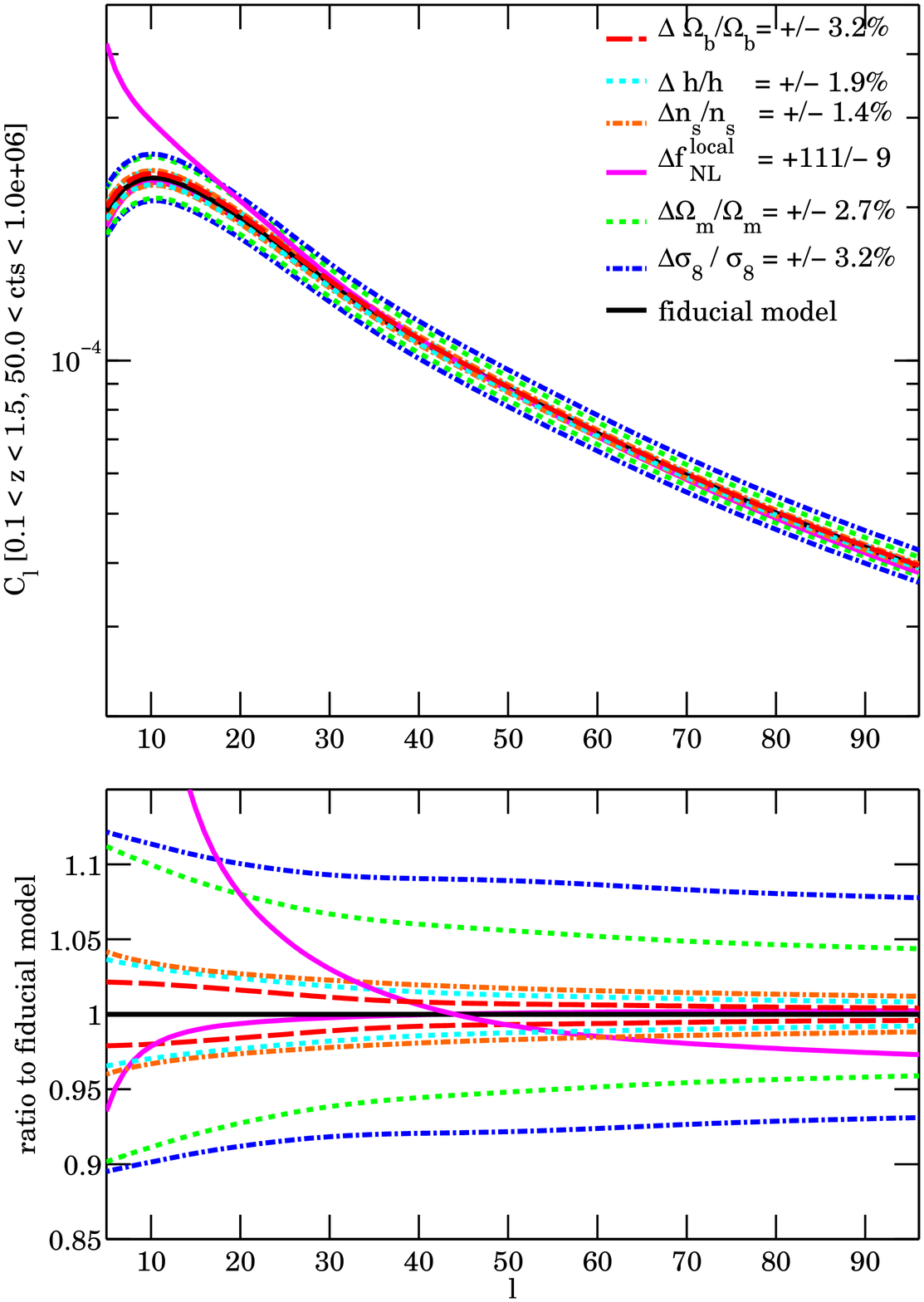}
\includegraphics[width=8cm]{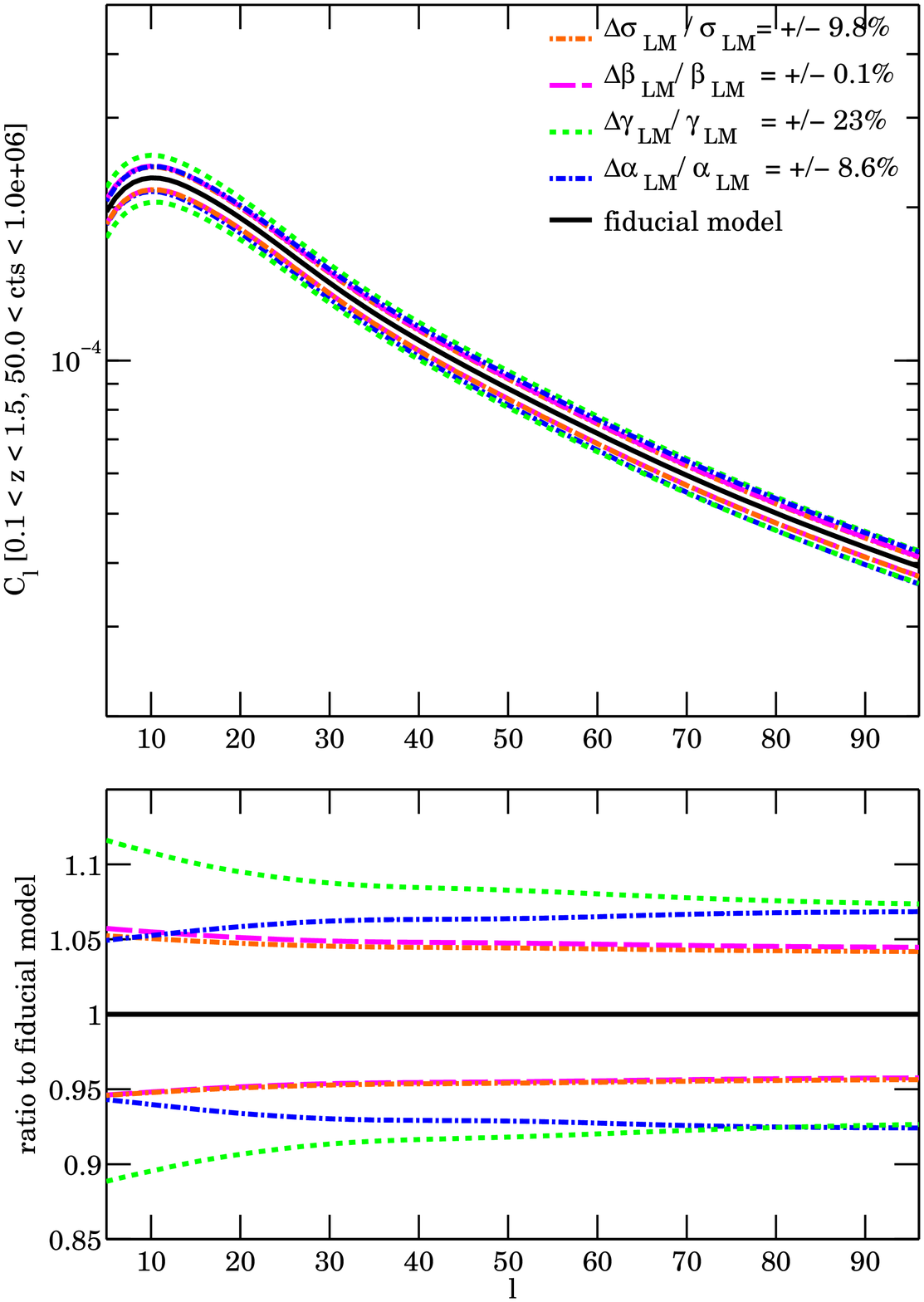}
\caption{\label{FIG:CLUSTERING} Angular power spectrum of $\eRO$ clusters and its dependence on the cosmological and ICM parameters. 
The signal refers to clusters detected with $\eta \geq 50$ photons ($T_{\rm exp}=$1.6ks) within the redshift range $0.1 < z< 1.5$.}
\end{center}
\end{figure*}
Current uncertainties in the parameters regulating the 
temperature-mass relation (not shown but discussed in Section \ref{SEC:DISC_TM}) affect the cluster counts at the per cent or sub per cent level
thus indicating that $\eRO$ counts will hardly be able to put interesting
constraints on those parameters.
Finally, we have checked that fitting errors in the parameters of the 
Tinker mass function 
(not shown but discussed in Section \ref{SEC:DISC_TMF}) also give sub per cent
modifications. This confirms that our theoretical framework is robust.\\

We now consider the spatial clustering of the detected clusters.
Imagine to split the $\eRO$ sample into bins (based on the cluster redshift
or photon counts). The angular cross spectrum between the pair of bins
$i,j$, can be written as
\BEA
C_\ell(i, j)  & = & \frac{2}{\pi} \int_0^{\infty} dk ~ k^2  \int_0^{\infty} dr_1~ W_i(r_1,k) ~ j_{\ell}(k r_1) \\
\nonumber
& &\times \int_0^{\infty} dr_2 ~W_j(r_2,k) ~ j_{\ell}(k  r_2) ~P_{NL}(k, r_1,r_2) .
\label{EQ:ANGCLUSTERING}
\EEA
where the functions $W_i(r,k)$ are defined as
\BE
W_i(r,k) = \frac{1}{N_i} ~\frac{dN_i}{dr}(r) ~b_i(k,z(r)),
\EE
and $N_i$ is the total number of objects within the $i$-th bin, $ j_{\ell}$ denotes the spherical Bessel function of the first type, and $b_i(k,z(r))$ is the effective bias of the cluster population (obtained by averaging  Eq.(\ref{EQ:NGBIAS}) over halo mass or photon counts and weighting by the number density) in the $i$-th bin. Here $P_{NL}$ denotes the cross-spectrum between the matter distribution at two different redshifts, $z_1$ and $z_2$, corresponding to the comoving distances, $r_1(z_1)$ and $r_2(z_2)$.

In Fig.~\ref{FIG:CLUSTERING} the dependence of the angular power spectrum 
on the cosmological and ICM parameters is shown for galaxy clusters with
$\eta \ge 50$ and lying within the broad redshift range [0.1-1.5]. 
Note that the largest modifications in the signal
are driven by the current uncertainties in $\SIGMA8$, $\OM$, $\FNLL$ and
$\GAMMALM$. An accurate determination of the cluster power spectrum
with $\eRO$ has therefore the potential to strongly constrain these
model parameters.

To speed-up calculations it is convenient to rewrite Eq.~(\ref{EQ:ANGCLUSTERING}) using the
Limber approximation:
\BE
C_\ell(i,j) \simeq 4\pi \int_0^\infty dz~ \frac{dV}{dz} ~P_{NL}
\left(\frac{\ell+1/2}{D_A},z\right)  ~W^i_{\ell,L}(z) \,W^j_{\ell,L}(z)
\label{EQ:ANGCLUSTERINGL}
\EE
with the Limber weight functions $W^i_{\ell,L}$ being defined as
\BE
W^i_{\ell,L}(z) = \frac{1}{N_i} ~\frac{dN_i}{dV}(z) \,b_i \left(\frac{\ell+1/2}{D_A},z \right),
\EE
$V$ is the comoving volume.
%However, as detailed in Section \ref{SEC:DISC_LIMBER}, the 
%Limber approximation is not very accurate in the presence of a scale-dependent
%bias induced by primordial non-Gaussianity of the local type.
%
%Depending on the thickness and the central value of the redshift bins,
%the deviation from the exact calculation can reach 20 per cent at
%$\ell \sim 5-10$ \textcolor{green}{for $\FNLL \sim 100$} and is not negligible up to multipoles of 60 to 100.
%These are the spatial scales at which primordial non-Gaussianity imprints
%the strongest signatures (Fig.~\ref{FIG:CLUSTERING}, left panel).
%Thus, we always use the exact calculation when necessary and switch to 
%the Limber approximation only at large multipoles.
We use Eq.~(\ref{EQ:ANGCLUSTERING}) for all multipoles with $\ell \lesssim 100$ and Eq.~(\ref{EQ:ANGCLUSTERINGL}) for larger values of $\ell$. A motivation for this choice and a critical discussion about the limitations of the Limber approximation in the presence of primordial non-Gaussianity will be given in Section \ref{SEC:DISC_LIMBER}.

\subsubsection{The choice of $\ell_{\rm min}$}
Primordial non-Gaussianity of the local type modifies the shape of the
angular power spectrum of galaxy clusters by introducing a scale-dependent bias
on very large scales (see Fig.~\ref{FIG:CLUSTERING}). Therefore, experimental
constraints on $\FNLL$ improve by measuring clustering at larger and larger
scales.   
However, the correlation function of biased tracers of the cosmic density field in scenarios with $\FNLL \neq 0$ formally diverges for every spatial separation if Fourier modes down to $k\to 0$ are considered \citep[e.g.][]{Wands:2009}. This is because the biased density field has infinite variance (as like as the gravitational potential which generates the scale-dependent bias \citep[see][]{Giannantonio:2010a}.
Any observed correlation function, anyway, will be finite because density fluctuations on scales larger than the survey are
not observable. Fluctuations are always defined with respect to the mean density as measured from the same survey, thus forcing their
average to zero. The expectation value of the observed correlation therefore departs from the underlying one by a constant shift
(sometimes known as the ``integral constraint'') which is finite only if $\FNLL =0$.
In practice, this effect is taken into account by including the window function of the survey in the calculation of $C_\ell$. This regularizes
all the integrals. For simplicity, here we approximate the influence of the window function by using a high-pass filter $k>k_{\rm min} \sim 10^{-3} h$ Mpc$^{-1}$ and only considering a minimum order for the spherical harmonics $\ell_{\rm min}$ so that the derivatives
in the Fisher matrix do not depend on $k_{\rm min}$ \citep[see also][]{Cunha:2010}.
In what follows, we will use $\ell_{\rm min}=5$ for the $\eRO$ all-sky survey,
and $\ell_{\rm min}=7-10$ for the deeper surveys presented in Section 
\ref{SEC:SURVEYCOMP}. These choices are motivated as follows.
First, multipoles with $\ell<10$ can not be precisely 
measured if $f_{\rm sky}$ substantially
departs from unity, since only a few modes are available in the whole sky.
Second, evaluating $C_\ell$ at such low multipoles requires 
knowledge of the halo bias at wavenumbers $k \ll 0.01 ~h\,$Mpc$^{-1}$, 
never probed by N-body simulations. Pushing the analysis to $\ell \sim 3$ would imply trusting the 
extrapolation of the non-Gaussian halo bias of Eq.~(\ref{EQ:NGBIAS}) 
down to $k\sim 10^{-4} ~h\,$Mpc$^{-1}$ where also general-relativity corrections
certainly become very important \citep[e.g.][]{Yoo:2010, Baldauf:2011}.

\begin{figure}
\BC
\includegraphics[width=8cm]{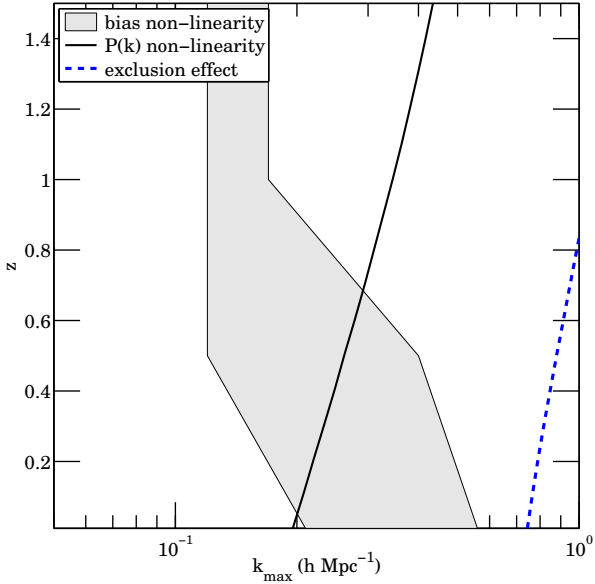}
\caption{\label{FIG:KMAX} 
Maximum wavenumber which can be robustly modeled in the power-spectrum study, as a function of redshift. The blue dashed line refers to exclusion effects and
indicates the minimum separation allowed for (spherical) objects of $\sim 10^{14} \HI \MSUN $. The black solid line marks the onset of the non-linear
regime for density perturbations (see main text for details). The grey area approximately traces the minimum wavenumber (as a function of redshift) at
which a scale-dependence in the bias of dark-matter haloes of mass $\sim 3\times 10^{13} \lesssim M \lesssim 2\times 10^{14} \HI \MSUN$  can be detected in the numerical simulations of Pillepich, Porciani \& Hahn 2010.
Note that, for galaxy clusters, the most severe requirement comes from the non-linearity of the bias. To overcome this problem, in our study we only consider wavenumbers such that $ k < k_{\rm max} = 0.1 h {\rm Mpc}^{-1}$.}
\EC
\end{figure}
\subsubsection{The choice of $\ell_{\rm max}$}
We also have to set a minimum angular scale that will be considered in
the clustering analysis. This is mainly dictated by the limitations of
the theoretical models for the angular power spectrum of galaxy clusters.  
%
%Within the Limber approximation, 
Angular multipoles and wavenumbers are 
related through the angular diameter distance in a redshift-dependent fashion.
The choice of the maximum multipole thus corresponds to selecting
the largest wavenumber that we want to consider, $k_{\rm max}$. 
Three issues have to be taken into account: cluster exclusion effects, 
dynamical non-linearity of density perturbations, 
and non-linearity of the DM halo bias. We compare the characteristic scales of these three effects in
Fig.~\ref{FIG:KMAX}.
Galaxy clusters are extended objects 
(at $z\sim 0.35$, a spherical $10^{14} \HI \MSUN$ cluster has a comoving
$R_{\rm  500}$ of about 0.6 $\HI$ Mpc)
and their spatial separation cannot
be much smaller than their characteristic size ($\sim 2R_{\rm  500}$).
This imprints a sharp drop in the amplitude of $C_\ell$ at small angular 
scales which is difficult to model accurately. The blue dashed curve in Fig.~\ref{FIG:KMAX} indicates 
the characteristic size (and thus, roughly,
the minimum separation) of objects with  $M\sim 10^{14} \HI \MSUN$. 
Regarding the non-linearity of the matter-density field, 
we do not want to rely on approximate 
prescriptions for the matter power spectrum on mildly non-linear scales 
which could compromise the accuracy of the fits for the cosmological parameters.
The onset of the non-linear regime for the matter power spectrum can be
approximately evaluated by determining $k_{\rm max}(z)$ such that
\BE
\SG^2(k_{\rm max},z) = \frac{1}{2 \pi^2} \int^{k_{\rm max}(z)}_{0} k^2 ~P_{\rm lin}(k,z)~ {d}k = 0.5\;.
\EE
The solid black line in  Fig.~\ref{FIG:KMAX}
shows the function $k_{\rm max}(z)$. 
Similarly, we are concerned with the linearity of cluster bias.
Eqs.~(\ref{EQ:PBSBIAS}) and (\ref{EQ:NGBIAS}) asymptote to a constant value for 
$k \rightarrow \infty$. However, N-body simulations show 
that on sufficiently small
scales (depending on redshift and mass), the bias of DM haloes departs from a 
constant thus becoming non-linear. 
%It is not currently known how the cosmological parameters of our interest affect the non-linear behavior of the halo bias: extrapolating with a constant the halo bias well within the non-linear regime might induce to artificially underestimate or overestimate our error forecasts, which has to be avoided. 
For haloes with mass $M\gtrsim 5\times 10^{13}\HI \MSUN$ at $z \sim 0$, 
the non-linear regime of the bias kicks in for wavenumbers above 
$k \sim 0.1-0.3 ~h \ {\rm Mpc}^{-1}$. 
This threshold decreases with redshift and mass. The grey band in Fig.~\ref{FIG:KMAX} indicates an approximate determination of the 
onset of non-linearities in the bias for DM haloes in the mass range 
$(3-20) \times 10^{13} \HI \MSUN$, as extracted from the N-body simulations of 
\cite{Pillepich:2010}. 
The numerical work by \cite{Manera:2010}, shows that for objects with $M \gtrsim 10^{14} \HI \MSUN$ the transition occurs at even slightly smaller wavenumbers. 
In summary,
at all redshifts greater than 0.1, the linearity of the bias gives the most 
stringent constraints, followed by the linearity of the density field.
%On the other hand, the accuracy of the angular power spectrum is such that departures from a constant bias cannot be distinguished before effective wavenumbers corresponding to smaller scales than the non-linear bias scales: $k^{\rm eff}_{\rm NL} \gtrsim k_{\rm NL}$.\\
%
Based on these arguments, we make a conservative choice and adopt a 
maximum wavenumber of $k_{\rm max} = 0.1 ~h \ {\rm Mpc}^{-1}$, 
which corresponds to $\ell_{\rm max} \sim 96$ at the median redshift of the all-sky survey. We relax this choice in Section \ref{SEC:DISC_KMAX}.

\section{Fisher-information formalism}
\label{SEC:FISHER}
The Fisher-matrix formalism 
\citep{Tegmark:1997} is a tool for
forecasting how well a future experiment will constrain some model parameters.
Given a dataset, $\boldsymbol{x}$, and a set of parameters $\boldsymbol{\theta}$ with fiducial
values, $\boldsymbol{\theta}_0$, we define the Fisher information matrix as 
\BE
F_{\alpha \beta} = \left\langle \frac{\partial^2 \mathcal{L}}{\partial \theta_{\alpha} \partial \theta_{\beta}} \right\rangle,
\label{EQ:FISHERMATRIX}
\EE
where the average is taken over an ensemble of realizations of the experiment and $\mathcal{L} = -\LN L$ with $L(\boldsymbol{x};\boldsymbol{\theta})$ being the likelihood of the data given a model.
For unbiased estimators, $\langle \hat{\boldsymbol{\theta}} \rangle = \boldsymbol{\theta}_0$ (where $\hat{\boldsymbol{\theta}}$ denotes an estimate
of the parameter set based on a single realization of the experiment),
the inverse of the Fisher matrix (evaluated at $\boldsymbol{\theta}=\boldsymbol{\theta}_0$) 
provides an approximation for the covariance matrix of the model parameters:
\BE
\Sigma^{\rm param}_{\alpha \beta} = (F^{-1})_{\alpha \beta}.
\EE
Under some weak regularity conditions,
the Cram\'er-Rao inequality assures that the marginal errors of the model parameters follow
\BE
\ERR \ge (F^{-1})^{1/2}_{\alpha \alpha}, 
\label{EQ:CRINEQ}
\EE
while the conditional errors (i.e. the errors that would be obtained keeping
all the other parameters fixed) will be larger or equal than $1/\sqrt{F_{\alpha \alpha}}$.
%(and the two minima coincide in the case of diagonal Fisher matrix).\\
%When forecasting, $\MUDATA$ and the covariance matrices are supposed to be evaluated at a fiducial choice of parameters, i.e. at the fiducial model.\\

For a (multi-variate) Gaussian likelihood function $L(\boldsymbol{x};\boldsymbol{\theta})$, 
the Fisher matrix can be explicitely written as 
\BE
F_{\alpha \beta} = \frac{1}{2} \rm{Tr}\Big[\ICOV (\MUDATA,_{\alpha} \MUDATA^T,_{\beta} + \MUDATA,_{\beta} \MUDATA^T,_{\alpha}) + \ICOV \COV,_{\alpha} \ICOV \COV,_{\beta}\Big]
\label{EQ:GAUSSFISHERMATRIX}
\EE
where 
$\MUDATA = \langle \boldsymbol{x} \rangle$, $\COV = \langle (\boldsymbol{x}-\MUDATA) (\boldsymbol{x}-\MUDATA)^T \rangle$, and commas
denote partial derivatives wrt the model parameters.
If ${\mathcal M}$ is the number of model parameters and ${\mathcal N}$ is the number of data points, the Fisher
matrix is a ${\mathcal M}\times {\mathcal M}$ matrix while $\COV$ is ${\mathcal N}\times {\mathcal N}$. The trace operator
in Eq.~(\ref{EQ:GAUSSFISHERMATRIX}) acts on the ${\mathcal N}-$dimensional space of data points.

If different experiments are independent, the total Fisher matrix is the sum of individual Fisher matrices. ``Adding priors'' can be regarded here as adding the Fisher matrix of a third experiment.
\subsection{Figure of merit}
In order to compare individual probes, optimize the experiments, 
and quantify their individual performance in constraining model parameters, 
we introduce the (total) figure of merit (FoM), defined as
\BE
\rm{FoM} = \rm{log}_{10} \left [  {\rm det} (F^{-1})\right]^{-1/2}\;.
\label{EQ:FOM}
\EE
This is inversely proportional to the volume of the ${\mathcal M}-$dimensional error
hyper-ellipsoid. The higher the FoM, the more suitable an experiment is to
constrain our selected parameter set. For example, halving uniformly all the parameter errors would correspond to an increment in the figure of merit of $\rm{log}_{10} \left ( 2^{\mathcal M} \right )$, which for ${\mathcal M = 10}$ reads $\sim 3$.
Similarly, if one wants to focus on a selected subset of parameters, 
a (partial) FoM can be defined by considering a Fisher matrix of lower
dimensionality.
\subsection{Number counts}
For an unclustered point process, 
the probability of counting $N_i$ objects in the i-$th$ bin is given by the Poisson distribution
\BE
p(N_i|\mu_i) = \frac{1}{N{_i}\!} \mu_i^{N_i}~{\rm e}^{- \mu_i}
\EE
where the mean $\boldsymbol{\mu} \equiv \langle \boldsymbol{N} \rangle$ is the average over an ensemble of realizations. 
Fluctuations in the counts are larger for point processes that display spatial clustering.
In this case the covariance matrix of the binned counts can be written as \citep{Lima:2004}
\BE
\COVCOUNTS_{ij} = S_{ij} + M_{ij}
\label{EQ:COVCOUNTS}
\EE
where $\boldsymbol{S}$ denotes the sample covariance encoding information from the two-point correlation function, and the second term is the diagonal Poisson noise $M_{ij} = \langle N_i \rangle \delta_{ij}$.
\cite{Hu&Kravtsov:2003} have shown that the sample covariance is subdominant wrt to Poisson errors when a cluster survey encompasses a large fraction of the sky, covers a broad redshift interval,
and high-threshold masses are adopted. We therefore neglect it in our calculations as 
%we do not want to extract information from the noise and} 
we will treat clustering separately as described
in the next section.
Note that this is different from considering a count-in-cells experiment where the covariance
of the counts is a direct observable \citep{Oguri:2009, Cunha:2010}. 
For negligible sample covariance, the Fisher matrix for the cluster counts can be written as 
\BEA
\FCOUNTS_{\alpha \beta} &=& \frac{1}{2} \rm{Tr}[\ICOV_{counts} (\MUDATA,_{\alpha} \MUDATA^T,_{\beta} + \MUDATA,_{\beta} \MUDATA^T,_{\alpha})] =\\
&=& \langle \boldsymbol{N} \rangle,_{\alpha}^T \boldsymbol{M}^{-1} \langle \boldsymbol{N}\rangle,_{\beta} = \\
&=& \sum_{i}^{\# \rm{bins}} \frac{\partial  \langle N_i \rangle}{\partial \theta_{\alpha}} \frac{1}{ \langle N_i \rangle} \frac{\partial  \langle N_i \rangle}{\partial \theta_{\beta}} \Big \vert_{\rm fiducial ~model}.
\label{EQ:FISHERCOUNT}
\EEA

\subsection{Two-point clustering}
Let us now consider the two-point clustering of the X-ray clusters. 
Our observables will be the angular cross and auto spectra $C_\ell(i,j)$ between clusters binned
either in redshift or in photon counts.
The measured signal is affected by Poissonian shot noise:
\BE
\tilde{C}_\ell(i,j) = C_\ell(i,j) + \frac{1}{N_i} \delta_{ij}
\EE
where $\delta_{ij}$ denotes the Kronecker delta. 
The covariance matrix for the angular power spectrum can be written as follows
\citep{Hu&Jain:2004, Cohn:2006, Huterer:2006, Yamamoto:2007}:
\BEA
\COV &\equiv& {\rm Cov}[\tilde{C}_{\ell}(i,j)~\tilde{C}_{\ell'}(m,n)] = \boldsymbol{\rm Cov}_{\ell} (ij)(mn)\\
\nonumber
& = & \frac{\delta_{\ell \ell'}}{(2\ell+1) f_{\rm sky}} \left[ \tilde{C}_{\ell}(i,m) ~\tilde{C}_{\ell}(j,n) + \tilde{C}_{\ell}(i,n) ~\tilde{C}_{\ell}(j,m) \right].
\label{EQ:COVCLUST}
\EEA
The covariance matrix is thus block-diagonal, where the number of blocks is given by the number of multipoles $\ell$ and where the dimension of every block is given by the number of distinct pairs which can be formed with the adopted number of bins, in redshift or photon counts.
Assuming a Gaussian likelihood 
function for the underlying density field, i.e. for the spherical coefficients $a_{\ell m}$ which define the angular power spectrum $ \langle a_{\ell m}a^*_{\ell' m'} \rangle = C_\ell ~\delta_{\ell \ell'} ~\delta_{m m'} $, we write the ``clustering'' Fisher matrix as follows:
\BE
\small{
\FCLUST_{\alpha \beta} = \sum_\ell \sum_{(i,j)(m,n)} ~\frac{ \partial C_{\ell}(i,j) }{\partial \theta_{\alpha} } ~\boldsymbol{\rm Cov}^{-1}_{\ell} (ij)(mn) ~\frac{\partial C_{\ell}(m,n)}{\partial \theta_{\beta}} .
}
\label{EQ:FISHERCLUSTERING}
\EE

%
%%%%%%%%%%%%%%%%%%%%%%%%%%%%%%%%%
\section{Results}
\label{SEC:ANALYSIS}

\subsection{Cluster counts}
\begin{figure*}
\BC
\includegraphics[width=8cm]{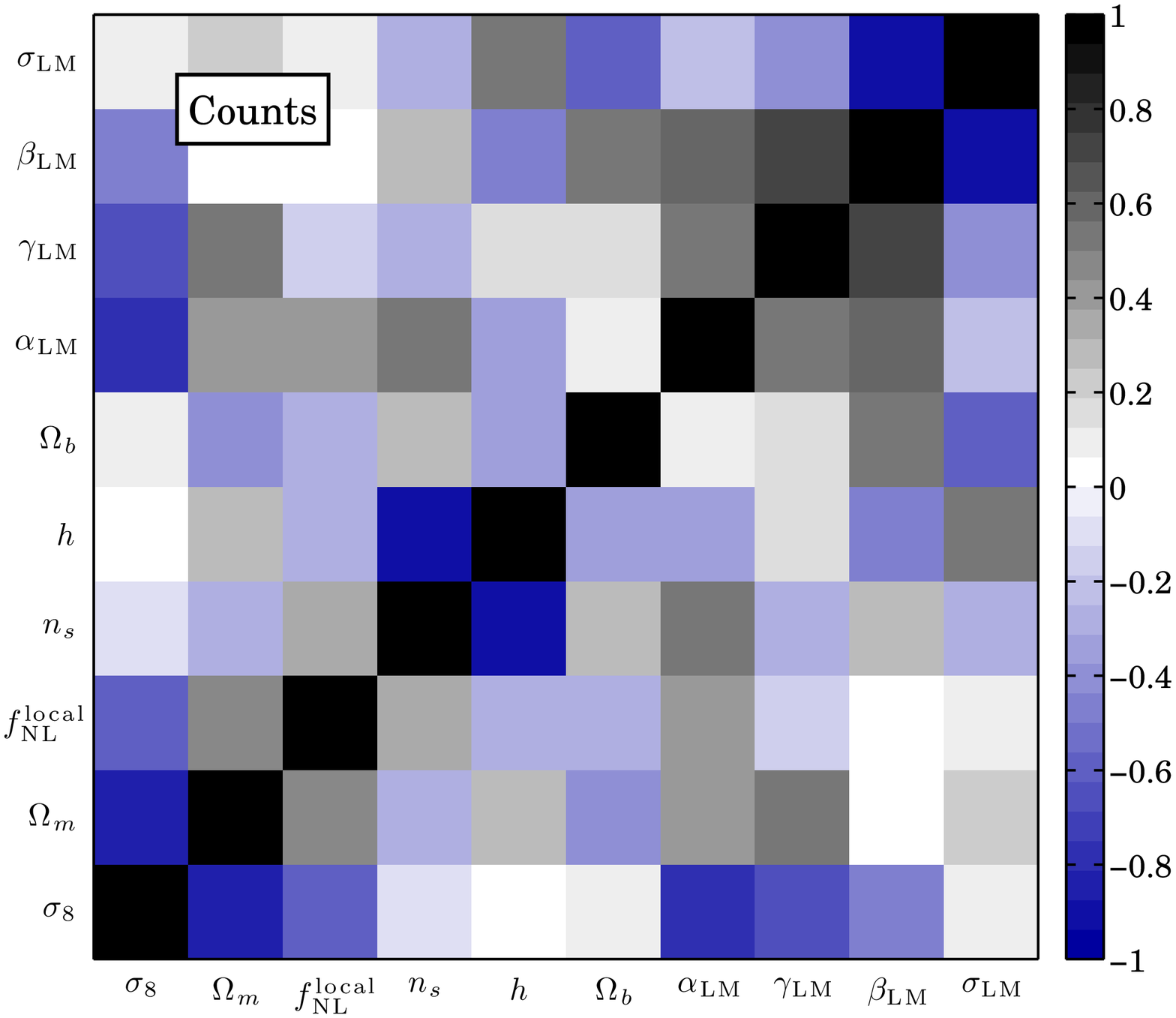}
\includegraphics[width=8cm]{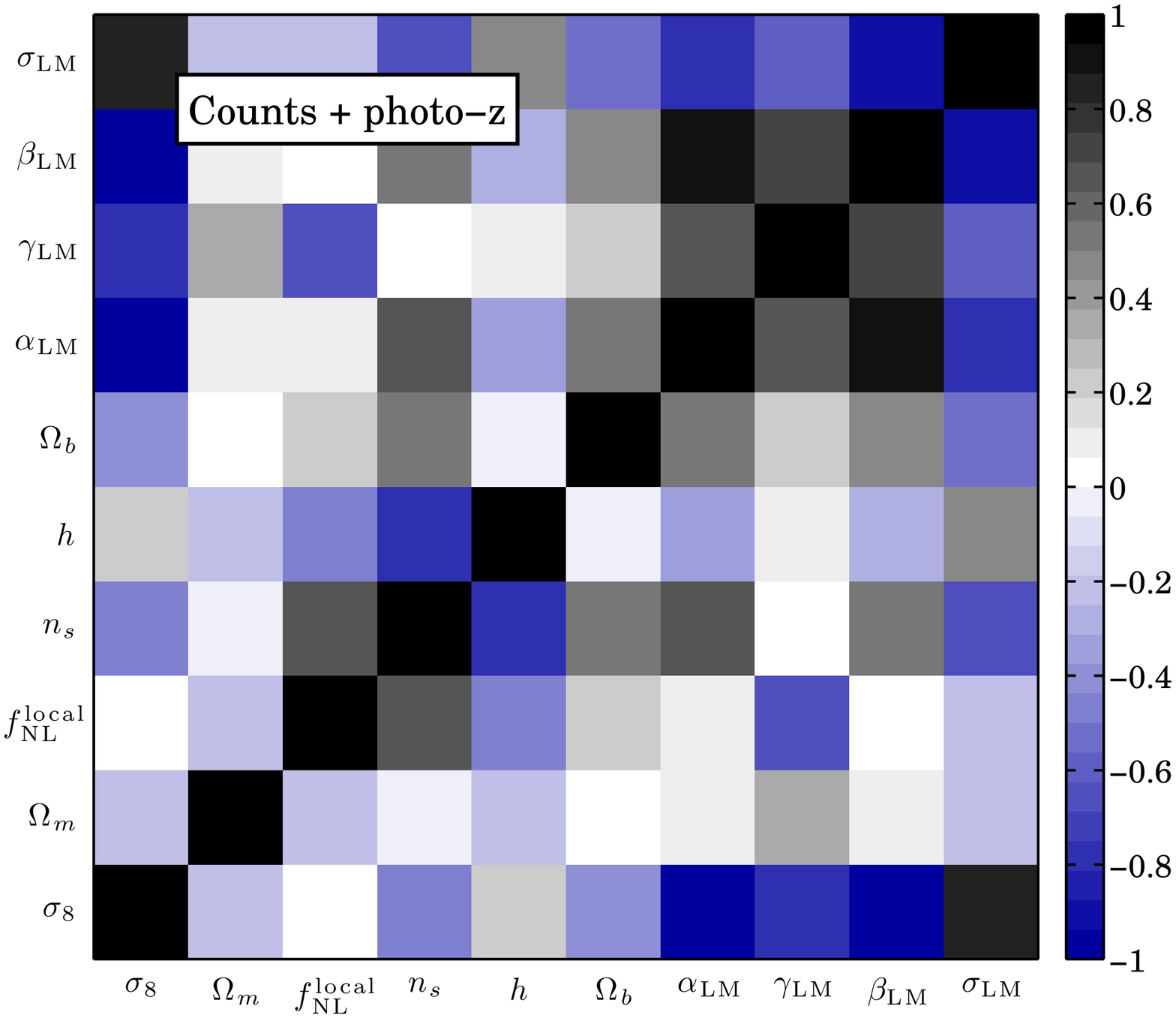}
\includegraphics[width=8cm]{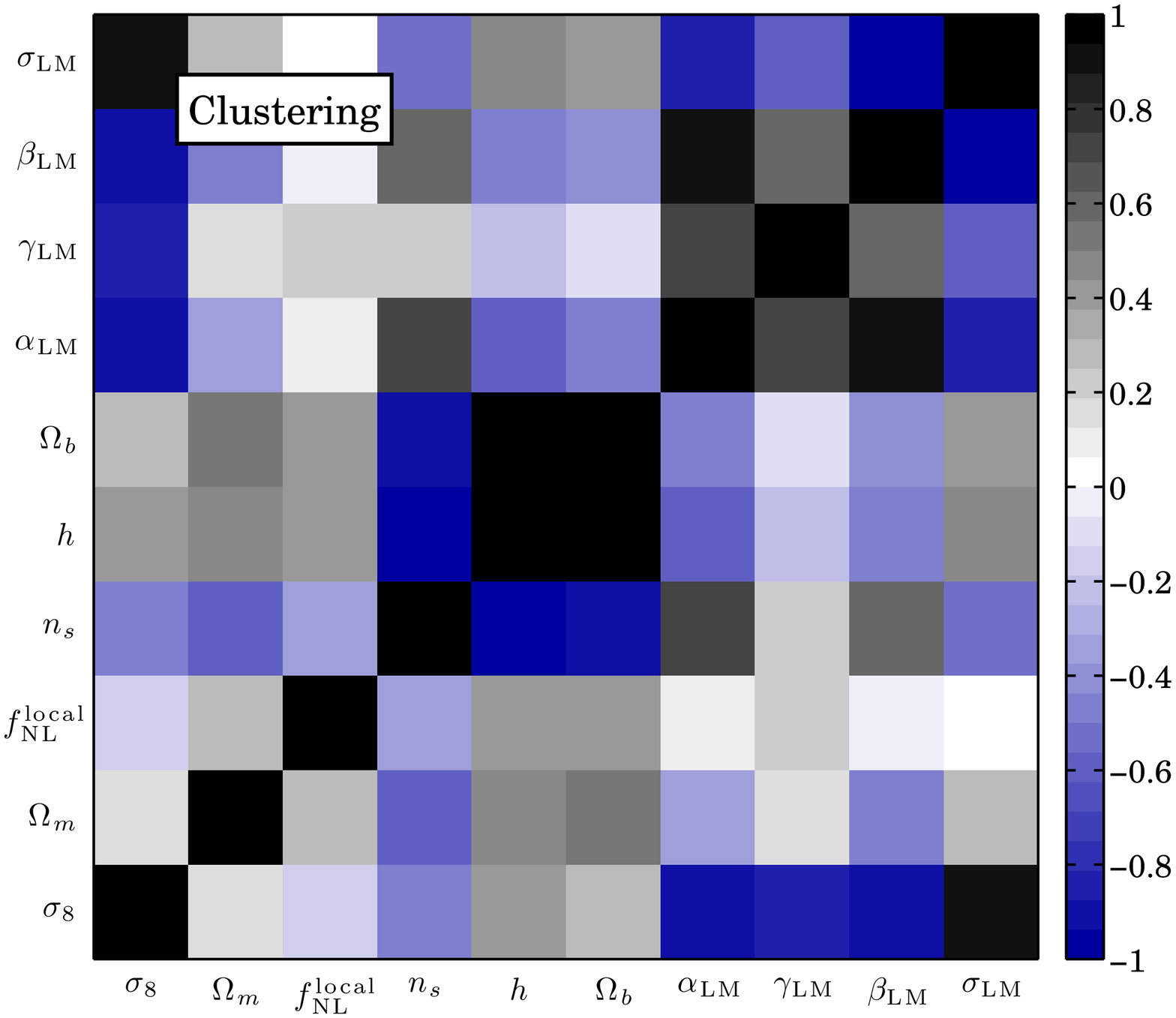}
\includegraphics[width=8cm]{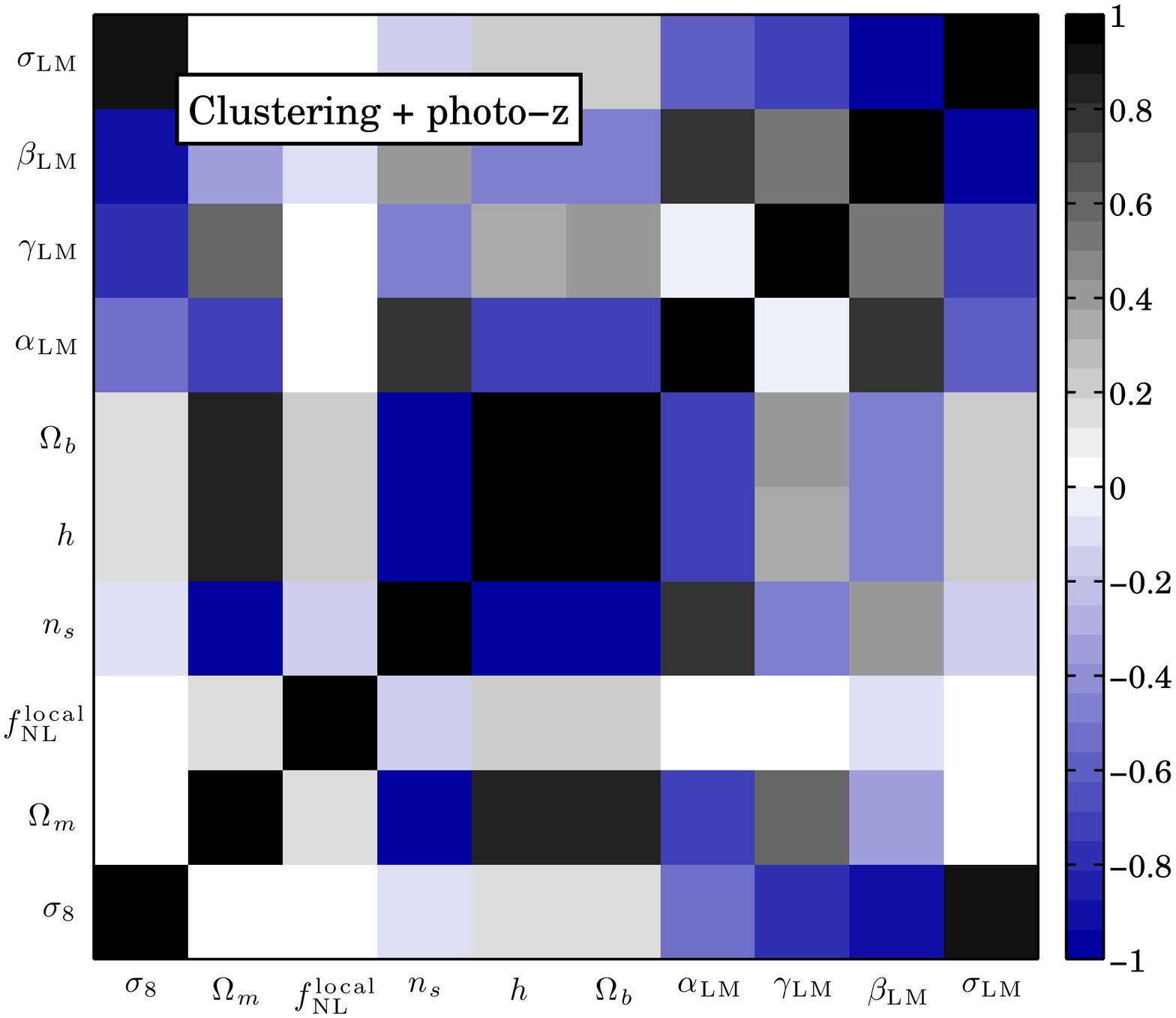}
\includegraphics[width=8cm]{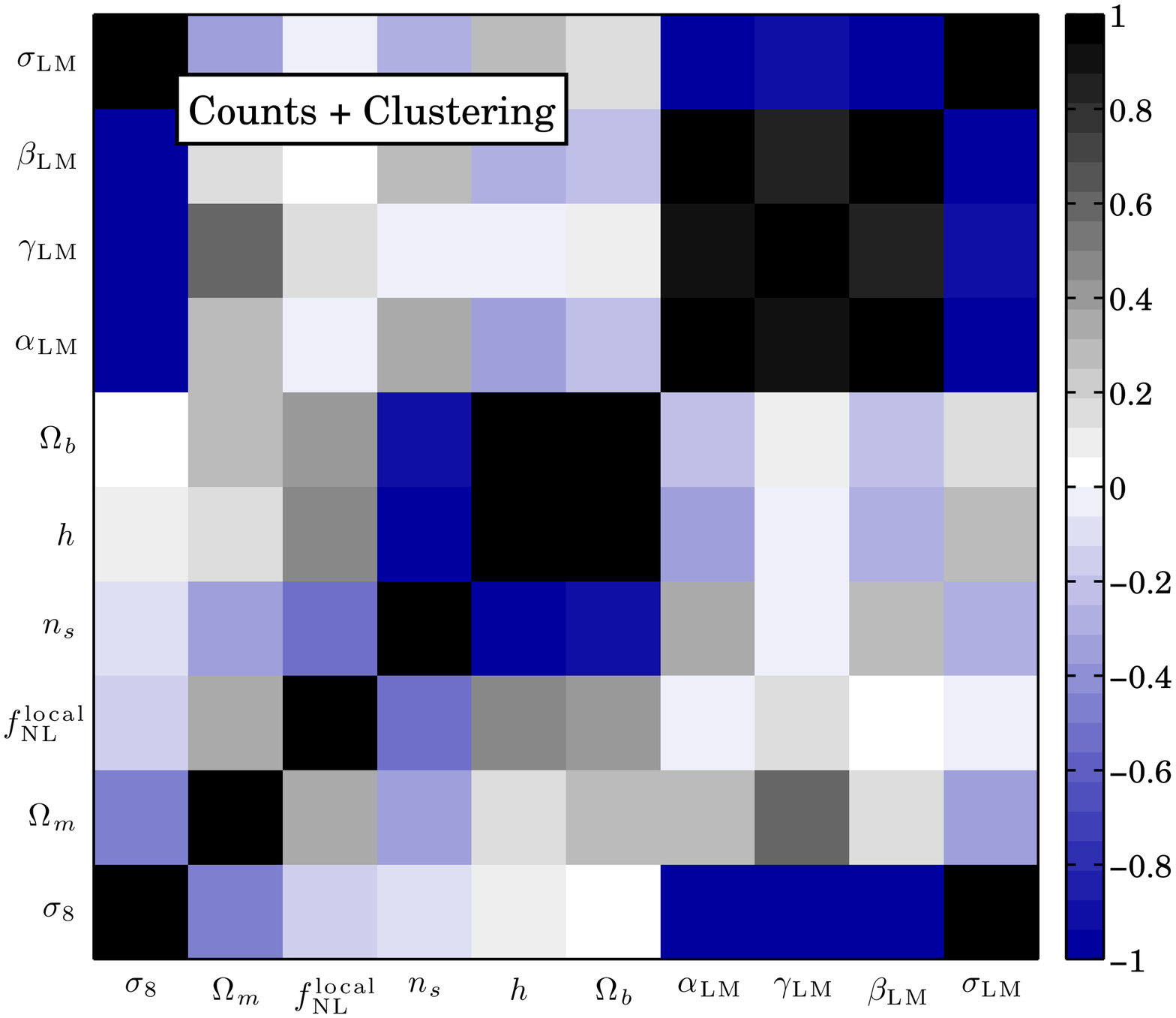}
\includegraphics[width=8cm]{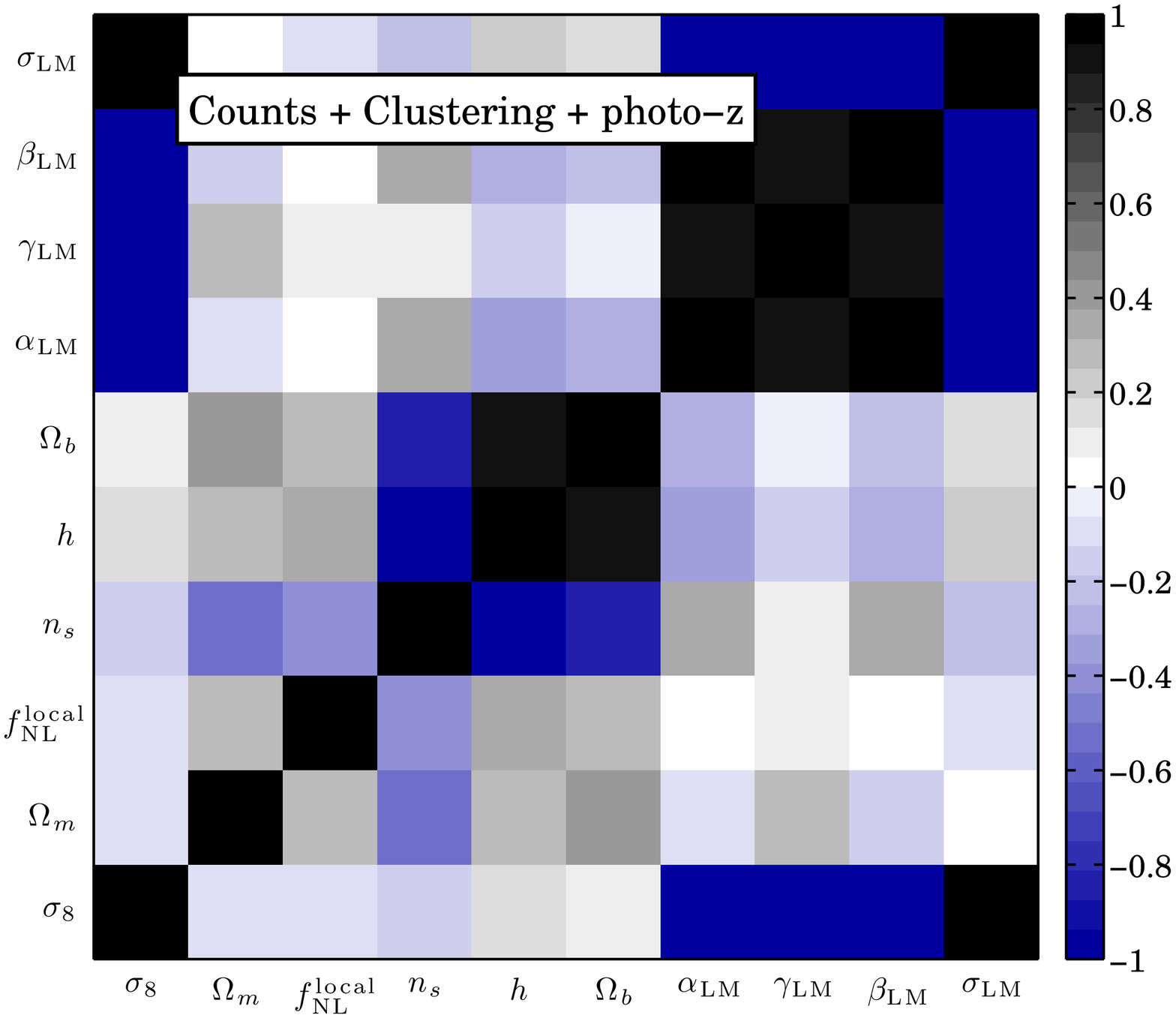}
\caption{\label{FIG:CORRELATION_MATRICES} Correlation coefficient between pairs of model parameters for the different experiments.}
\EC
\end{figure*}

We first assume having no information on the cluster redshifts and study
the constraining power of a number-count experiment. 
The first row in Table \ref{TAB:ERRORS} summarizes our (optimal) results, obtained with about 20 logarithmically-spaced bins in the range $50 \lesssim \eta \lesssim 10^5$ plus one bin extending to infinity:  
1-$\sigma$ errorbars often exceed the fiducial values of the parameters 
themselves and are thus of little interest.
We only quote errors obtained marginalizing over the entire parameter set.
Note that
the conditional errors could be as good as $\Delta \FNLL = 9$ and $\Delta \SIGMA8 \simeq 10^{-4}$. 
This indicates that covariances among parameters are strong (see Fig.~\ref{FIG:CORRELATION_MATRICES}, upper-left panel).
The most degenerate pairs are $\OM$--$\SIGMA8$, $\FNLL$--$\SIGMA8$, $\FNLL$--$\OM$, $\BETALM$--$\SIGMALM$, and $\ALPHALM$--$\SIGMA8$.
It is exactly these covariances that are reduced by optimizing the 
binning scheme in $\eta$, while conditional errors are rather insensitive
to it.

Redshift information on the individual clusters must be added if we want
to constrain the cosmological model and the physics of the ICM 
from a number-count experiment with $\eRO$. 
We thus study how the 1-$\sigma$ errorbars
for the various parameters 
improve with the accuracy of the redshift estimates.
For this calculation
binning is implemented both in photon counts and in redshift.
We consider $\sim 20$ logarithmically-spaced bins in photon counts as detailed earlier, 
while redshift bins have size $\Delta z  (1+z_{bin}) $, 
where $\Delta z$ is a parameter we vary 
and $z_{bin}$ is the median redshift in a bin. 
As soon as some redshift information is available, even as rough as with
$\Delta z\sim 0.1$, uncertainties on the parameters shrink by factors of ten (e.g. for $\SIGMA8$). This applies to the marginal errors, indicating that redshift information breaks degeneracies. The conditional errors, on the other hand, do not show any dependence on the number and size of the redshift bins. 
In Table \ref{TAB:ERRORS} we list the uncertainties that $\eRO$ should be able to place if a redshift binning with $\Delta z \sim 0.05$ is
 performed, which approximates 
 what could be achieved from photometric estimates (``Counts + Photo-$z$ '').

Spectroscopic-redshift measurements ($\Delta z \sim 0.01$) would result 
in significantly better constraints (by 30-60 per cent) on parameters like 
$n_s$, $h$, $\OB$, $\ALPHALM$ and $\SIGMALM$ with respect to what is
achievable using photometric redshifts: 
yet, $\eRO$ counts alone would not improve upon current constraints which have been
however obtained by combining different probes.
Note that the current constraints on the cosmology sector listed in Table \ref{TAB:ERRORS} 
refer to other probes than
galaxy clusters, and the ones on the ICM sector are at fixed cosmology.
Our results instead have always been obtained after 
marginalizing over both the cosmology and ICM sectors simultaneously.
Therefore a direct comparison is not completely meaningful and the entries  ``Current Errors'' should be simply considered as a reference. Note that,
if the luminosity--mass scaling-relation parameters were exactly known, 
$\eRO$ cluster counts alone would constrain $\SIGMA8$ and $\OM$ down to 
$\Delta \SIGMA8 = 0.017$ (2.1 per cent) and $\Delta \OM = 0.0086$ (3.1 per cent;  see Table \ref{TAB:ERRORS_FIXEDICM} in Appendix \ref{APP:ERRORS}).

Finally, it is worth stressing that, even with some redshift information, 
errors in parameter estimates display strong correlations.
As shown in the upper-right panel of Fig.~\ref{FIG:CORRELATION_MATRICES} (for photo-$z$ bins and a few tens of bins in photon counts),
this is particularly evident for
the triplet $n_s$, $h$, and $\FNLL$ within the cosmology sector, 
all the LM parameters among themselves, 
and for $\SIGMA8$ with all the LM parameters (see also Appendix \ref{APP:FM}).

%%%%%%%%%%%%%%%%%%%%%%%%%%%%%%%%%
\subsection{Angular clustering}
\label{SEC:ANGULARCLUSTERING}
\begin{figure}
\BC
\includegraphics[width=4.2cm]{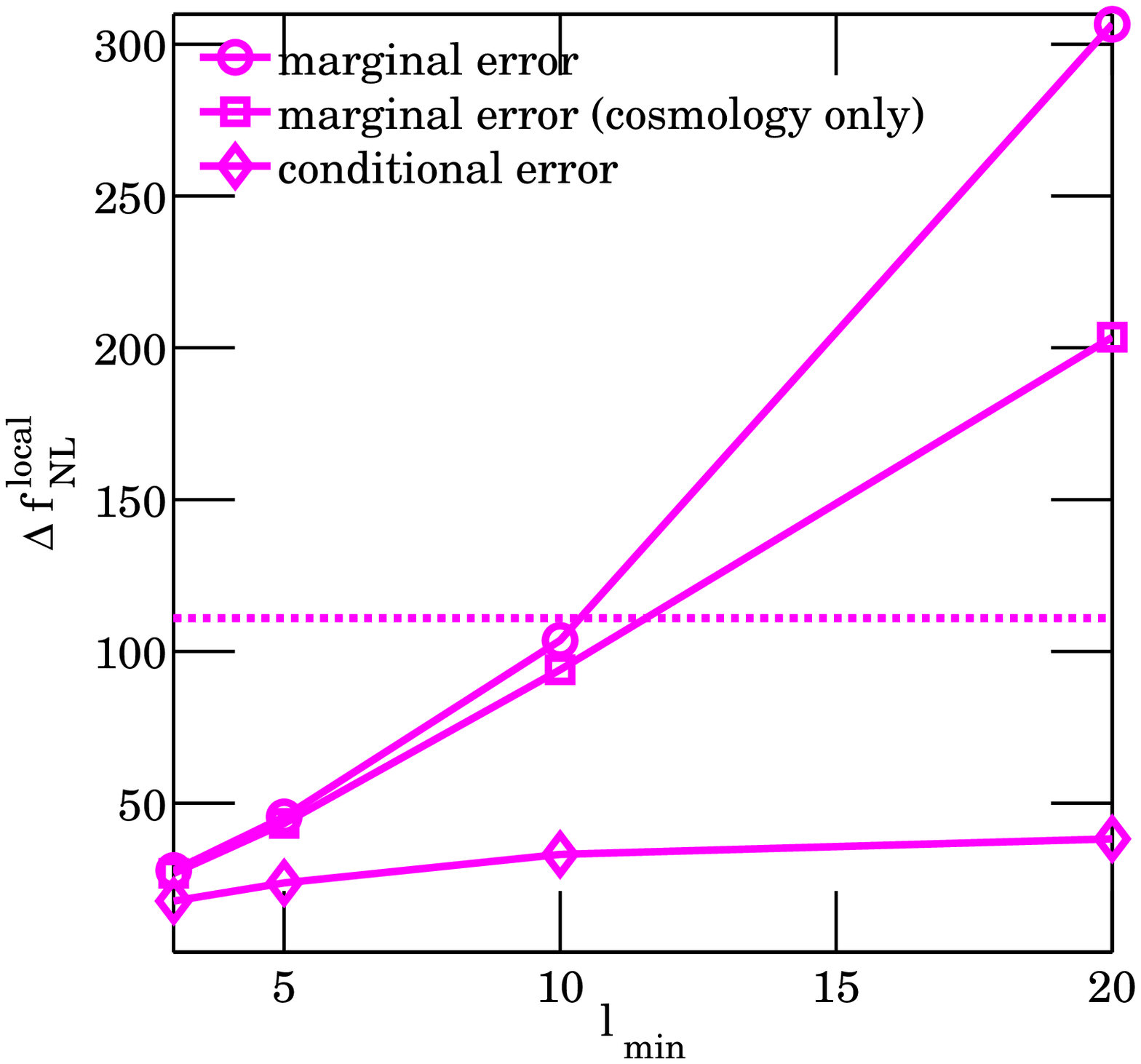}
\includegraphics[width=4.1cm]{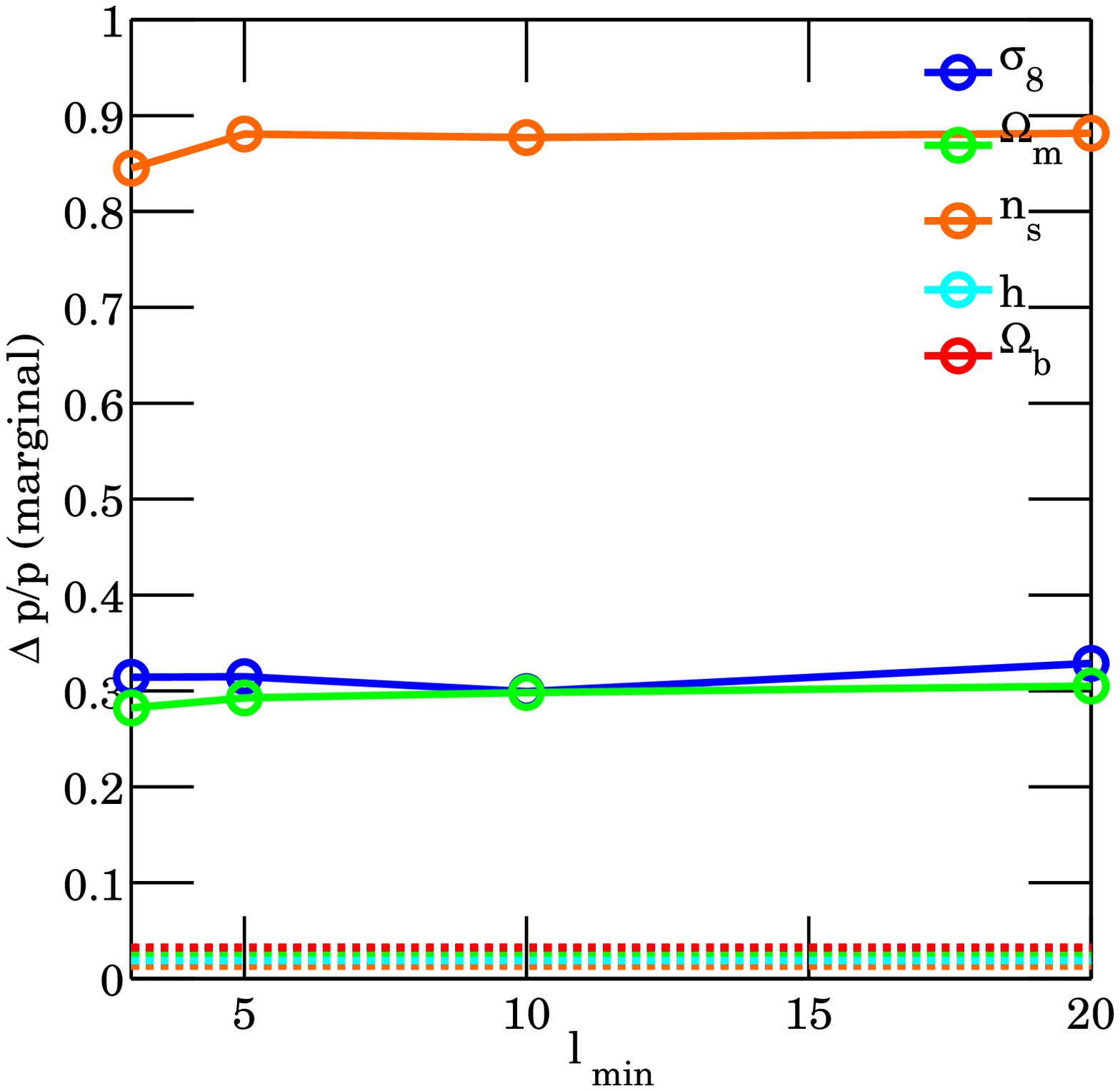}
\includegraphics[width=4.2cm]{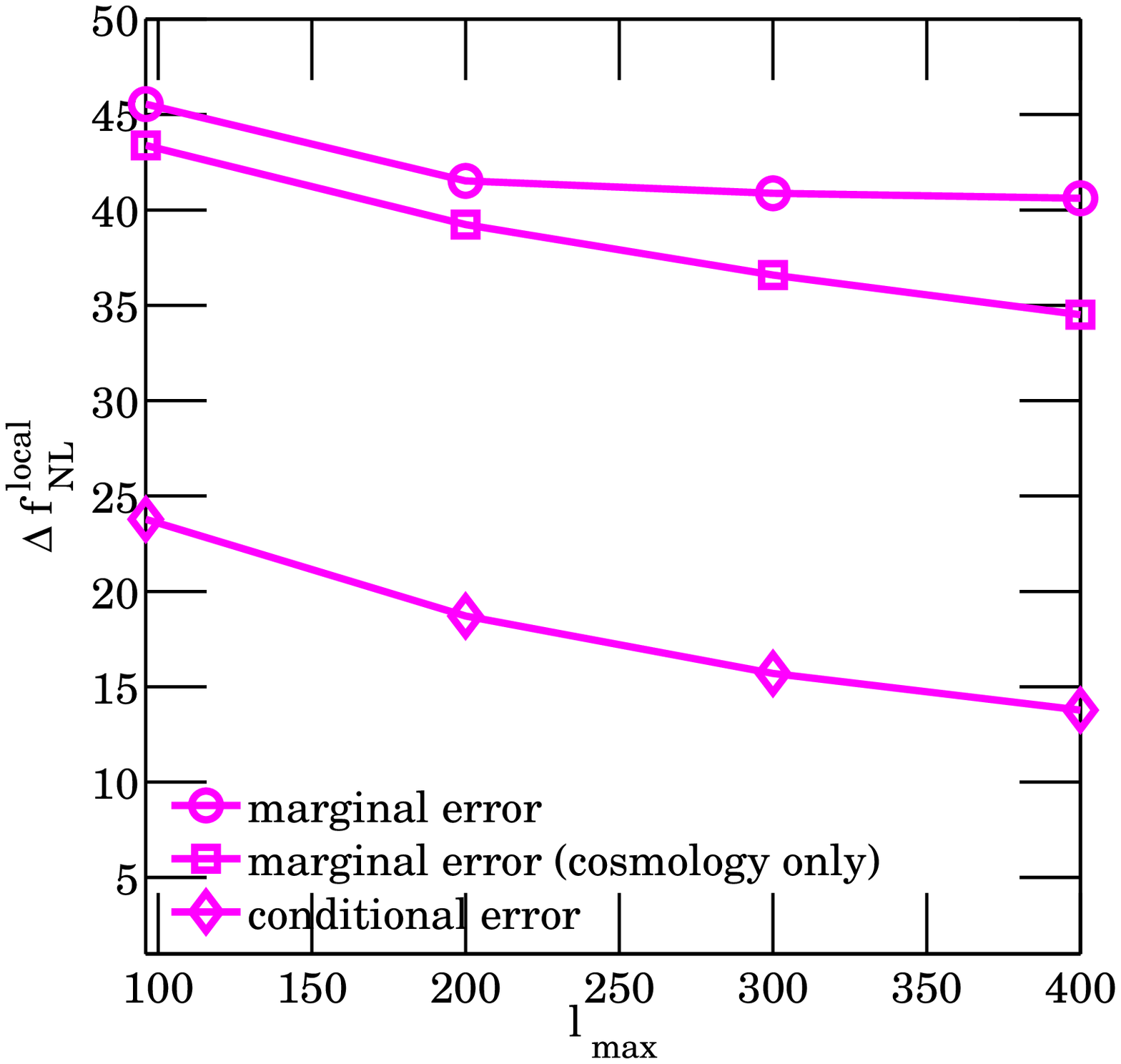}
\includegraphics[width=4.1cm]{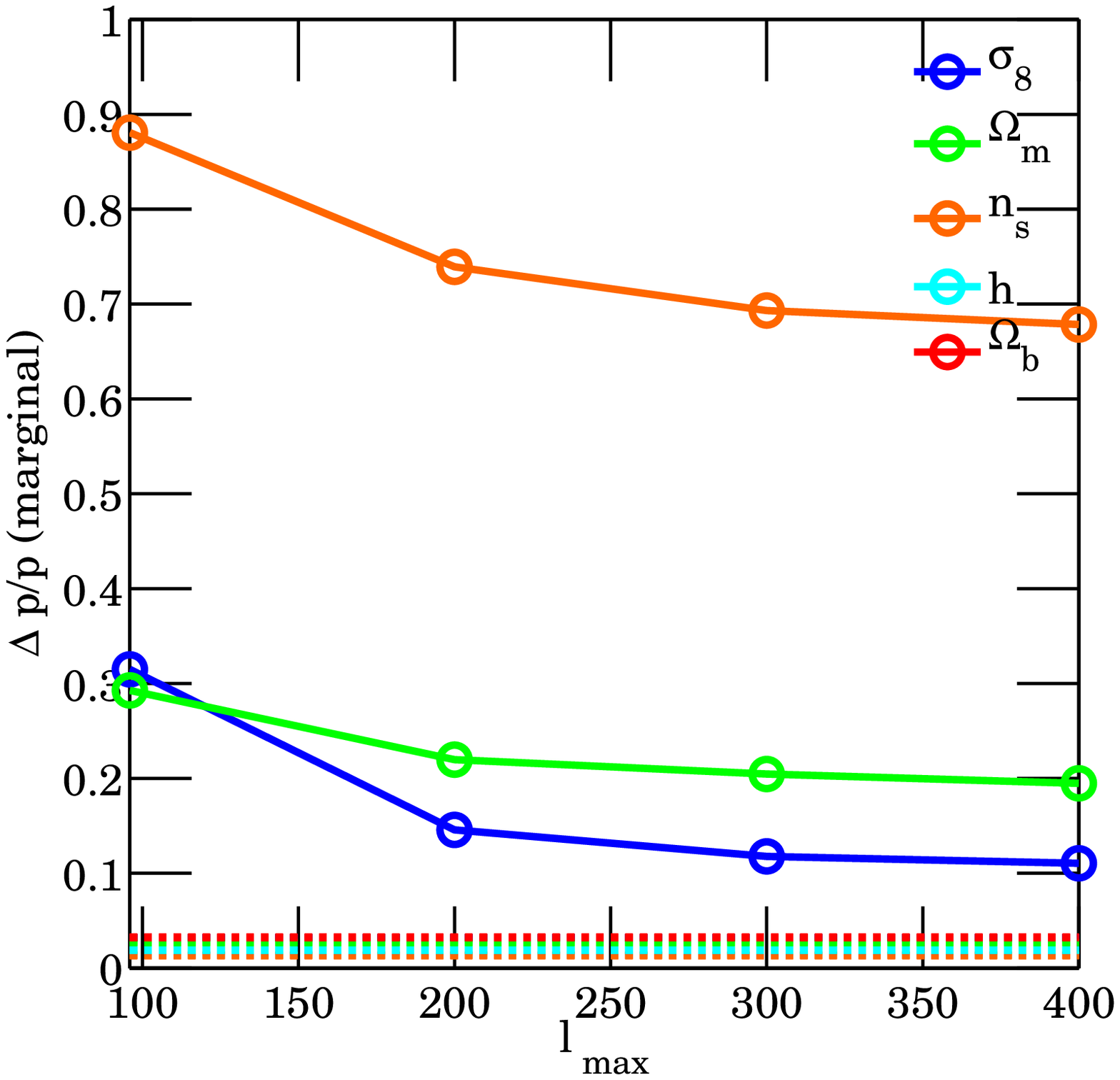}
\caption{\label{FIG:ERRORS_CLUSTERING_L} Choice of $\ell_{\rm min}$ and $\ell_{\rm max}$ in the clustering measurements:
1-$\sigma$ errors on the different parameters are shown 
as a function of $\ell_{\rm min}$ (at fixed $\ell_{\rm max} = 96$, upper panels) 
and as a function of $\ell_{\rm max}$ (at fixed $\ell_{\rm min}$ = 5, lower panels). 
The entire $\eRO$ sample is projected on the same sphere and split in 12 bins based on the photon counts $\eta$.}
\EC
\end{figure}
\subsubsection{$C_\ell$ without redshift information}
Starting from the reference configuration summarized in Table \ref{TAB:PARAM},
we want to optimize the measurement of the cluster 2-point function to extract maximal information, for the all-sky survey and with multipoles in the range $5\leq \ell \leq 96$.
We start by assuming no knowledge about the redshifts of the individual clusters.
The detected clusters will be binned only in terms of the observed photon counts, such that an approximately equal number of clusters are assigned to each bin.
Distinguishing in photon counts is fundamental to tighten the errorbars both on $\FNLL$ and all the other parameters.
The gain on $\FNLL$ due to the increment in the number of bins already converges after 6 bins: 
in fact, shot noise is not negligible already with a few $10^4$ objects per bin. 
The best constraints from the angular clustering are summarized in Table \ref{TAB:ERRORS} where we report the results obtained using 12 bins (``Angular clustering''):
while the conditional error on $\FNLL$ could be as good as $\Delta \FNLL \sim 24$, the rather poor
constraints on the other parameters increase the corresponding 
marginal constraint to $\pm 46$ at 1-$\sigma$.
Not surprisingly, the non-Gaussianity parameter presents very weak correlations with the other ones, once a few bins in photon counts are considered. On the other hand, the pairs $n_s$--$h$ and $h$--$\OB$ are extremely correlated even with the tightest binning scheme (see Fig.~\ref{FIG:CORRELATION_MATRICES}, center-left panel). Moreover the measurement of the angular power spectrum in a single redshift bin cannot break the severe degeneracies which plague the ICM parameters, as predictable by simple inspection of Fig.~\ref{FIG:CLUSTERING} (see also Appendix \ref{APP:FM}). 

Note that the minimum multipole which is considered in the analysis of the angular power spectrum impacts the estimation of the different cosmological parameters (Fig.~\ref{FIG:ERRORS_CLUSTERING_L}, where the effect of $\ell_{\rm max}$ is also shown).
After marginalizing over the whole set of parameters (and considering multipoles up to $\ell_{\rm max}= 96$), the uncertainty on $\FNLL$ improves by a factor of 6 if the minimum multipole is reduced from $\ell \sim 20 $ to $\ell \sim 5$. On the other hand, the constraints on the other cosmological parameters  are insensitive to $\ell_{\rm min}$. 
%Finally, these estimates on the impact of $\ell_{\rm min}$ on $\FNLL$ may be less severe when selecting the sample in a redshift range at higher redshift or equivalently for a sample of objects with a different redshift distribution (as with galaxy surveys like EUCLID, \textcolor{blue}{Giannantonio et al. 2011}, submitted).\\
%
\subsubsection{$C_\ell$ tomography}
How much does the constraining power of $C_\ell$ measurements improve if the redshifts of the 
clusters were available? 
To answer this question we divide the $\eRO$ sample into redshift bins of size
$\Delta z  (1+ z_{bin})$ and, additionally, into 12 bins based on the photon counts.
In the rest of the paper we will refer to this configuration as ``tomography''.
By slicing in redshift, the marginal errorbars of all parameters drop significantly, 
even by orders of magnitude. 
Four are the factors which make the tomography more powerful than measuring $C_\ell$ from the
full sample: 
i) the maximum multipole can be raised to much higher values, reaching $\ell_{\rm max} > 200$ for redshift bins above $z\sim 0.8$ with $k_{\rm max} = 0.1~h\ {\rm Mpc}^{-1} $;
(ii) the $C_\ell$ signal itself increases for thinner bins due to reduced projection effects; 
(iii) the scale-dependent bias induced by $\FNLL$ is more pronounced at high redshift and not averaged out by the populations of lower redshift less-biased objects; 
(iv) since we always consider all cross-spectra among different redshift slices, 
the number of data points scales as $n(n+1)/2$, $n$ being the number of redshift slices.
The proper inclusion of all the cross-spectra (wrt to the case where only auto-spectra are included) tightens the error budget up to 33 per cent (for $\FNLL$) when $\Delta z \sim 0.05$ 
and up to 40-60 per cent (for $\FNLL$, $n_s$, $h$, $\OB$) when $\Delta z \sim 0.01$. Note that even with $\Delta z \sim 0.01$, there are still more than a thousand objects per $z$-slice in half of the bins.

By combining the angular clustering signal from more than 10 different redshift slices ($\Delta z \sim 0.1$), it is possible to constrain primordial non-Gaussianity of the local type with a marginal error $\lesssim 10$
(the corresponding conditional error on the other hand is $\sim 5-6$ at 1-$\sigma$). On the other hand, the gain in the information about $\FNLL$ by increasing the number of redshift slices flattens much faster than almost any other parameter (see also Section \ref{SEC:DISC_SPECZ} and Table \ref{TAB:Z}).
In Table \ref{TAB:ERRORS} we
give as a reference the constraints obtained with $\Delta z \sim 0.05$
 (``Angular clustering + Photo-$z$ ''), consistently with what we adopted for the number-count experiment when photometric-redshift estimates are available.
%
%Notice that for small numbers of redshift slices, i.e. for thick redshift slices, the output result for some parameters is extremely sensitive to the scheme adopted to choose $\ell_{max}$, whether according to the median or minimum redshift of the bin. 
%
For $\FNLL$ we find a forecasted marginal error of $\sim 10$ which is very interesting and could potentially rule out entire classes of inflationary mechanisms. With an exposure of 1.6 ks, a tomography experiment is more efficient than a ``Counts + Photo-$z$ '' measurement to constrain the $LM$ parameters, but we have noticed that the situation is reversed for higher exposure times (i.e. 3 ks). On the other hand,
$\eRO$ tomography alone cannot place constraints on the other cosmological and ICM parameters 
which are by any means competitive to what is already known nowadays by the combination of CMB, BAO and SN data.
Note, however, that
assuming perfect knowledge of the luminosity--mass relation, the constraints 
imposed on $\FNLL$, $\SIGMA8$, and $\OM$ by
the $\eRO$ tomography alone would be competitive with the current ones (see Table \ref{TAB:ERRORS_FIXEDICM} in Appendix  \ref{APP:ERRORS}). If viceversa one could assume perfect knowledge of the cosmological parameters, tomography alone would largely improve our knowledge on the slope and evolution of the $LM$ relation (see in Table \ref{TAB:ERRORS} the results with Planck priors for an upper 
limit of the uncertainties).

It is worth mentioning that estimates on $\FNLL$ based on the $C_\ell$ tomography  are nearly perfectly uncorrelated from any other model parameter
(Fig.~\ref{FIG:CORRELATION_MATRICES}, center-right panel)  
although strong correlations are still present 
in the complementary parameter space. 
\begin{figure}
\BC
\includegraphics[width=9cm]{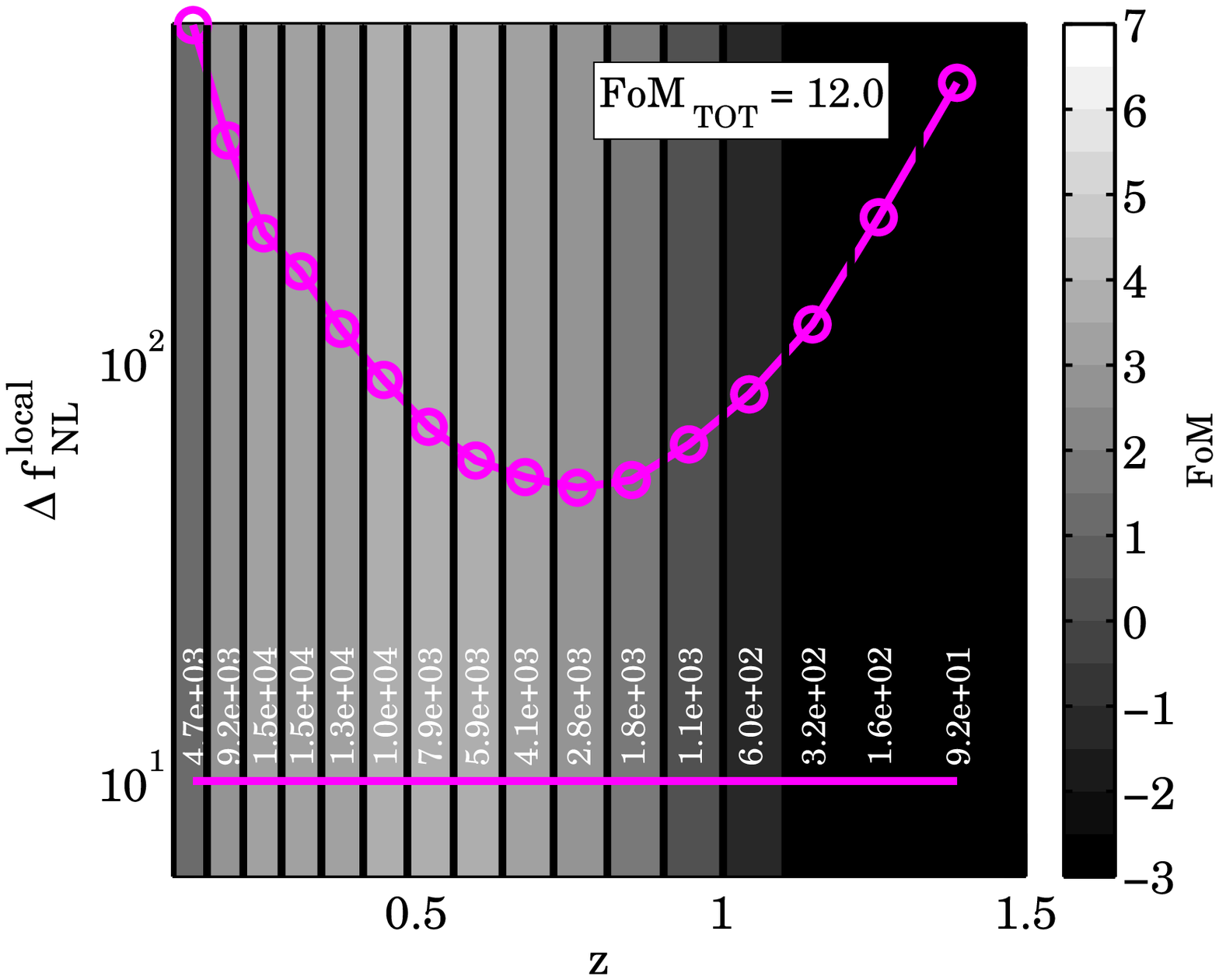}
\caption{\label{FIG:FOM_TOMOGRAPHY} Marginal error on $\FNLL$ and total Figure of Merit from different individual redshift bins. The bin width is $\Delta z  (1+z)$ with $\Delta z=0.05$ for a total of 16 bins up to $z\simeq 1.5$.
Magenta circles indicate the constraints on $\FNLL$ obtained from every individual redshift slice, the horizontal magenta line is the global constraint on $\FNLL$ obtained considering all the bins together, including cross-correlations among redshift bins. The Figure Of Merit of Eq.~(\ref{EQ:FOM}) is color coded in shades of gray for the individual $z$-bins, while the total FoM of the tomographic measurement is indicated in the upper-right box.}
\EC
\end{figure}

Fig.~\ref{FIG:FOM_TOMOGRAPHY} allows us to stress the importance of high-redshift objects: there we show how the total figure of merit (as defined in Eq.~(\ref{EQ:FOM}) and which takes into account the errorbars and correlations among the whole set of 10 parameters) and the marginal constraints on $\FNLL$ depend on the individual $z-$bin, within our binning scheme; the magenta horizontal line refers to the marginal constraint on $\FNLL$ given by the combination of all redshift slices, including cross-correlation among bins; also the number of objects in each redshift bin is indicated. The final constraints are the results of at least three factors: the thickness and the redshift of  the $z$-bin, and the number of objects in each bin. Considering only the slice with $0.76<z<0.85$ (which contains 
$\sim 2800$ clusters) gives already $\Delta \FNLL=51$.  In general, the sample of clusters with $z>z_{\rm med}$ 
better constrains the cosmological parameters (by a factor of a few)
wrt the identically sized sample with $z<z_{\rm med}$.

\subsection{Combining counts and clustering}
\begin{figure*}
\BC
\includegraphics[width=13cm]{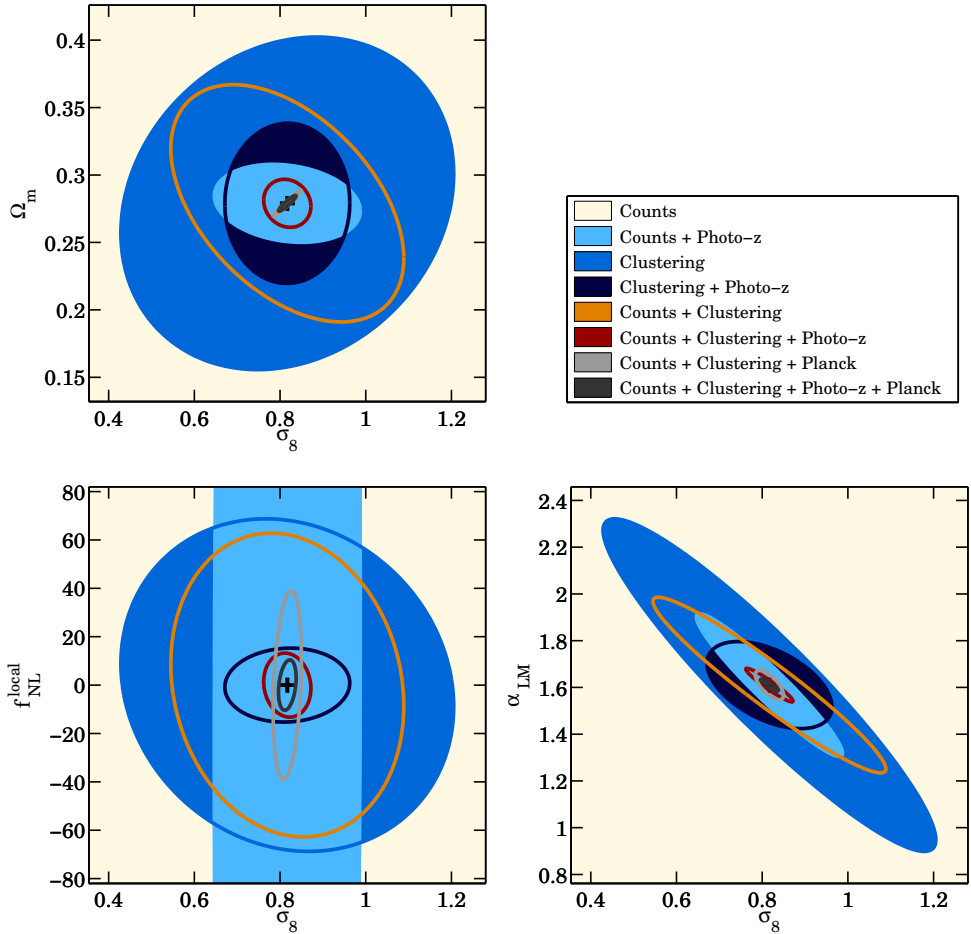}
\caption{\label{FIG:FM_SOMEPAIRS} Joint 1-$\sigma$ error ellipses for a selection of parameter pairs obtained by marginalizing over all the 
other model parameters (no priors assumed). 
Individual experiments and their combinations for the all-sky survey (described in Table \ref{TAB:PARAM}) are indicated
with different colors. Contours for the number-count experiment with no redshift information vastly exceed the area in the plot.
The complete set of error ellipses is given in Appendix \ref{APP:FM}, with priors oh $h$ and $\OB$.}
\EC
\end{figure*}

We want now to combine our results on the cluster counts and on the angular power spectrum to get tighter constraints on the model parameters.
This requires accounting for
the cross covariance between the two probes, which is proportional to 
the angular bispectrum of the clusters \citep[see e.g.][]{Takada:2007}.
For Gaussian initial conditions the bispectrum vanishes on large scales.
Cross covariances remain small also for the weakly non-Gaussian models
we consider, in spite of the large cluster bias factors. 
This is because we only consider
very large scales and rather heavily projected data.
For this reason we ignore cross covariances between the cluster number counts
and the measurement of the angular power spectrum. As a result, the total
Fisher matrix is obtained by summing up the single Fisher matrices of the
individual experiments.
The results obtained considering the joint datasets are summarized 
in Table \ref{TAB:ERRORS}, in Fig.~\ref{FIG:FM_SOMEPAIRS}, where error contours for a selection of parameter pairs are shown, and in Fig.~\ref{FIG:CORRELATION_MATRICES} where we quantify the correlations among parameter pairs for all the different probes.
Once again, we distinguish the cases in which
redshift estimates for the individual clusters are available or not.
For all parameters, measuring photometric redshifts (with $\Delta z \sim 0.05$)
shrinks the marginal errorbars by a factor of 4 or 5.
Adding information on the number counts to tomographic $C_\ell$ measurements
has little impact on $\FNLL$ but significantly shrinks the confidence interval
of the other cosmological and ICM parameters. On the other hand, ``Counts + Photo-$z$ '' alone gives better results than ``Counts + Angular clustering'' without photo-$z$ for all parameters but $\FNLL$ and $n_s$.
Properly taking into account the uncertainties on the $LM$ parameters strongly degrades the cosmological constraints which could be ideally achieved if the cluster observable--mass mapping was perfectly known (e.g. ``Counts + Angular clustering + Photo-$z$'' vs ``Counts + Angular clustering + Photo-$z$ + LM fixed'', and Table \ref{TAB:ERRORS_FIXEDICM}).

\begin{table*}
\begin{center}
\begin{minipage}{180mm}
\caption{Marginal $1-\sigma$ errors for the cosmological and the LM parameters obtained using the Fisher-matrix formalism. The label ``+Priors'' indicates results obtained adopting external priors on $h$ and $\OB$ as detailed in Section \ref{SEC:PRIORS}.
The shaded row marks the best constraints obtained using eROSITA data only. We refer to the following survey: 
$T_{\rm exp} = 1.6 {\rm ks}, ~ \eta_{\rm min} = 50,~ M_{\min} = 5 \times 10^{13} \HI \MSUN , ~ f_{\rm sky} = 0.658, ~\Delta z = 0.05$.}
\label{TAB:ERRORS}
\scriptsize
%\footnotesize
%\rowcolors{1}{}{yellow}%gray!10}{}
\BC
\begin{tabular}{l r c c c c c c c c c c }
\hline
$\eRO$ data& FoM & $\Delta\FNLL$ & $\Delta\SIGMA8$ & $\Delta\OM$ & $\Delta n_s $& $\Delta h$ & $\Delta\OB$  & $\Delta\ALPHALM$ & $\Delta\GAMMALM$ & $\Delta\BETALM$ & $\Delta\SIGMALM $\\
\hline
Counts	 											& 1.0	 	&  $\sim 9\times 10^3$	&$\sim 1.6 $ 	& $\sim .5$	& $\sim 4$ 	&$\sim 4 $	&$\sim .3$ 	& $\sim1.8$	&$\sim 7  $ 	& $\sim 9$	&$\sim 3$ \\
Counts + Priors	 										& 4.6	 	&  $\sim 8\times 10^3$	&$\sim 1.5 $ 	& $\sim .4$	& $\sim 2$ 	&.080		&.0113		& $\sim1.7$	&$\sim 7  $ 	& $\sim 7$	&$\sim 2$ \\
Counts + Photo-$z$ 										& 10.7	& 423				&.113 		& .0191  		& .559		&.558		&.0649 		& .20		&.75 		& $\sim1$		& .277 \\ 
Counts + Photo-$z$ + Priors								& 12.8	& 360				&.100 		& .0188  		& .205		&.078		&.0110 		& .16		&.73 		& $\sim1$		& .202 \\ 
\hline
Angular clustering				 						& 7.1	 	& 46					&.257		& .0817		& .845		&$\sim 1 $	&.0974 		& .47		&$\sim 1$		&$\sim3$		&$\sim1$\\
Angular clustering + Priors	 							& 9.1	 	& 42					&.226		& .0693		& .256		&.068		&.0100 		& .36		&$\sim 1$		&$\sim3$		&$\sim1$\\
Angular clustering + Photo-$z$ 							& 12.0 	& 10.1 				&.097		& .0393		& .264		&.299		&.0232		& .12		&.47			&$\sim1$ 		&.247\\
Angular clustering + Photo-$z$ + Priors	 					& 13.2 	& 9.8 				&.095		& .0207		& .076		&.028		&.0033		& .08		&.43			&$\sim1$ 		&.242\\
\hline
Counts + Angular clustering 								& 10.6 	& 42					&.180 		& .0582		& .530		&.967		&.0736		& .24		&$\sim 1$		&$\sim2$		&.621\\
Counts + Angular clustering + Priors							& 12.5 	& 37					&.169 		& .0531		& .154		&.064		&.0089		& .22		&$\sim 1$		&$\sim2$		&.557\\
\rowcolor[gray]{.85}
Counts + Angular clustering + Photo-$z$ 						& 16.3	& 8.8					&.036 		& .0118		& .088		&.153		&.0114		& .05		&.20 			&.397		&.117\\
Counts + Angular clustering + Photo-$z$  + Priors				& 17.2	& 8.2					&.036 		& .0111		& .033		&.027		&.0030		& .05		&.20 			&.384		&.114\\
\hline
Counts + Angular clustering + LM fixed						& - 		& 36					&.016 		& .0099		& .172		&.461		&.0464		& -		&-			&-			&-\\
Counts + Angular clustering + Photo-$z$ + LM fixed				& -		& 8.4					&.003 		& .0029		& .055		&.110		&.0092		& -		&- 			&-			&-\\
\hline
Counts + Angular clustering + Planck						& 19.7 	& 26					&.022 		& .0065		& .004		&.005		&.0005		& .04		&.16			&.348		&.137\\
Counts + Angular clustering + Photo-$z$ + Planck				& 22.1	& 6.9					&.014 		& .0039		& .003		&.003		&.0003		& .02		&.07 			&.173		&.045\\
\hline
\hline
Current Errors\footnote{WMAP7+BAO+SN for the Cosmology sector independent of the ICM sector, \cite{Komatsu:2011}; results by \cite{Vikhlinin:2009b} for the $LM$ parameters, at fixed Cosmology. 95.4 per cent CI for $\FNLL$}		& - 		& [-10,+74] & .024 		& .0061 		& .012 		&.014 		& .0016 &.14 		& .42 		& .085 		& .039 		 \\
Planck Errors \footnote{Error estimates for future power-spectrum measurements with Planck: marginalization solely over the cosmology sector, excluding $\FNLL$}					& - 		& -					&.024 		& .0071		&.004		&.006		&.0006		& -			&- 			&-			&-\\ 
\hline
\end{tabular}
\EC
\end{minipage}
\end{center}
\end{table*}
\subsection{Adding external priors}
\label{SEC:PRIORS}
X-ray cluster counts and power spectra (with and without photometric redshifts) cannot simultaneously determine {\it all} the model parameters (see Table 3 and Fig. 11): in particular the triplet $n_s$, $h$, and $\OB$ are left unconstrained by $\eRO$ data. We therefore complement $\eRO$ with strong standard priors on the Hubble parameter \citep[from the Hubble Key Project $\Delta h = 0.08$, ][]{Freedman:2001} and on the mean baryon density of the universe \citep[from primordial nucleosynthesis, $\Delta (\OB~ h^2) = 0.002$, ][]{Kirkman:2003}, as commonly done in X-ray studies \citep[e.g.][]{Mantz:2010a}. The sensitivity of our results to these priors is shown in Table \ref{TAB:ERRORS} (``+ Priors ''). The parameter that benefits the most is the spectral index, while all the other cosmological and ICM parameters are rather insensitive to them (at least for the joint constraints with photo-$z$). While constraints on $h$ and $\OB$ are dominated by the imposed priors for the abundance experiments and the clustering without redshifts, $\eRO$'s tomography (and thus the combinations of the experiments) significantly contributes to further shrink those errors.

%So far we have just used $\eRO$ data to constrain the model parameters.
%It makes then sense to explore how much we would gain by combining them with
%additional cosmological probes.
We focus now on the CMB analysis performed by the Planck satellite
whose results will be available by the time the $\eRO$ 
all-sky survey will be completed.
We consider the Planck Fisher matrix for a measurement of the power spectrum
of temperature anisotropies calculated by the Dark Energy Task Force \citep{Albrecht:2009}, kindly made available to us by Dragan Huterer and Wayne Hu.
This constrains all cosmological parameters but $\FNLL$ (we do not consider
the CMB bispectrum in this work). 
The results obtained using the Planck forecast as a prior for our $\eRO$ 
analysis are
shown in Table \ref{TAB:ERRORS}:
the constraints on all parameters are strongly improved, in particular for $n_s$, $h$, $\OB$ (where the improvement reaches a factor of 30!).
Yet, $\eRO$ significantly contributes to the determination of all parameters. Cosmological constraints on $\sigma_8$ (which is not a ``natural'' choice to fit CMB data) and $\OM$ will be known at $\sim 1.5-2$ per cent accuracy, while constraints such as  $\Delta \FNLL \sim 7$
will be comparable to the results of the CMB three-point statistics (but on different spatial scales). Similarly, 
all the ICM parameters will be more accurately determined than they are
nowadays (see error ellipses in Figs.~\ref{FIG:ELLIPSES_COSMO},~\ref{FIG:ELLIPSES_NUISA}, and \ref{FIG:ELLIPSES_INTER}). Note that in terms of $LM$ parameters only, adding Planck priors to the cosmological sector (``Counts + Angular clustering + Planck'') is almost equivalent to measuring photometric redshifts for the whole cluster sample (``Counts + Angular clustering + photo-$z$'').
A comparison between the constraining capabilities of Planck and $\eRO$ 
is given in Table \ref{TAB:ERRORS_COSMONONG} in Appendix \ref{APP:ERRORS}, where the analysis is repeated exclusively for the cosmology sector and 
assuming Gaussian initial conditions (5-parameter fit).
Note that, if we could perfectly characterize the X-ray cluster scaling-relations, $\eRO$ would outperform Planck in the determination of $\sigma_8$ while, even in this idealistic scenario, Planck would still do significantly better for $n_s$, $h$, and $\OB$.
%

%%%%%%%%%%%%%%%%%%%%%%%%%%%%%%%%%
\section{Discussion}
\label{SEC:DISCUSSION}

\subsection{Wide or deep surveys?}
\label{SEC:SURVEYCOMP}
Is there any advantage in undertaking a deeper survey than the all-sky one analyzed so far? 
We consider here an hypothetical survey with average exposure time of 
7.5 ks. Accordingly, we assume a sky coverage of 6,000 deg$^2$ (with $\ell_{\rm min}\sim 9$), which gives approximately the same
amount of observing time used for the all-sky survey.
Although the higher exposure time enables the detection of fainter 
clusters at high redshift, the limited sky coverage significantly reduces
the total number of sources. With a detection limit of 50 photons and 
a minimum mass of $5 \times 10^{13}~ \HI  \MSUN$, 
the deeper and less extended survey should identify
$6.85 \times 10^4$ clusters with a median redshift of 0.56. 

These changes impact the cosmological constraints in different ways.
For a number-count experiment, the deeper survey better constrains some
parameters such as $\FNLL$, $n_s$, and $\GAMMALM$ (15-30 per cent improvements) but 
is not optimal overall.
On the other hand, clustering studies clearly benefit from the all-sky
coverage (parameter constraints improve by many tens per cent), 
especially if one wants to measure $\FNLL$ 
through the scale-dependent bias which is only evident on large projected angular scales. 
\begin{table*}
\begin{center}
%\small
\begin{tabular}{l c c c c c c }
\hline
Surveys 				& area (deg$^2$) & $T_{\rm exp}$(ks) 	&$N_{\rm objects}$ & $\Delta\FNLL$ & FoM$^{\rm Cosmo} $&FoM$^{\rm ICM}$   \\
\hline
$\eRO$ 				& $27`000$ & 1.6	& $9.32 \times 10^4$	& 8.8		& 8.0	 & 5.3\\
Deeper $\eRO$ 		& $27`000$ & 3.0 	& $1.61 \times 10^5$         & 6.5         & 8.6 &  5.6\\           
%Deeper $\eRO$ II		& [$27`000$; 7.5]	& $3.10 \times 10^5$         & 4.6         & 9.4 &  6.1\\
Focussed $\eRO$		&  $6`000$ & 7.5	& $6.85 \times 10^4$         & 21.1       & 6.4 & 4.3\\
1/2 $\eRO$			& $13`500$ & 1.6	& $4.66 \times 10^4$         & 18.9       & 6.5 & 4.6\\
(1/2 + 1/2) $\eRO$ 		& $27`000$ & 1.6	& $9.32 \times 10^4$         & 13.4       & 7.3 & 5.1\\
					& 					& 	        	                            &                &        &\\
$z \le 1 ~\eRO$ 		& $27`000$ & 1.6	& $9.19 \times 10^4$         & 9.5         & 7.8 & 5.2\\
``The magnificent 1000''	& $27`000$ & 1.6	& $\sim$1000                      & 41          & 2.0 & 1.1\\								
\hline
\end{tabular}
\caption{ \label{TAB:SURVEY} 
Forecasted performance of different survey strategies (no priors assumed). The reference case (labelled $\eRO$) refers to the all-sky survey extensively discussed in Table 
\ref{TAB:ERRORS}. 
In all cases we consider a detection limit of 50 photons and a minimum cluster
mass of  $5 \times 10^{13}~ \HI  \MSUN$.
Results are given for the joint analysis of number counts and angular 
clustering with photometrically determined cluster redshifts,
i.e. the data are split in redshift bins of width $0.05 (1+z)$. 
We use here separate figures of merit to distinguish the cosmological parameters from those of the LM relation.}
\end{center}
\end{table*}
In Table \ref{TAB:SURVEY}, 
we report the total number of detected clusters and 
some constraints on the model parameters that can be obtained
from the joint analysis of cluster counts and $C_\ell$ tomography (both with redshifts of photometric quality) for a series of survey strategies, where the covered sky fraction and the average exposure time are changed with respect to our reference choice.
For example, the 3 ks survey over 27'000 deg$^2$ could be realized if, instead of performing three years of pointed observations after the first four years of the $\eRO$ mission, the all-sky survey would be continued for the remaining lifetime of the satellite. This, of course, would have the drawback
that the planned pointed cluster follow-up observations could not be performed. The mild improvement in the resulting constraints 
might not justify this possible extension of the all-sky survey.
\subsection{Does 1/2+1/2 equal 1?}
It has been agreed that $\eRO$ data will be equally split between the Russian 
and the German consortia: each of them will own an equal fraction of the sky.
We want to investigate what would be the impact of performing separate analyses
for the two half-sky surveys.
Two effects have to be considered: halving the number of objects and 
the loss of the largest angular spatial separations where the effect of 
primordial non-Gaussianity is larger.
As an example, we report how passing from $f_{\rm sky} = 0.658$ 
to $f_{\rm sky} = 0.33$ (with $\ell_{\rm min}\sim 7$)
impacts the constraints on $\FNLL$ and $\SIGMA8$.
We find that the joint marginal errors for the all-sky survey
$\Delta \FNLL = 8.8$ and $\Delta \SIGMA8 = 0.036$
degrade to
$\Delta \FNLL = 18.9$ and $\Delta \SIGMA8 = 0.062$. 
Combining the results from the two halves of the sky a posteriori 
would then give: $\Delta \FNLL = 13.4$ and $\Delta \SIGMA8 = 0.044$
(where we have optimistically assumed that the two halves of the sky are statistically independent).
\subsection{The case with no clusters beyond $z \sim 1$ and\\ ``the magnificent 1000''}
Many parameters of interest exhibit stronger effects on the cluster number counts and $C_\ell$ at higher redshifts than at lower redshifts: this is particularly pronounced for $\FNLL$, but also evident for some of the ICM parameters.
Measurements at high redshifts may be difficult or even out of reach for at least two reasons: i) the higher the redshift, the more problematic the AGN contamination might be; ii) the higher the redshift, the lower the chances of 
successfully measuring a (photometric or spectroscopic) redshift are.
We briefly compare here the results discussed so far with the pessimistic scenario where no clusters beyond redshift $z \sim 1$ can be included in the analysis. The total number of objects which would be lost beyond redshift 1 is about 1200 (for $T_{\rm exp} = 1.6$ ks; three times higher with a double exposure time). With redshift bins of width $0.05 (1+z)$, the constraints on $\FNLL$ would deteriorate to $\Delta \FNLL = 517$ and $\Delta \FNLL = 11.1$ for a number-count and a  tomography experiment, respectively (see also Table \ref{TAB:SURVEY} for the
joint constraint). 
Similarly,
the loss of accuracy in measuring all the other parameters would be more pronounced for number counts, with relative deteriorations of about 10-20 per cent.
%
%Such a deterioration may look like less dramatic than expected: on the other hand, the sample up to redshift 1 is already optimally exploited. Moreover, such comparison does not say anything about the constraining power of the high-redshift objects alone.

Let us now imagine to only use the most massive $1000$ clusters at $z \ge 1$ for our analysis (``the magnificent 1000''). This sample includes objects with masses above $\sim 2.2 \times 10^{14} \HI \MSUN$, possibly similar to what Sunyaev--Zel'dovich surveys might yield in the future. 
The joint analysis of number counts and $C_\ell$ tomography with this subsample would give a marginal error of $\Delta \FNLL \sim 41$ (with photometric-redshift information). The attractive conditional error of $\Delta \FNLL \sim 12$ is in practice uninteresting because plagued by huge systematics and parameter degeneracies: with the ``magnificent 1000'' only the exponential cut off of the mass function is probed, where the models are more uncertain and where every parameter is in practice degenerate with another. Note that for the analysis with the whole $\eRO$ sample, the comparison between conditional and marginal error is much less dramatic, reading $\Delta \FNLL \sim 4.4$ vs $\Delta \FNLL \sim 8.8$ (see Table \ref{TAB:ERRORS}).
In any case, the marginal constraints on all the other parameters would be much less appealing (Table \ref{TAB:SURVEY}).
Combining the analysis of the magnificent 1000 with Planck priors 
would still give $\Delta \FNLL \sim 26$ and $\Delta\SIGMA8 = 0.024$ (corresponding to a relative error of 2.9 per cent), a figure which is destined to further improve with multi-wavelength follow-ups of the cluster sample and consequently better mass proxy measurements. 
\begin{table}
\begin{center}
\scriptsize
\begin{tabular}{lccc}
\hline
$\eRO$ Data &FoM  & $\Delta \FNLL $ & $\Delta \SIGMA8$ \\
\hline
\hline
Counts	 						& 1.1		& $\sim 9\times 10^3$ 	& $\sim1.6$ \\
Counts + Photo-$z$  					& 10.7	& 423 			& .113\\ 
Counts + Spectro-$z$ 				& 11.2	& 370 			& .095\\
\hline 
Angular clustering				 	& 7.1 	& 46				& .257\\
Angular clustering + Photo-$z$   			& 12.0 	& 10.1 			& .097\\
Angular clustering + Spectro-$z$  			& 14.5	& 7.8 			& .059\\
\hline
Counts + Angular clustering 			& 10.6 	& 42 				& .180\\
Counts + Angular clustering + Photo-$z$  	& 16.3	& 8.8				& .036\\
Counts + Angular clustering + Spectro-$z$ 	& 17.7	& 7.0 			& .024\\
\hline
\end{tabular}
\caption{\small \label{TAB:Z} Forecasted total figure of merit, and 1-$\sigma$ marginal errors on $\FNLL$ and $\SIGMA8$ 
obtained with the $\eRO$ cluster sample. First, no redshift information is assumed. Then the data are sliced in redshift bins of width $\Delta z  (1+z)$, with $\Delta z$ = 0.05 and 0.01 (mimicking photometric vs. spectroscopic redshift estimates). No priors assumed.}
\end{center}
\end{table}
%
%\subsection{Binning with spectroscopic-quality redshifts}
\subsection{Spectroscopic redshifts}
\label{SEC:DISC_SPECZ}
Our reference forecasts assume that photometric redshifts of the individual clusters will become available. However, given that spectroscopic follow-ups of the $\eRO$ survey are being proposed, in Table \ref{TAB:Z} we present new results that refer to the optimistic scenario where spectroscopic redshifts for all the detected clusters will be available. In this case we use redshift bins of thickness $0.01(1+z)$. For all model parameters improvements are much more pronounced for the $C_\ell$ tomography rather than in a count experiment.
Marginal constraints on all the cosmological parameters but $\FNLL$ and $\SIGMA8$ exhibit an improvement of more than 50 per cent when passing from photometric- to spectroscopic-redshift quality with a tomography measurement alone; all parameters but $\FNLL$ get constrained 30-40 per cent tighter when joint constraints with spectroscopic redshifts are considered wrt to the photo-$z$ case. Overall, the total Figure of Merit would improve substantially if spectroscopic redshifts were available. 
\subsection{The mass cut}
\label{SEC:DISC_MCUT}
So far we have always applied a
cut in the $\eta$--$z$ plane which corresponds
to a cluster mass of $M=5 \times 10^{13} ~h^{-1}\MSUN $ in the fiducial model.
Moving this threshold significantly changes the selected cluster population
both in terms of their sheer numbers and redshift distribution
(see Table \ref{TAB:NTOT} for details).
Of course there are more objects when the mass threshold is lowered
and this formally produces better parameter constraints from counting experiments.
On the other hand, a lower threshold reduces the effective bias of the
cluster population (and thus the non-Gaussian correction) and moves also down 
the median redshift (and thus $\ell_{\rm max}$ for the clustering measurement): for the clustering experiments, trends are different for different parameters.
The joint constraints for $\FNLL$ are rather insensitive to the mass threshold, while the parameters which are most affected by the exact value of the mass cut are
$\SIGMA8$ and $\GAMMALM$ 
(their constraints improve by 30 per cent when $M_{\rm cut}=10^{13} ~h^{-1}\MSUN$).

A note of a caution is in order here.
In a meaningful Fisher-matrix study the sample definition (and thus
the mass cut) must not change with 
the model parameters. We identify the locus in the
$\eta$--$z$ which corresponds to a cluster with the threshold mass 
in the fiducial model and never modify it when we compute the derivatives
with respect to the parameters.
This mimics what would be done in practice to select high-mass objects 
out of an observational sample. The sample selection is done before computing
the statistical quantities.
For this reason it would be incorrect to impose a cut based on the actual
mass or X-ray luminosity of the clusters as sometimes it has been done in the literature 
(unless mass or luminosity are a direct observable, which is not the case for $\eRO$).
We have checked that adopting this procedure 
leads to unrealistically optimistic forecasts.
\section{Other non-Gaussian models}
\label{SEC:NONLOCALFNL}
Primordial non-Gaussianity of the local type is only one (the simplest) of the countless ways to 
perturb the primordial gravitational potential around the Gaussian assumption 
\citep[e.g][]{Chen:2010}. 
%It is also the template that exhibits the strongest features in terms of modifications to the dark matter halo mass function and bias, among the most popular alternatives in the literature.\\
Here we present the error budget that $\eRO$ should achieve shall the data be fitted with two other models/templates for early non-Gaussianity, 
the so-called orthogonal and equilateral types.
These models differ from the local one in the sense that the bispectrum of the Bardeen's potential peaks at different triangle configurations 
and are characterized by a single parameter,  $\FNL^{\rm ortho}$ or  $\FNL^{\rm equi}$, which quantifies the amplitude of such bispectrum \citep{Senatore:2010,Creminelli:2006}.
While the local shape is expected in all models where non-linearities develop outside the horizon (e.g. in multi-field inflation), equilateral non-Gaussianity 
is produced in single-field inflationary models with a non-minimal 
Lagrangian where the
mode coupling is created by non-canonical kinetic terms or 
higher-derivative operators (e.g. in ghost inflation or DBI\footnote{Dirac-Born-Infeld	} inflation).
Finally, the orthogonal template reproduces (among the others) non-Gaussian features arising from assuming non-standard vacuum choices.
These three shapes are ``orthogonal'' to each other \citep{Babich:2004, Senatore:2010} and linear combinations of them can arise in physically-motivated models of the early universe.

\begin{table*}
\begin{center}
\begin{tabular}{lccc}
\hline
$\eRO$ Data  & $\Delta \FNLL $ &  $\Delta \FNL^{\rm ortho} $ &  $\Delta \FNL^{\rm equil} $ \\
\hline
\hline
Counts	 						& $9\times 10^3$		& $4\times 10^4$ 	& $2\times 10^4$ \\
Counts + Photo-$z$  					& 423				& $2\times 10^3$ 	& $1\times 10^3$\\ 
Angular clustering				 	& 46 					& 461 			& $1.4\times 10^3$\\
Angular clustering + Photo-$z$   		& 10.1 				& 102 			& $1.3\times 10^3$\\
\hline
Counts + Angular clustering 			& 42 					& 317 			& $1.1\times 10^3$\\
Counts + Angular clustering + Photo-$z$  & 8.8				& 36				& 144\\
\hline
Counts + Angular clustering + Planck			& 26 			& 168 			& 740\\
Counts + Angular clustering + Photo-$z$  + Planck 	& 6.9			& 19				& 115\\
\hline
\hline
WMAP7, 95 per cent C.I.		& [-10, +74]		                   & [-410, +6]                   & [-214, +266]\\
\cite{Komatsu:2011} 			& 					& 				&\\
\hline
\end{tabular}
\caption{\label{TAB:FNL} Marginal 1-$\sigma$ constraints enabled by $\eRO$ for different models of primordial non-Gaussianity. 
Assumptions and the survey strategy are as in Table \ref{TAB:ERRORS} with no external priors unless explicitly stated.}
\end{center}
\end{table*}
In terms of late-time observables, the striking difference between these models lies in the scale-dependence of the bias for DM haloes: 
at large separations, the bias scales as $\FNLL\,k^{-2}$ and 
$\FNL^{\rm ortho} k^{-1}$ for the local and the orthogonal types respectively, while it is asymptotically constant (but, for fixed halo mass, numerically different than in the Gaussian case) for the equilateral model. The different shapes of the primordial bispectra imply differences in the skewness of the linear density field and, consequently, different modifications to the Gaussian halo mass function too, via Eqs.~(\ref{EQ:NGMF}) and (\ref{EQ:NGMFLOVERDE}). We adopt the formulae for the skewness from \cite{Taruya:2008} and \cite{Schmidt:2010}, and the improved formulae for the non-Gaussian halo bias by \cite{Desjacques:2011b}: these have shown better agreement with N-body simulations than previous analytical derivations.
However, these expressions are only valid asymptotically as $k\to 0$ and cannot be used at much larger $k$ where the halo bias is basically constant and the provided formulae do not reproduce such behavior.
In order to solve this problem,
we have manually imposed this asymptotic behavior for large $k$ by removing the scale dependencies due to the linear transfer function $T(k)$ at the wavenumbers where the factor $1/T(k)$ starts to dominate.

As detailed in Table \ref{TAB:FNL}, given the less prominent scale-dependence of the halo bias, the constraints on  $\FNL^{\rm ortho} $ and  $\FNL^{\rm equil} $ are weaker than for the local case; the constraints on the other cosmological parameters and those for the $LM$ relation also exhibit a non-negligible modification. Counts and tomography joint constraints read 0.053 and 0.049 for $\SIGMA8$ in the orthogonal and equilateral models, respectively, instead of  0.036 with the local template. $LM$ parameters worsen by 10-20 per cent and 20-40 per cent with the orthogonal and equilateral models, respectively, while the triplet $n_s,~h,~\OB$ improve in both cases by a few per cents.

\section{Additional remarks}
\label{SEC:ADDREMARKS}
In this Section we critically discuss some of the assumptions and the
methods we have been using to derive our main results.
\subsection{Theoretical models for the halo mass function}
\label{SEC:DISC_TMF}
Since the parameters of the mass function are determined from N-body simulations,
they are known with some intrinsic uncertainty 
that we have ignored in our main study.
One possible way of gauging this uncertainty is to introduce
in the Fisher-matrix analysis
a set of nuisance parameters (as many as the mass-function parameters)
with their own  
prior distribution (which embodies the covariance matrix from the
fits to the simulation) and marginalize over them
\citep[see ][for dark-energy surveys]{Wu:2010,Cunha:2010b}.
We have implemented this scheme and 
repeated our calculations using the covariance matrices for the 
parameters $ A^{\rm TMF}_0,  a^{\rm TMF}_0,  b^{\rm TMF}_0$ and $ c^{\rm TMF}_0$ 
kindly made available to us by Jeremy Tinker.
Unfortunately, owing to the procedure used to fit the simulations,
a covariance matrix including also the 3 evolutionary parameters
 $A^{\rm TMF}_z, a^{\rm TMF}_z$ and $\alpha$ is not available. 
Anyway our analysis should at least give an order-of-magnitude estimate of
the  effect. The errors provided by Jeremy Tinker on his $z$=0 mass function parameters at fixed cosmology translate in an overall uncertainty of 1 per cent in the shape of the mass function: it is the variation in Cosmology which is responsible for the 5 per cent accuracy level quoted in \cite{Tinker:2008}.
In order to approximately take into account possible larger uncertainties,
we have rescaled the covariance matrix so that to obtain a marginal uncertainty of about 10 per cent in the abundance of objects with $M\sim 10^{15}\, \HI \MSUN$ at 
$z=0$ before using it as a prior on the mass-function parameters.
We find that, within this setup, the constraints on the parameters both of the cosmology and ICM sector are degraded only by up to 10 per cent when 4 additional mass function parameters are included in the fitting analysis and marginalized over. 
These constraints are destined to further degrade when adding the 3 evolutionary parameters. Moreover, this picture does not take into account possible systematic errors (up to 20 per cent) that 
baryonic physics seems to induce in the halo mass function \citep{Stanek:2009, Cui:2011}.

Imperfect knowledge on the large-scale bias of the DM haloes should have a smaller effect and will not be discussed here \citep[see][]{Wu:2010}.
\subsection{Mass-luminosity relation}
\label{SEC:DISC_LM}
Our forecasts rely on a series of assumptions for converting DM halo masses 
into X-ray photon counts. 

1) We have assumed a power-law $LM$ relation. While there is, as of now, no very
 compelling evidence that this assumption breaks down \citep[see e.g.][]{Sun:2009, Eckmiller:2011OK}, the scatter has been shown to increase for low-mass clusters \citep{Eckmiller:2011OK}.
For this reason
we have introduced a minimum mass threshold and varied it within 
a reasonable range to test the stability of our results.

2) We have assumed that the scatter around the power-law relation is lognormal, but no consensus has been reached regarding the amplitude
and nature of this noise component \citep[see e.g.][]{Vikhlinin:2009b, Pratt:2009, Stanek:2010}. As shown by \cite{Shaw:2010}, an unknown amount of asymmetry in the scatter of the observable-mass relation can affect cosmological constraints in the same manner as additional uncertainties in the fiducial value of the scatter $\SIGMALM$ itself.

3) We have used the current limited knowledge on the $LM$ relation to fix the 
parameters that define our fiducial model. 
For instance, we used a scatter in the $LM$ relation of 40 per cent \citep{Vikhlinin:2009b}. However, more recent observational scaling relations for core-excised clusters 
show a reduced dispersion, although spanning somehow smaller redshift ranges \citep{Mantz:2010b}. We have repeated our analysis adopting the evolving scaling-relations measured by \cite{Mantz:2010b}, converted in the (0.5-2.0) keV band: we find that the number of clusters detected by $\eRO$ increases by more than 30 per cent, with a total population of $1.24 \times 10^5$ objects (more massive than $5\times 10^{13} \,\HI \MSUN$) and a slightly lower median redshift, $z_{\rm median}\sim 0.31$. Consequently, the joint constraints on all the cosmological parameters are improved by 10-50 per cent, with the exception of $\FNLL$ and $\OM$ which show a smaller improvement. On the other hand, the marginal constraints on the $LM$ parameters themselves (but not on the $LM$ scatter) are tighter by a factor of a few. The choice of the fiducial prescription for the $LM$ relation changes more the predictions for the cluster counts than for the clustering measurements, and $\FNLL$ is the parameter which exhibits the smallest dependence on it. 
In any case, also in the light of our (possibly conservative) predictions, one of the major results of this paper is that $\eRO$ and Planck results combined together will be able to constrain the slope and the evolution factor of the $LM$ relation at per cent level (see Table \ref{TAB:ERRORS}).

4) We assumed that the slope and the scatter of the $LM$ relation do not
depend on redshift. This is basically Occam's razor applied to current
data but might not be adequate for $\eRO$ accuracy.

5) We always assumed no prior knowledge for the measurements of the ICM parameters
which, in our study, are fully determined by $\eRO$ data. Adding priors in fact
would require assuming a covariance matrix for the parameters. Results adopting perfect prior knowledge of such parameters are given as a reference in Table \ref{TAB:ERRORS_FIXEDICM}.
\subsection{Temperature-mass relation}
\label{SEC:DISC_TM}
Fig. \ref{FIG:EROSITA} and the discussion of Figs. \ref{FIG:COUNTS_ALL} and \ref{FIG:CLUSTERING} not only show that the photon counts depend very weakly on the X-ray cluster temperature but also that
the exact $TM$ relation for galaxy clusters has a little effect
on our observables. 
We have checked anyway what happens if 4 additional parameters regulating
this scaling relation are included in the analysis of Table \ref{TAB:ERRORS} (10+4 parameter fit).
We found that this leads to a deterioration of the constraints in the cosmology
sector (by an amount ranging from a few per cent to a factor of a few for
$\FNLL$) as the new parameters are poorly constrained by the data.
It is worthwhile mentioning that the deterioration on the cosmological constraints (6 parameters) due to 4 additional temperature--mass scaling-relation parameters is up to three times less severe than the effect of 4 additional parameters regulating the $LM$ relation and its redshift evolution (see Appendix \ref{APP:ERRORS} and Table \ref{TAB:DEGRADATION_SIGMA8_C}).
\subsection{Limber approximation}
\label{SEC:DISC_LIMBER}
The Limber approximation is often used to forecast the constraining power
of the angular power spectrum for tracers of the cosmic large-scale structure.
It is well known, however,  that this approximation, for Gaussian initial conditions, becomes 
less and less accurate when projections
are taken within thinner and thinner redshift bins
\citep{Simon:2007, Loverde:2008b}.
We find that this effect becomes even more important in the presence of primordial non-Gaussianity of the local type or whenever a scale-dependent DM halo bias changes 
the shape of the angular power spectrum on large scales.  
With $\FNLL\sim100$, the Limber approximation deviates from the exact 
calculation by up to 20 per cent for $\ell\sim5$.
Remarkably, the sign of the discrepancy reverts depending on the actual
value of $\FNLL$ \citep[see also][]{Giannantonio:2011}.
Therefore the inaccuracy of the Limber approximation is particularly damaging 
to compute the Fisher matrix for a clustering experiment where 
derivatives of $C_{\ell}$ with respect to the individual parameters must be 
taken.
We have noticed that using the Limber approximation all over the considered multipole range to predict
the performance of a $\eRO$ $C_{\ell}$ study
can underestimate the marginal error on $\FNLL$ by a factor of 4.
\subsection{The mildly non-linear regime}
\label{SEC:DISC_KMAX}
As discussed in Section \ref{SEC:ANGULARCLUSTERING}, the current accuracy to which we model non-linearities both in the matter-density field and in the cluster bias prevents us to trust predictions including information from scales with $k>k_{\rm max} = 0.1 ~h~ {\rm Mpc}^{-1}$. 
We have formally tested that if one could (with more robust models for the mildly non-linear regime) increase this limit to $ k_{\rm max} = 0.3 ~h ~{\rm Mpc}^{-1}$, the joint $\eRO$ constraints (with photometric redshifts) might tighten by 20-50 per cent for all parameters but $\FNLL$.
\subsection{Other issues} 
%Correlation with cosmology, detection threshold, and uniform exposure time and hydrogen column density}
In our analysis, the cosmology and ICM sectors are not fully independent.
In fact, Eq.~(\ref{EQ:VIKHLININLM}) introduces a correlation between the $LM$-evolution parameter and the cosmology sector through the term 
$\GAMMALM~{\rm ln}E(z)$.
On the other hand, in
Section \ref{SEC:DISC_LM}, we have presented forecasts based on the alternative $LM$ relation by \cite{Mantz:2010b}, where the evolution scales as 
$\GAMMALM~{\rm log_{10}}(1+z)$ and does not couple with the cosmological
parameters.
Also $\Omega_{\rm b}$ 
should correlate with the $LM$ relation since the cosmic baryon fraction relates
to the total gas mass of a cluster and thus to its observed $X$-ray brightness.
We postpone to future work the choice of a proper parameterization for
this effect.

Throughout our study, we have assumed a detection limit for the galaxy clusters
of 50 photon counts, assuming the total flux gets detected.
We defer to future work the exploration of different selection criteria, e.g. based on the signal-to-noise ratio.
Moreover, we will address the impact of changing the photon-count limit and compare our findings with the output of the $\eRO$ sub-sample with strong observed
fluxes (say above 1000 photon counts) and thus better mass proxies than $L_X$.

Finally, we have assumed a constant exposure time $T_{\rm exp}$ and a uniform hydrogen column density $N_H$, instead of taking into account the possible variations of these quantities across the survey area. 
A more detailed treatment of the distribution of $T_{\rm exp}$ and $N_H$ may introduce corrections to our forecasts: for example, within Sunyaev--Zel'dovich cluster count measurements, \cite{Khedekar:2010} have shown that the combination of a wide with a deep survey, keeping fixed the total exposure time, gives better constraints than a longer-exposure wide survey only. Further study is required to assess whether this is the case also for X-ray cluster count and clustering experiments, and when fluctuations of $T_{\rm exp}$ and $N_H$ along different lines of sight are not averaged over large patches of the sky.
\section{Conclusions and summary}
We have forecasted the accuracy with which the X-ray telescope $\eRO$
will constrain cosmological parameters 
based on the number counts and spatial distribution of galaxy clusters.
Our reference case considers 10 model parameters
6 of which characterize 
the cosmological model (vanilla, flat, $\Lambda$CDM model) 
while the remaining 4 describe the luminosity--mass relation
for galaxy clusters. 
Special attention is dedicated to the primordial non-Gaussianity parameter 
$\FNLL$.

Galaxy clusters are sorted in terms of the photon counts that will be detected
by the $\eRO$ telescope. We convert the masses of DM haloes (for which
we can predict abundance and clustering properties as a function of the
cosmological parameters) into this observable by taking into account 
observationally-motivated scaling relations, 
the properties of the X-ray detector, and the integration time of the
observations. 
We find that, in an all-sky survey with a typical exposure time of 1.6 ks,
$\eRO$ will observe 9.3$\times10^4$ galaxy clusters
(more massive than $5 \times 10^{13} \HI \MSUN$) 
above the detection limit of 50 photons in the (0.5-2.0) keV band.
Their redshift distribution will be broad, 
extending up to $z \sim 1.5$ with a median of 0.35.

Our forecast is based on measuring 
the abundance of galaxy clusters as a function of the X-ray photon counts
and the corresponding angular power spectra, via a Fisher matrix approach.
We combine the two experiments 
without assuming any prior knowledge of the X-ray scaling-relation 
parameters (the so-called self-calibration technique). 
We distinguish two cases based on whether redshift information on the individual
clusters is available or not.
Finally, we integrate $\eRO$ data with priors on the Hubble parameter and on the mean baryon fraction, and with future constraints from the Planck satellite; we analyze different survey strategies, study the impact of model uncertainties, 
and explore how constraints degrade when extra
nuisance parameters are considered.
Our main findings are reported in Tables \ref{TAB:ERRORS}, \ref{TAB:SURVEY}, \ref{TAB:Z}, and \ref{TAB:FNL}.
They can be summarized as follows.
\begin{itemize}
\item 
Despite the unprecedented size of the $\eRO$ all-sky sample,
without any knowledge of the redshifts of the individual clusters, 
it is not possible to simultaneously improve currents constraints on  
cosmology (with the exception of $\FNLL$)
and on the $LM$ scaling relation.
Redshift information is vital
to break the strong degeneracies among the parameter estimates. 
The availability of
photometric redshifts, with an accuracy better than $0.05 (1+z)$, 
already shrinks the constraints on all parameters by a factor of many times 
(see Table \ref{TAB:ERRORS}).
\item 
Binning the data in X-ray photon counts ($\eta$) already removes 
some degeneracies but,
in general, binning in redshift is more efficient in tightening the errorbars 
on the model parameters, 
although much more expensive in terms of observing time.
Anyway, binning in $\eta$
should be always adopted to achieve the optimal constraints independently 
of the accuracy to which the cluster redshifts are known.
\item 
A tomographic 
study of the angular power spectrum with at least 10 redshift slices 
(i.e. with $\Delta z \lesssim 0.1(1+z)$, where the data are binned also in photon counts) gives exceptional 1-$\sigma$ marginal errorbars on the non-linearity parameter of 
$\Delta \FNLL \lesssim 10$. 
This also provides tighter constraints than any number-count experiment
in (photometric-) redshift bins for all cosmological 
parameters with the exception of $\OM$. 
The combination of 1- and 2-point statistics in tomographic slices
based on photometric redshifts is optimal, 
with $\Delta \FNLL \simeq 9$, $\Delta \SIGMA8 \simeq 0.036$ (4.4 per cent)
and $\Delta \OM \simeq 0.012$ (4.2 per cent)  (see Table \ref{TAB:ERRORS}).

\item 
Also for the $LM$ scaling relation, $C_\ell$ tomography gives tighter constraints than cluster number counts, although this result is reversed with a deeper exposure ($T_{\rm exp}=3$ ks).
The joint analysis of the two $\eRO$ experiments
improves the current errorbars on the slope of the $LM$ relation by a factor 
of 3 (see Table \ref{TAB:ERRORS}).
\item 
Measuring spectroscopic redshifts for the individual clusters 
would further tighten the marginal error on each parameter by 
an additional $\sim 30-40$ per cent wrt to the photo-$z$ case.
This applies to all parameters but $\FNLL$, which exhibits somewhat 
smaller improvements (see Table \ref{TAB:Z}).
\item 
Yet, even the optimal combination of $\eRO$ measurements with redshift information only loosely constrains parameters such as $n_s$, $h$, and $\OB$. Standard priors on $h$ and $\OB$ are necessary to reduce uncertainties for this subset of the parameter space (see Section \ref{SEC:PRIORS}).
\item 
If we could perfectly characterize the $LM$ scaling relation with
prior data, $\eRO$ (with photometric redshifts) 
would constrain the vanilla $\Lambda$CDM 
in line with the future CMB mission Planck, 
with unrivaled constraints on the amplitude of the linear DM power spectrum 
$\SIGMA8$ (to 0.4 per cent accuracy, see Tables \ref{TAB:ERRORS_FIXEDICM} and \ref{TAB:ERRORS_COSMONONG}).
In fact, while accounting for primordial non-Gaussianity does not sensibly degrade the best constraints on $\SIGMA8$ and $\OM$, 
the $LM$ relation is the main source of noise in the measurement of cosmological parameters out of the $\eRO$ cluster sample (see Table \ref{TAB:ERRORS} vs Table \ref{TAB:ERRORS_FIXEDICM}).
\item 
In turn, combining $\eRO$ and Planck data gives sensational 
constraints on both the cosmology and the ICM sectors 
(see Table \ref{TAB:ERRORS}).
Note that $\Delta \FNLL \simeq 7$, $\Delta \SIGMA8 = 0.014$ 
and $\Delta \OM =0.0039$ as well as 
$\Delta \ALPHALM=0.02$ and $\Delta \GAMMALM =0.07$.
\item Other models of primordial non-Gaussianity are analyzed in addition to the local shape: the best constraints (with no priors) read $\Delta \FNL \simeq 9,~ 36,~144$ for the local, orthogonal, and equilateral shape, respectively, and 7,~19,~115 when Planck priors are applied to the other cosmological parameters (see Table \ref{TAB:FNL}).
\item 
Regarding the survey strategy, we find that
the all-sky survey ($f_{\rm sky} = 0.658$, $T_{\rm exp} = 1.6$ ks) is the optimal choice with respect to a possible deeper and smaller survey 
($f_{\rm sky} = 0.15$, $T_{\rm exp} = 7.5$ ks) in terms of parameter estimation (see Table \ref{TAB:SURVEY}).
\item
All the results summarized above are obtained
considering galaxy clusters with mass $M>5 \times 10^{13} \HI \MSUN$ 
(in the fiducial model).
Lowering this threshold would increase the constraining power 
of the experiments in the Fisher-matrix analysis.
This formal improvement, however, would be achievable only if robust knowledge
of the X-ray scaling-relations at low masses were available, which nowadays
is not the case.
\item 
The forecasted error estimates for the cosmological parameters are affected
by the exact form and redshift evolution of the fiducial $LM$ scaling relation. 
Switching from the observationally motivated relation by \cite{Vikhlinin:2009b} (our reference choice) to the one by \cite{Mantz:2010b} shrinks the
forecasted errors by a factor ranging between 10 to 50 per cent, 
except for $\FNLL$ and $\OM$ (whose constraints are barely affected, 
see Section \ref{SEC:DISC_LM}).
This shows that our main results might possibly be conservative.
%Note that the combination of $\eRO$ and Planck data will determine the $LM$
%parameters with unprecedented accuracy (WE ALREADY SAID THIS ABOVE).
%
\item 
It is impossible to constrain the temperature-mass ($TM$) scaling relation on top of our standard 10 parameters (Section \ref{SEC:DISC_TM}).
\item 
The Limber approximation for the calculation of the angular power spectrum is less accurate in the presence of a scale-dependent halo bias induced by primordial
non-Gaussianity than previously established using Gaussian initial conditions.
The erroneous use of the Limber approximation on large angular scales would 
optimistically underestimate the marginal constraints on $\FNLL$ by a factor 
of 4.
\end{itemize}

In conclusion, together with the Planck satellite and upcoming photometric galaxy surveys, $\eRO$ will substantially contribute to the simultaneous determination of the cosmological parameters and of the X-ray luminosity-mass relation. 
In particular, it will shed new light on the physics of the primordial universe by constraining the non-linearity parameter
with remarkable accuracy, and possibly rule out entire classes of inflationary models.
Further studies will focus on the design of dedicated follow-up campaigns to further exploit the $\eRO$ cluster sample.

\section*{Acknowledgments}
We thank Peter Kalberla for providing Galactic hydrogen data tables from
the LAB survey, Frank Haberl for {\it eROSITA} response matrices, and Jan Robrade for eROSITA exposure maps.
We express our gratitude to Alexey Vikhlinin for providing the 
luminosity--mass relation expression with the mass pivot, to Dragan Huterer and Wayne Hu for kindly providing the Planck priors, to Jeremy Tinker for the mass-function parameter covariance matrices, and to Andres Balaguera Antolinez for carefully reading the manuscript.
A.P. thanks Julien Carron for the amusing and useful conversations about Fisher matrices and Tommaso Giannantonio for many 
detailed exchanges. 
A.P. acknowledges support from the Swiss Science National Foundation.
C.P. and T.H.R. acknowledge support from the German Research Association (DFG)
through the Transregional Collaborative Research Center ``The Dark Universe'' 
(TRR33) and (for T.H.R.) through Heisenberg grant RE 1462/5 and grant RE 1462/6.
{\it eROSITA} is funded equally by the DLR and the
Max-Planck-Gesellschaft zur F\"orderung der Wissenschaften (MPG).

\appendix

\section{Error ellipses}
\label{APP:FM}
We show here the joint (marginal) 1-$\sigma$ error ellipses for all the parameter pairs of our 
main analysis. Results from different experiments are color-coded as in 
Fig.~\ref{FIG:FM_SOMEPAIRS}, but differently from there results are given adopting the external priors on $h$ and $\OB$, as detailed in Section \ref{SEC:PRIORS}. Note that the contours obtained 
from the number-count 
experiment with no redshift information exceed the area in the plot in all cases but for the pair $\OB$ and $h$, where they are indistinguishable from the results of the number counts with photometric redshifts.
\begin{figure*}
\BC
\includegraphics[width=17cm]{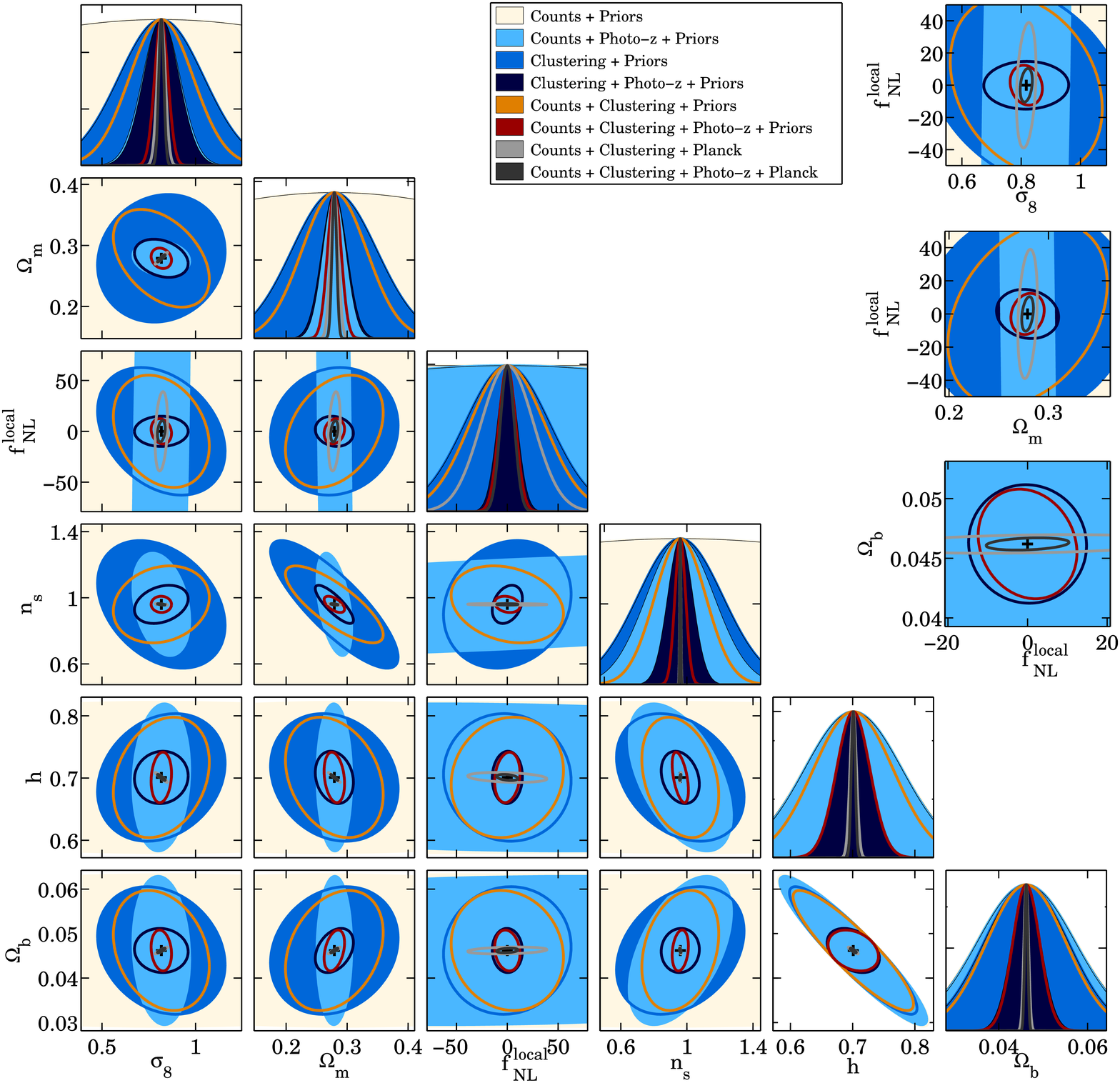}
\caption{\label{FIG:ELLIPSES_COSMO} Joint 1-$\sigma$ error ellipses for the cosmology sector. 
Color coding is as in Fig.~\ref{FIG:FM_SOMEPAIRS}. The three top-right panels are a zoomed-in version of the corresponding panels in the matrix. }
\EC
\end{figure*}
\begin{figure*}
\BC
\includegraphics[width=14cm]{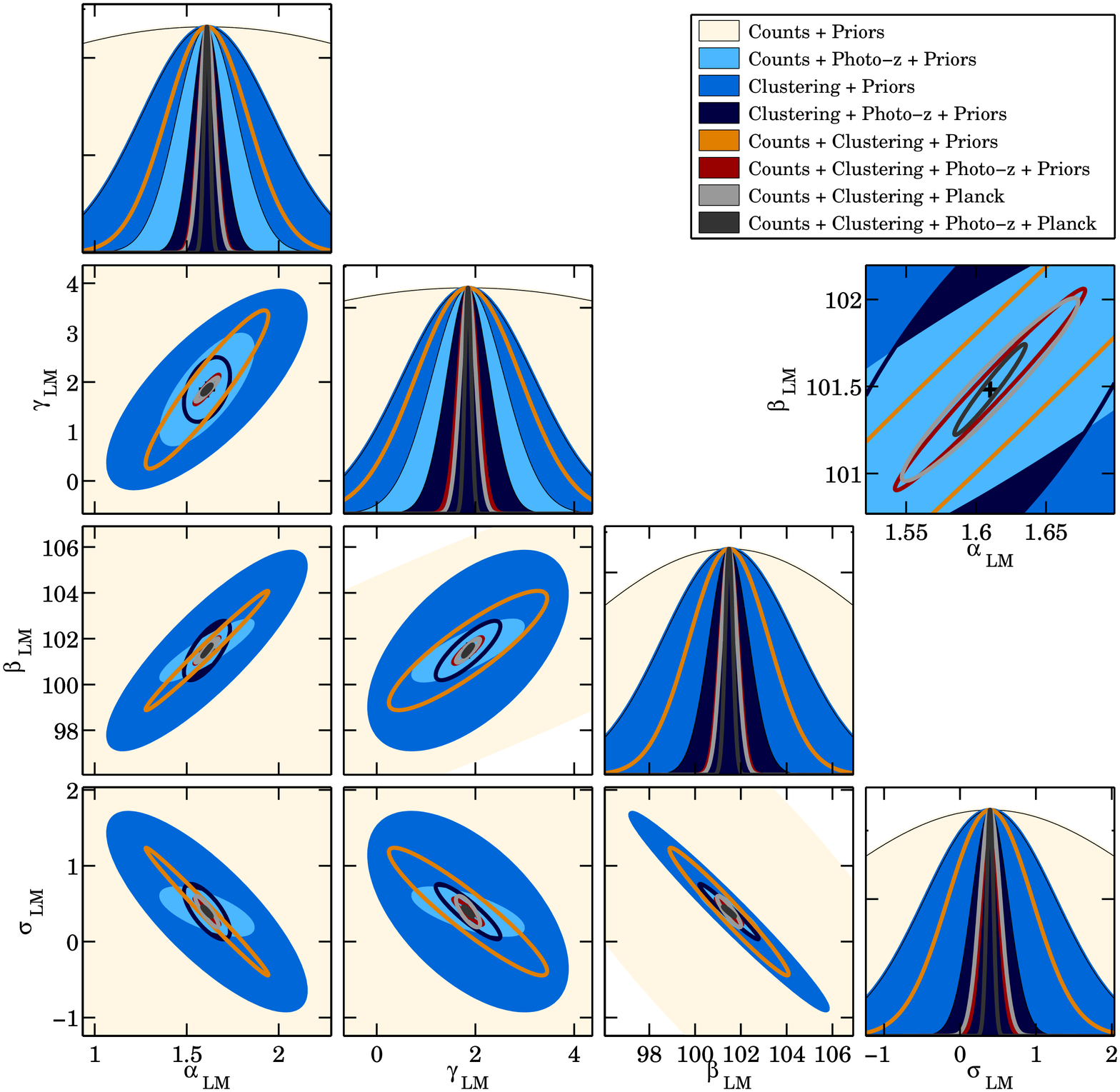}
\caption{\label{FIG:ELLIPSES_NUISA} As in Fig.~\ref{FIG:ELLIPSES_COSMO} but for the parameters of the $LM$ scaling relation. The top-right panel is a zoomed-in version of the corresponding panel in the matrix. }
\EC
\end{figure*}
\begin{figure*}
\BC
\includegraphics[width=17cm]{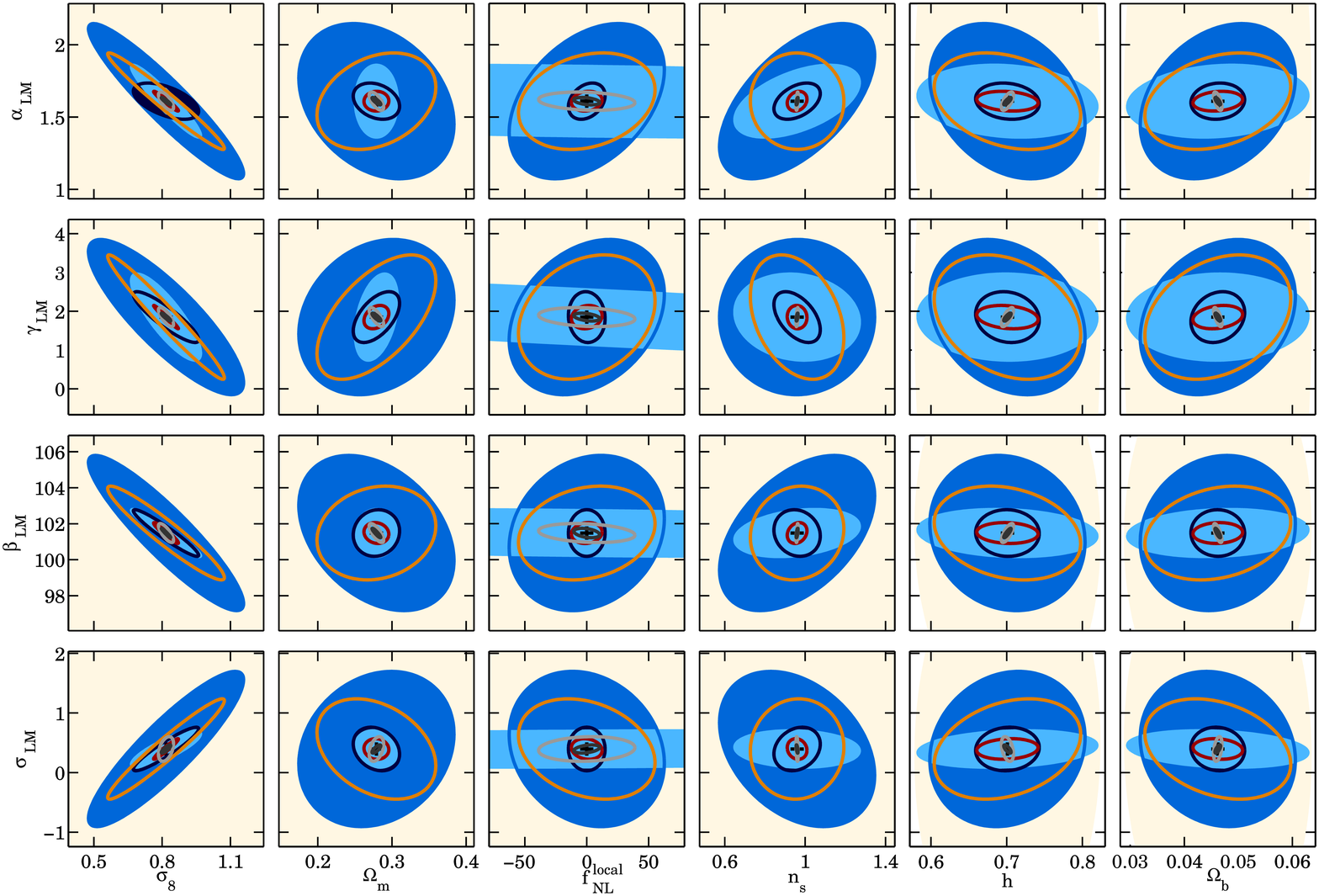}
\caption{\label{FIG:ELLIPSES_INTER} As in Fig.~\ref{FIG:ELLIPSES_COSMO} but for pairs of parameters belonging to different sectors.}
\EC
\end{figure*}

\section{Additional results}
% and degradation of constraints by adding parameters to the analysis}
We report here some additional results which may still be of interest for the community and shall facilitate comparisons with analog calculations in the literature. First of all, 
we show in Table \ref{TAB:ERRORS_FIXEDICM} 
the $\eRO$ constraints on the cosmological parameters that would be obtained
if the 4 parameters in the $LM$ relation were perfectly known before performing
the observations (6-parameter fit).
These are the most optimistic constraints that can be achieved using $\eRO$ 
data. 
Comparing these results with Table \ref{TAB:ERRORS} shows a manifest 
degradation of the results due to the inclusion of the 4 additional
$LM$ parameters:
marginal errors on $\SIGMA8$ and $\OM$ are tighter by a factor of 10 
when the ICM parameters are kept fixed.
This suggests that extracting cluster subsamples for which precise
determinations of the observable-mass relation could be obtained 
will be a convenient strategy.

In order to assess the impact of adding $\FNLL$ to the set of free parameters,
in Table \ref{TAB:ERRORS_COSMONONG} we repeat the analysis by assuming that
both the $LM$ relation and $\FNLL$ are perfectly known in advance 
(5-parameter fit).
This shows that accounting for primordial non-Gaussianity of the local type
does not sensibly 
degrade the constraints on the $\Lambda$CDM parameters, although details
depend somewhat on the individual experiments which are considered. 
Table \ref{TAB:ERRORS_COSMONONG} shows a comparison between $\eRO$ and Planck, if $LM$ and $\FNLL$ parameters were perfectly known and where cosmological results for Planck come from a 5-parameter fit of the power spectrum of the temperature anisotropies, for Gaussian initial conditions.
Finally, in Table \ref{TAB:DEGRADATION_SIGMA8_C} we show how 
increasing the number of free parameters in the fit to the data degrades
the marginal constraints on $\SIGMA8$ from the $\eRO$ cluster number counts. 
In this exercise, the parameters which are not fitted are assumed to be perfectly known. No prior knowledge of any sort is instead considered for the fitted 
parameters.
It is worthwhile mentioning that
adding the four parameters of the $TM$ relation to our ``cosmology-only''
6-parameter fit gives a marginal error on $\SIGMA8$ which is three times
smaller than what is obtained by adding the 4 parameters regulating the 
$LM$ relation and its redshift evolution.  This occurs also for all the other cosmological parameters, although at lower extents.
Once again this shows that  some knowledge of the $LM$ relation is the most important piece of information on the ICM physics
which is necessary to accurately determine the cosmological parameters. 
This holds true also for the measurements of the angular power spectrum.
\label{APP:ERRORS}
\begin{table*}
\begin{center}
\begin{tabular}{lccccccc}
\hline
 $\eRO$ Data & FoM & $\Delta\FNLL$ & $\Delta\SIGMA8$ & $\Delta\OM$ & $\Delta n_s $& $\Delta h$ & $\Delta\OB$\\
\hline
Counts 								& 2.7 	&  $\sim 7\times 10^3$	&0.804 		& .2725		& $\sim 1$ 	&$\sim 2 $	&$\sim .2$  \\
Counts + Photo-$z$  						& 8.2		& 85					&.017 		& .0086  		& .172		&.371		&.0514 \\ 
Angular clustering 						& 6.1 	& 43					&.028		& .0241		& .467		&$\sim 1$		&.0797\\
Angular clustering + Photo-$z$  			& 8.8 	& 9.3 				&.016		& .0101		& .084		&.151		&.0113\\
\hline
Counts + Angular clustering 				& 8.0		& 36					&.016 		& .0099		& .172		&.461		&.0464\\
Counts + Angular clustering + Photo-$z$  	& 11.0 	& 8.4					&.003 		& .0029		& .055		&.110		&.0092\\
\hline
\hline
Current Errors 							& - 		& [-10,+74] 			& .024 		& .0061 		& .012 		&.014 		& .0016 \\
Planck Errors  							& - 		& -					&.024 		& .0071		&.004		&.006		&.0006\\ 
\hline
\end{tabular}
\caption{As in Table \ref{TAB:ERRORS} with no priors
but marginalizing only over the cosmological sector, fixing the ICM parameters. }
\label{TAB:ERRORS_FIXEDICM}
\end{center}
\end{table*}
\begin{table*}
\begin{center}
\begin{tabular}{lrcccccc}
\hline
$\eRO$ Data  & FoM &  $\Delta\SIGMA8$ & $\Delta\OM$ & $\Delta n_s $& $\Delta h$ & $\Delta\OB$\\
\hline
Counts  							& 6.5 		&.196 		& .1213		& $\sim 1$ 	&$\sim 2 $	&$\sim .2$  \\
Counts + Photo-$z$  					& 10.1 		&.003 		& .0031  		& .143		&.364		&.0492 \\ 
Angular clustering 	 				& 7.7			&.025		& .0228		& .369		&.783		&.0688\\
Angular clustering + Photo-$z$ 		& 9.8			&.015		& .0100		& .078		&.138		&.0107\\
\hline
Counts + Angular clustering 			& 9.5			&.015 		& .0094		& .156		&.439		&.0459\\
Counts + Angular clustering + Photo-$z$  & 11.9		&.003 		& .0027		& .050		&.104		&.0090\\
\hline
Counts + Angular clustering + Planck 			& 16.1		&.002 		& .0010		& .002		&.001	&.0002\\
Counts + Angular clustering + Photo-$z$  + Planck	& 16.3		&.001 		& .0008		& .002		&.001	&.0002\\
\hline
\hline
Current Errors  		& - 			& .024 		& .0061 		& .012 		&.014 		& .0016 \\
Planck Errors  		& 14.3 		&.024 		& .0071		&.004		&.006		&.0006\\ 
\hline
\end{tabular}
\caption{As in Table \ref{TAB:ERRORS} with no priors unless explicitly stated but marginalizing only over 5 parameters of the cosmological sector, 
fixing both the ICM parameters and $\FNLL \equiv 0$.}
\label{TAB:ERRORS_COSMONONG}
\end{center}
\end{table*}
\begin{table*}
\begin{center}
\begin{tabular}{lccc}
\hline
 Parameter Set & \# parameters& $\Delta\SIGMA8$ ($\Delta\SIGMA8/\SIGMA8$) & Most correlated parameter\\
\hline
$\SIGMA8 $									&1	& .0005 (0.06 per cent)                   & - \\
$\SIGMA8 + \OM + n_s + h + \OB = \Lambda$CDM 		&5	& .003 (0.4 per cent)                        & $\OM$\\
$\Lambda$CDM + $\FNLL$	 					&6	& .017 (2.1 per cent)                        & $\OM$,$\FNLL$\\
$\Lambda$CDM + LM Sector 						&9	& .113 (14 per cent)                        & $\GAMMALM$\\
\rowcolor[gray]{.85}
$\Lambda$CDM + $\FNLL$ + LM Sector 				&10	& .113 (14 per cent)                         & $\BETALM$\\
$\Lambda$CDM + $\FNLL$ + TM Sector 				&10	& .038 (4.6 per cent)                         & $\OM$\\
%$\Lambda$CDM + $\FNLL$ + LM Sector + TMF Sector	&17	& .231 (28 per cent)		               &$\BETALM$\\
$\Lambda$CDM + $\FNLL$ + LM Sector + TM Sector	&14	& .192 (23 per cent)		               &$\BETALM$\\
\hline
\end{tabular}
\caption{The marginal error on $\SIGMA8$ as determined from the number-count experiment with photometric redshifts is 
shown as a function of the number of free parameters used to fit the $\eRO$ data.
The last column lists the parameter showing the largest linear correlation coefficient with $\SIGMA8$.
Note that $LM$ and $TM$ 
%and TMF 
stand respectively for the luminosity-mass relation and the temperature-mass relation. The shaded row indicates the reference set-up adopted throughout this work.
%and the halo mass function in 
%Eq.~(\ref{EQ:TINKERMFPARAM}).
}
\label{TAB:DEGRADATION_SIGMA8_C}
\end{center}
\end{table*}
%
%
%\bibliographystyle{mn2eFixed}
%\bibliography{Pillepich2011_v0901}

\end{document}